\documentclass[aps,amsmath,amsfonts,amssymb,11pt,nofootinbib,dvips]{revtex4}
\usepackage[dvips]{graphicx,color}
\usepackage{verbatim,rotating,pstricks}
\usepackage[utf8]{inputenc}
\usepackage[dvips]{hyperref}
\hypersetup{pdftitle={Recovering the state sequence of hidden Markov models using mean-field approximations},colorlinks=true,linkcolor=blue,citecolor=blue}

\newcommand{\captionfonts}{\small}

\makeatletter
\long\def\@makecaption#1#2{%
  \vskip\abovecaptionskip
  \sbox\@tempboxa{{\captionfonts #1: #2}}%
  \ifdim \wd\@tempboxa >\hsize
    {\captionfonts #1: #2\par}
  \else
    \hbox to\hsize{\hfil\box\@tempboxa\hfil}%
  \fi
  \vskip\belowcaptionskip}
\makeatother

\def\uS{{\underline{S}}}
\def\uY{{\underline{Y}}}
\def\cS{{\cal S}}
\def\hS{\widehat{S}}
\newcommand{\argmax}[1][]{\arg\!\max_{#1}}

\def\E{\mathbb E}
\def\prob{{\mathbb P}}
\def\var{\text{Var}}

\def\0t{{\tt 0}}
\def\1t{{\tt 1}}
\def\u0t{{\tt \underline{0}}}

\def\v0{\vec{\tt 0}}

\def\uA{\underline{A}}
\def\uY{\underline{Y}}
\def\uX{\underline{X}}

\newcommand{\makePDFTitles}[2]{%
	\pscustom{
		\code{SDict begin [
  			/Title (#1)
  			/Subject (#1)
  			/Creator (Latex)
  			/Author (#2)
  			/Producer (Latex)
  			/Keywords ()
  			/DOCINFO pdfmark
			end}}
}

\begin{document}

\title{Recovering the state sequence of hidden Markov models using mean-field approximations}

\author{Antoine Sinton}
\affiliation{Laboratoire de Physique Th\'eorique de l'Ecole Normale Sup\'erieure,\\
 24 rue Lhomond 75231 Paris Cedex 05, France}

\begin{abstract}
Inferring the sequence of states from observations is one of
the most fundamental problems in Hidden Markov Models. 
In statistical physics language, this problem is equivalent to computing the
marginals of a one-dimensional model with a random external field.
While this task can be accomplished through transfer matrix methods, 
it becomes quickly intractable when the underlying state space is large.

This paper develops several low-complexity approximate algorithms to 
address this inference problem when the state space becomes large.
The new algorithms are based on various mean-field approximations of
the transfer matrix. Their performances are studied in detail on
a simple realistic model for DNA pyrosequencing.
\end{abstract}

\maketitle
%
%
\section{Introduction}

Hidden Markov Models (HMM's) are a workhorse of modern statistics and 
machine learning, with applications ranging from speech recognition
to biological sequence alignment, to pattern classification
\cite{rabiner.tutorial,KBH,Cappe}. 
An HMM defines the joint distribution over a sequence of states
$\uS =(S_1,S_2,\dots,S_i,\dots, S_t)$, $S_i\in\cS$, and observations 
$\uY=(Y_1,Y_2,\dots,Y_i,\dots,
Y_t)$, whereby the states form a Markov chain and the observations 
are conditionally independent given the sequence of states. 
In formulae we have
\begin{eqnarray}
\prob \left[ \uS,\uY \right]= P_1(S_1)\prod_{i=2}^{t}P_i(S_{i}|S_{i-1}) \prod_{i=1}^{t}Q_i(Y_i|S_i)\, .
\end{eqnarray}

The most fundamental algorithmic task related to HMM's is
arguably the problem of inferring the sequence of states 
$(S_1,S_2,\dots, S_t)$ from the observations. The conditional 
distribution of the state sequence given the observations
is, by Bayes theorem,
\begin{eqnarray}
\prob \left[ \uS|\uY \right] =\frac{1}{Z(\uS)}  P_0(S_0)\prod_{i=2}^{t}P_i(S_{i}|S_{i-1})\prod_{i=1}^{t}Q_i(Y_i|S_i)\, ,\label{eq:Conditional}
\end{eqnarray}
where $Z(\uS) = \prob \left[ \uY \right]$ can be thought as a normalization constant.
The state sequence can then be estimated by the sequence of most likely states
(symbol maximum a posteriori probability -MAP- estimation)
\begin{eqnarray}
\hS_i(\uY) = \argmax\left\{\sum_{\{Y_{j}\}_{j\neq i}}\prob \left[ \uS|\uY \right] \right\}\, .
\end{eqnarray}
This reduces the inference problem to the problem of computing marginals of
$\prob \left[ \uS|\uY \right]$.

From a statistical physics point of view \cite{Zuk}, the 
conditional distribution (\ref{eq:Conditional}) can be regarded as the
Boltzmann distribution of a one dimensional system with variables
$S_1,S_2,\dots,S_t$ and energy function
\begin{eqnarray}
E(\uS) =  -\log P_1(S_1)-\sum_{i=2}^{t} \log P_i(S_{i}|S_{i-1}) - \sum_{i=1}^{t} \log Q_i(Y_i|S_i) \, ,
\end{eqnarray}
at temperature $\beta=1$. The sequence of observations thus act 
as a quenched external field. As suggested by this analogy, the 
marginals of $\prob \left[ \uS|\uY \right]$ can be computed efficiently using a
transfer matrix algorithm. In the present context this is also
known as the Bahl-Cocke-Jelinek-Raviv (BCJR) algorithm \cite{bcjr.bcjr}.

The BCJR algorithm has complexity that is linear in the sequence
length and quadratic in the number of states $|\cS|$. 
More precisely, the complexity in $|\cS|$ is the same as 
multiplying an $|\cS|\times|\cS|$ matrix times an
$|\cS|$ vector. While this is easy for simple models with a few states,
it becomes intractable for complex models. A simple mechanism leading
to state space explosion is the presence of memory in the underlying
Markov chain, or the dependence of each observation on multiple
states. In all of these cases, the model can be reduced to
a standard HMM via state space augmentation, but the augmented 
state space becomes exponential in the memory length.
This leads to severe limitations on the tractable memory length.

This paper proposes several new algorithms for addressing this problem.
Our basic intuition is that, when the memory length gets large,
the transfer matrix can be accurately approximated using mean field
ideas. We study the proposed method on a concrete model used in
DNA pyrosequencing. In this case, one is interested in inferring the 
underlying DNA sequence from an absorption signal that carries 
traces of the base type at several positions. The effective memory length
scales roughly as the square root of the sequence length, thus
making plain transfer matrix impractical.

The paper is organized as follows. The next section will define the 
concrete model we study, and Section \ref{section_pyrosequencing} describes the connection with DNA pyrosequencing
to motivate it. Section \ref{section_algorithms} describes the transfer matrix algorithm
and several low complexity approximation schemes. After describing a few bounds in \ref{section_discussion}, numerical
and analytical results are collected in Section \ref{section_results}.\\

%
%

\section{Model and definitions}

\subsection{Definition of the model} \label{section_model}

Consider a sequence of $t$ positive integers $\underline{A} = \{ A_1, \ldots , A_t \}$. Each entry $A_a$ of this sequence
is generated randomly and independently from the others with probability distribution $\beta(A_a)$. This distribution has finite support, i.e. we introduce a positive integer $c$ such that $\beta(x) = 0$ if $x>c$.

This sequence is observed through a non-recursive linear filter, i.e. each observation does not depend on any previous observation. That is, we observe the sequence $\underline{Y} = \{ Y_1, \ldots , Y_t \} \in \mathbb{R}^t$ defined by
\begin{equation}
Y_a = \sum_{i=1}^a \alpha (i,a) A_i +  \eta_{a } \,, \label{eq_y_a_def}
\end{equation}
where $\alpha (i,a)$ is what we call the memory function and $\eta_{a}$ is a Gaussian random variable with mean $0$ and variance $\sigma^2$ which is drawn independently for each position $a$. The memory function also has a finite support. We introduce an integer $n$ which represents the total memory length, i.e. we assume $\alpha(i,a) = 0$ when $a-i \geq n$. Therefore the sum on the right hand side of Eq. (\ref{eq_y_a_def}) effectively starts at $\max (1,a-n)$. There is no restriction on the sign of $\alpha(i,a)$.

\begin{figure}[h!]
\begin{center}
\setlength{\unitlength}{0.25pt}
\begin{picture}(1500,550)(0,0)
\thicklines
\put(200,150){\circle{50}}
\put(400,150){\circle{50}}
\put(600,150){\circle{50}}
\put(800,150){\circle{50}}
\put(1000,150){\circle{50}}
\put(1200,150){\circle{50}}

\thinlines
\put(575,125){\line(0,-1){10}}
\put(575,115){\line(1,0){650}}
\put(1225,125){\line(0,-1){10}}
\put(900,40){\vector(0,1){60}}
\put(850,0){$n+1$}
\put(1190,515){$a$}

\put(175,450){\rule{12.5pt}{12.5pt}}

\put(200,450){\line(-2,-1){100}}
\put(200,450){\line(-4,-3){100}}
\put(200,450){\line(-2,-3){100}}
\put(200,450){\line(0,-1){300}}

\put(375,450){\rule{12.5pt}{12.5pt}}

\put(400,450){\line(-2,-1){300}}
\put(400,450){\line(-4,-3){300}}
\put(400,450){\line(-2,-3){200}}
\put(400,450){\line(0,-1){300}}

\put(575,450){\rule{12.5pt}{12.5pt}}

\put(600,450){\line(-2,-1){500}}
\put(600,450){\line(-4,-3){400}}
\put(600,450){\line(-2,-3){200}}
\put(600,450){\line(0,-1){300}}

\put(775,450){\rule{12.5pt}{12.5pt}}

\put(800,450){\line(-2,-1){600}}
\put(800,450){\line(-4,-3){400}}
\put(800,450){\line(-2,-3){200}}
\put(800,450){\line(0,-1){300}}

\put(975,450){\rule{12.5pt}{12.5pt}}

\put(1000,450){\line(-2,-1){600}}
\put(1000,450){\line(-4,-3){400}}
\put(1000,450){\line(-2,-3){200}}
\put(1000,450){\line(0,-1){300}}

\put(1175,450){\rule{12.5pt}{12.5pt}}

\thicklines
\put(1200,450){\line(-2,-1){600}}
\put(1200,450){\line(-4,-3){400}}
\put(1200,450){\line(-2,-3){200}}
\put(1200,450){\line(0,-1){300}}
\thinlines

\put(800,150){\line(2,1){500}}
\put(1000,150){\line(4,3){300}}
\put(1200,150){\line(2,3){100}}

\put(1000,150){\line(2,1){300}}
\put(1200,150){\line(4,3){100}}

\put(1200,150){\line(2,1){100}}

\put(100,150){\ldots}
\put(100,475){\ldots}
\put(1350,150){\ldots}
\put(1350,475){\ldots}
\end{picture}
\caption{Portion of a factor graph representation of the relationship between the original sequence $\underline{A}$ (circles) and the sequence $\uY$ (squares) and where $n=3$.}
\label{fig_factor_graph}
\end{center}
\end{figure}
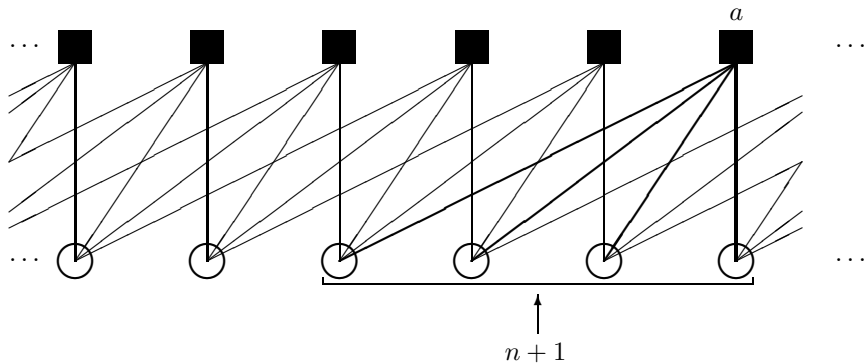

The relationship between the sequences $\underline{A}$ and $\underline{Y}$ can be described by a factor graph representation. This is a bipartite graph including one \emph{function node} for every $Y_a$ in $\underline{Y}$ and one \emph{variable node} for every $A_a$ in $\underline{A}$. Except for the first $n$, every function node is connected to exactly $n+1$ variable nodes by as many edges. A schematical representation is presented in Fig. \ref{fig_factor_graph}.

Our goal is to develop an efficient algorithm to recover the sequence $\uA = \{ A_1, \ldots , A_t \}$ from the observed noisy sequence $\uY = \{Y_a, \ldots , Y_t \}$.\\

The sequence $\uA$ is a chain of i.i.d. variables therefore it can be regarded as zero-order Markov chain. The output $Y_a$ is observed from the underlying state $A_a$ as well as the all the states preceding $A_a$. This problem can therefore be thought of as a variable length higher-order hidden Markov model (HMM) where the hidden states are i.i.d. and the observations depend on all previous hidden states (\cite{rabiner.tutorial,lee_lee,wang}).

We will denote by $\uX = \{x_1, \ldots, x_t \} \in [1,c]^n$ an estimate of $\uA$. Using Bayes rule, the posterior probability of $\underline{X}$ knowing $\underline{Y}$ is
\begin{equation}
\mathbb{P} \left[ \underline{X} | \underline{Y}  \right] = \frac{ \prob \left[ \underline{Y} | \underline{X} \right] \mathbb{P} \left[ \underline{X} \right]}{\mathbb{P} \left[ \underline{Y} \right]} \,. \label{eq_posterior_prob}
\end{equation}
We use the maximum a posteriori (MAP) method \cite{mackay.information_theory} to produce the estimation.
The probability $\mathbb{P} \left[ \underline{Y} | \underline{X}  \right]$ is known as the \emph{likelihood function} and takes the form
\begin{equation}
\prob \left[ \underline{Y} | \underline{X} \right] = \prod_{a=1}^t \Psi_a (x_{a-n}, \ldots ,   x_a ) \,, \label{eq_likelihood_function}
\end{equation}
where, if the index of $x_i$ is such that $i \leq 0$ then we set $x_i = 0$. This sets the boundary condition for small $a$. For $a > t$, the variables $x_a$ are not present in any equation and are considered free. The probability $\Psi_a (x_{a-n}, \ldots ,   x_a )$ in Eq. (\ref{eq_likelihood_function}) is the density of $\eta_a$ defined in Eq. (\ref{eq_y_a_def}) and is written
\begin{equation}
\Psi_a (x_{a-n}, \ldots ,   x_a ) = \prob \left[ Y_a | \{ x_{a-n}, \ldots , x_a \} \right] = \frac{1}{\sqrt{2 \pi \sigma^2}} \exp \left[ - \frac{1}{2 \sigma^2} \left(   Y_a -  \sum_{i=a-n}^a \alpha(i,a) x_i \right)^2 \right] \, , \label{eq_psi}
\end{equation}
where $\sigma^2$ is the variance of $\eta_{a}$. The other terms in Eq. (\ref{eq_posterior_prob}) are
\begin{equation}
\prob \left[ \underline{X} \right] = \prod_{a=1}^t \beta (x_a) \,, 
\end{equation}
which is the prior distribution, and
\begin{equation}
\prob \left[ \underline{Y}  \right]  = \sum_{\underline{X}}  \prob \left[ \underline{X} \right]  \prod_{a=1}^t \Psi_a (x_{a-n}, \ldots ,   x_a )  = \mathcal{N} \,,
\end{equation}
which is a normalization constant. In fine, we construct the marginal distribution 
\begin{equation}
\nu (x_a) = \sum_{ \{ x_b / b \neq a \}} \mathbb{P} \left[ \underline{X} | \underline{Y}  \right] \,,
\end{equation}
which yields the decoded sequence as
\begin{equation}
x_a^*  = \argmax[x_a] \left\{ \nu ( x_a ) \right\} \,.
\end{equation}
which is the maximum likelihood estimate. We use here a symbol MAP decoding, with the hope of minimizing the error for each single $x_a$, instead of a block MAP decoding which would be to minimize the error over the sequence as a whole.

The direct computation using this method entails a summation over $\Theta (c^t)$ terms which rapidly becomes unpractical when $c$ and/or $t$ grow. 
In section \ref{section_algorithms} we introduce four separate algorithms with various levels of approximation to overcome this limitation.\\

\subsection{Specific forms of the prior and memory functions} \label{section_approaches}

In this section we give some details about the sequences we use in our numerical simulations. To generate the integer sequence $\uA$ we use several different probability distributions $\beta$. The details of these will be described in subsection \ref{subsection_beta_distributions}. There are also several $\alpha$ functions we will be using, these are described in subsection \ref{subsection_alpha_functions}.

\subsubsection{The distribution $\beta$} \label{subsection_beta_distributions}

Consider a sequence of i.i.d. Bernoulli variables taken in $\{0, 1\}$ with success probability $q$. We then construct the integer sequence $\uA$ by counting the number of repetitions of $0$s or $1$s in this Bernoulli sequence. For instance, if the Bernoulli sequence is $000101100011111$ this will correspond to $\uA = \{ 311235 \}$. We can generate this sequence directly using the distribution
\begin{equation}
\beta_g (l) = q (1-q)^{l-1} ~~~~~~ ,\forall l \in \mathbb{N}\,. \label{equation_prob_a_priori}
\end{equation}
 We will give more details on this particular distribution in section \ref{section_pyrosequencing}. The sequence $\uY$ is then generated using Eq. (\ref{eq_y_a_def}).
The distribution $\beta_g$ does not admit a finite support but it decays rapidly as $l \to \infty$, thus we can still introduce an effective cutoff parameter $c$.

An alternate approach is to truncate the distribution at large enough $c$. The corresponding \emph{truncated geometrical} distribution is then defined as
\begin{eqnarray}
\beta_t (l) &=& \gamma_t \, q (1-q)^{l-1} ~~~~~~~~~~~\textrm{ if } l \in [1,c] \, , \nonumber \\
&=&0 ~~~~~~~~~~~~~~~~~~~~~~~~~~\textrm{ otherwise}, \label{eq_beta_t}
\end{eqnarray}
where $\gamma_t^{-1}= \beta_g (l \leq c) =  \sum_{ l=1}^c  q(1-q)^{l-1} = 1 -(1-q)^{c} $ is very close to one if $c$ is large enough. This expression gives us a behavior similar to the one of $\beta_g$ while enabling us to use the same distribution for coding and decoding.

The uniform distribution, called $\beta_u$, is 
\begin{eqnarray}
\beta_u (l) &=& \frac{1}{c}  \textrm{~~~~~if } 1 \leq l  \leq c \,, \nonumber \\
&=& 0 \textrm{~~~~~~otherwise}. \label{eq_beta_u}
\end{eqnarray} \\

\subsubsection{The function $\alpha$} \label{subsection_alpha_functions}

Here we will give the expressions we use in our different test cases for the function $\alpha$ defined in Eq. (\ref{eq_y_a_def}).\\

The first expression we use for the function $\alpha$ is
\begin{equation}
\alpha_r(i,a) = { a - \left\lfloor \frac{a+1-i}{2} \right\rfloor \choose \left\lfloor \frac{a-i}{2} \right\rfloor}  (1-p)^{\left\lfloor \frac{a+1-i}{2} \right\rfloor} p^i \,,
\label{eq_real_alpha}
\end{equation}
if there is a $k \in \mathbb{N}$ such that $i = a - 2k$, otherwise $\alpha_r(i,a)=0$. Furthermore, we have $p \in [0,1]$. The derivation of this expression, as was the expression of $\beta_g$ in the previous section, will be described in more detail in section \ref{section_pyrosequencing}. The value of $p$ effectively tunes the memory length, i.e. $n \simeq t p$.

We define two other functions that will be useful for test cases, as were Eqs. (\ref{eq_beta_t}) and (\ref{eq_beta_u}),
\begin{eqnarray}
 \alpha_f (i,a) & = & 1 \,,  \label{eq_alpha_flat}\\
\alpha_h (i,a) & = & \frac{1}{a-i+2} \,, \label{eq_alpha_hyper}
\end{eqnarray}
for $(a-i) \leq n$ and are equal to zero otherwise. These functions are useful as they are non-zero in the range $[a-n-1,a]$ only and thus we have a better control over the precision of our approximation since we decide of the value of $n$ for Eq. (\ref{eq_y_a_def}). 

When there is no subscript to $\alpha$ it can take any of these three values.\\

\section{The Pyrosequencing example} \label{section_pyrosequencing}

The original inspiration of this work comes from the Sequence-by-Synthesis technique, called pyrosequencing, originally introduced in \cite{ronaghi_orig} and described in \cite{eltoukhy_gamal,elahi_ronaghi,ronaghi,ronaghi2,ewing_green} with a full review in \cite{nyren.history_pyroseq} which gives a concrete and precise description of the method and its history. It is one way of sequencing DNA strands by a repeated set of chemical tests. Since its introduction, it has become an industry standard for low cost high efficiency sequencing. This paper being, in the end, not directly connected to the technique we will simply give an introduction to its fundamentals. A more detailed review of the usage of our approach on experimental pyrosequencing data is planned as a future publication.

\begin{figure}[h!]
\includegraphics[width=0.5\linewidth]{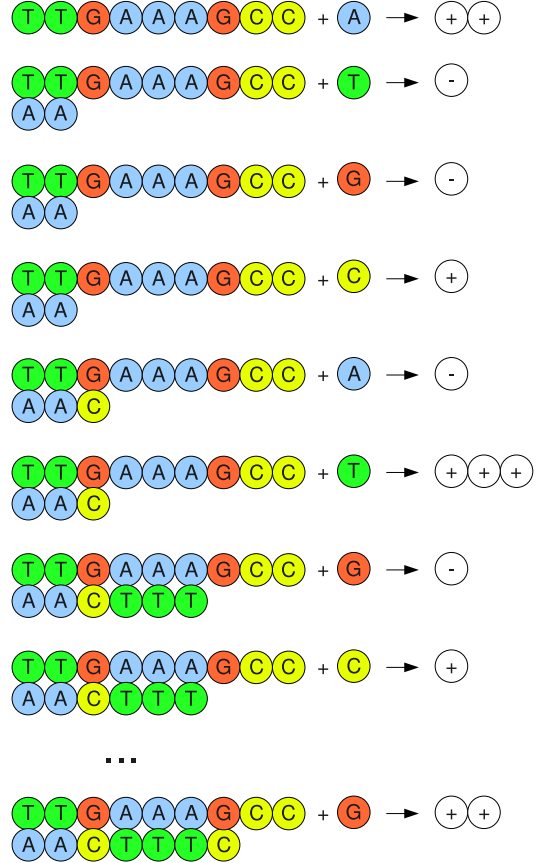}
\caption{Example of pyrosequencing test cycles to decode sequence $TTGAAAGCC$. The cycles follow the order  $A \to T \to C \to G$. A ``$-$'' sign means no reaction occurred, every sequence of ``$+$'' signs means a reaction occurred and its amplitude was multiplied by as many times.}
\label{fig_pyrosequencing}
\end{figure}

 In pyrosequencing, the initial solution contains many copies of the same strand, which is called the base sequence, and a set of enzymes that will catalyze and react with the by product of the main reaction to emit light. Tests are done by incorporating a repeated cycle of the $4$ base type nucleotides into the solution. When a nucleotide is introduced it will react with the DNA sequences if the first available position of the base sequence is its complementary base (that is bases $A$ and $T$ on one side and bases $G$ and $C$ on the other). The reaction will happen for as long as this same base is repeated. Furthermore, the reaction will produce a readable response (a pulse of light) that is proportional to the total number of repetitions that were encountered, this is called a homopolymeric (HP) sub-sequence. Finally, all positions on the DNA strands that reacted will now be obstructed to subsequent tests and thus freeing the next available base in the base sequence for reaction. As an example, Fig. \ref{fig_pyrosequencing} shows a series of cycles applied to the sequence $TTGAAAGCC$.

When a test is positive, the chemical reaction is incomplete, that means only a fraction of all the strand copies react. We call the average of this fraction the incorporation rate $p \in [ 0,1 ]$. Furthermore, it means there is a fraction $1-p$ of all the copies that did not react to this test and that a fraction $p$ of this fraction will react only at the next cycle, and so on for each test. Finally, it results that the responses will depend on this incomplete incorporation and which dependency can be simulated by the use of a memory function. In full rigour, there is an additional parameter called the non-specific incorporation rate \cite{eltoukhy_gamal}. It measures the average fraction of strands that react when the test is {\it negative}. We do not take this element into account in our model since its value is usually very small.

An original mathematical description was introduced in \cite{svantesson.math_pyroseq} as a biochemical model. Using kinetic considerations, it investigates the differential equations describing the single pulse due to a single incorporation as well as a succession of pulses linked to as many incorporations. The model that is developed gives an effective description of pyrosequencing without approximation. 

If we denote by $\uA = \{ A_1 , \ldots, A_t \}$ the HP sub-sequences and $\uY = \{ Y_1 , \ldots , Y_t \}$ the response sequence, then in fact this work can be considered a slight adaptation of pyrosequencing in which we consider base sequences to be made up of only two base types. This model being binary, we naturally call them $0$ and $1$. This also explains the inspiration for the definition of the distribution $\beta_g$ in Eq. (\ref{equation_prob_a_priori}) if we take $q=\frac{1}{2}$. In this case, we usually take the value of this cutoff parameter to be $c=15$ since the probability for $l$ being bigger than this value of $c$ is $P(l > c) \simeq 3.05 \times 10^{-5}$.

\begin{figure}[h!]
\includegraphics[width=0.5\linewidth]{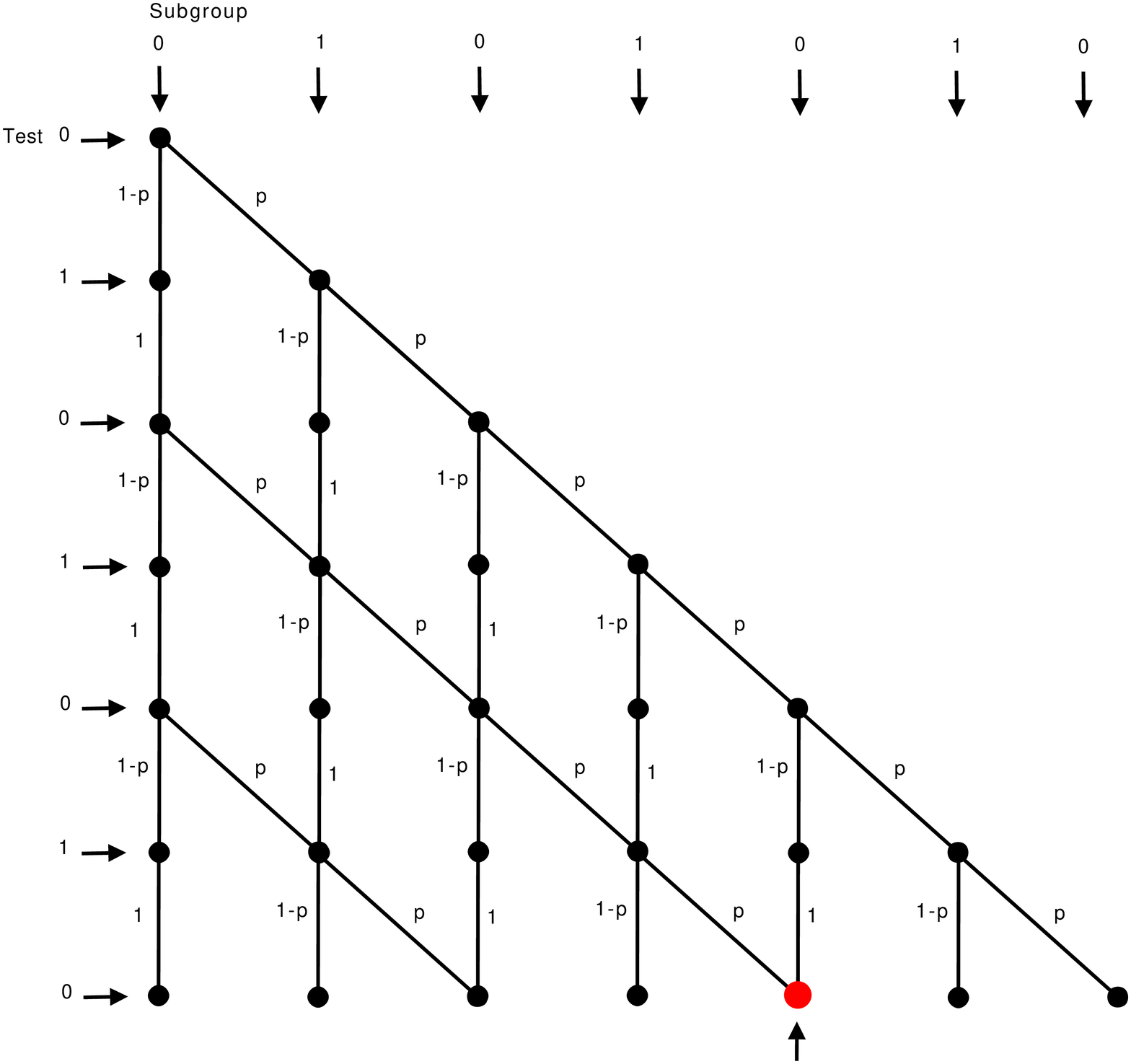}
\put(-100,-10){$\vdots$}
\put(-80,5){$\alpha(4,6)$}
\caption{Graph showing evolution of subgroups in response to tests when the first base in the sequence is a $0$ and this is also the first test performed.}
\label{fig_pyro_diag}
\end{figure}

We keep the same definition for the incorporation rate $p$. The sequence of incomplete reactions can thus be plotted onto a directed acyclic graph (DAG) which can look like Fig. \ref{fig_pyro_diag} (when the first tested base is represented by $0$).

For each test in our two base sequence, we have a fraction $p$ which reacts and presents the other base in the next HP sub-sequence for the following test and a fraction $1-p$ which does not react and thus will not react with the subsequent test. We then count the vertices on the graph to obtain $\alpha$. For instance, in Fig. \ref{fig_pyro_diag} we have singled out the position of $\alpha(4,6)$. Its value is obtained by adding the lengths of all the direct paths that lead from the starting point to its position. That is, there are $5$ possible paths and they all have the same length of $(1-p) p^6$ and therefore $\alpha(4,6) = 5(1-p) p^6$. By generalizing this to any position we obtain $\alpha_r(i,a)$ expressed in Eq. (\ref{eq_real_alpha}) and by adding noise the pyrosequencing equivalent response $Y_a$ of test $a$ is expressed in Eq. (\ref{eq_y_a_def}).

\begin{figure}[h!]
\includegraphics[width=0.45\linewidth]{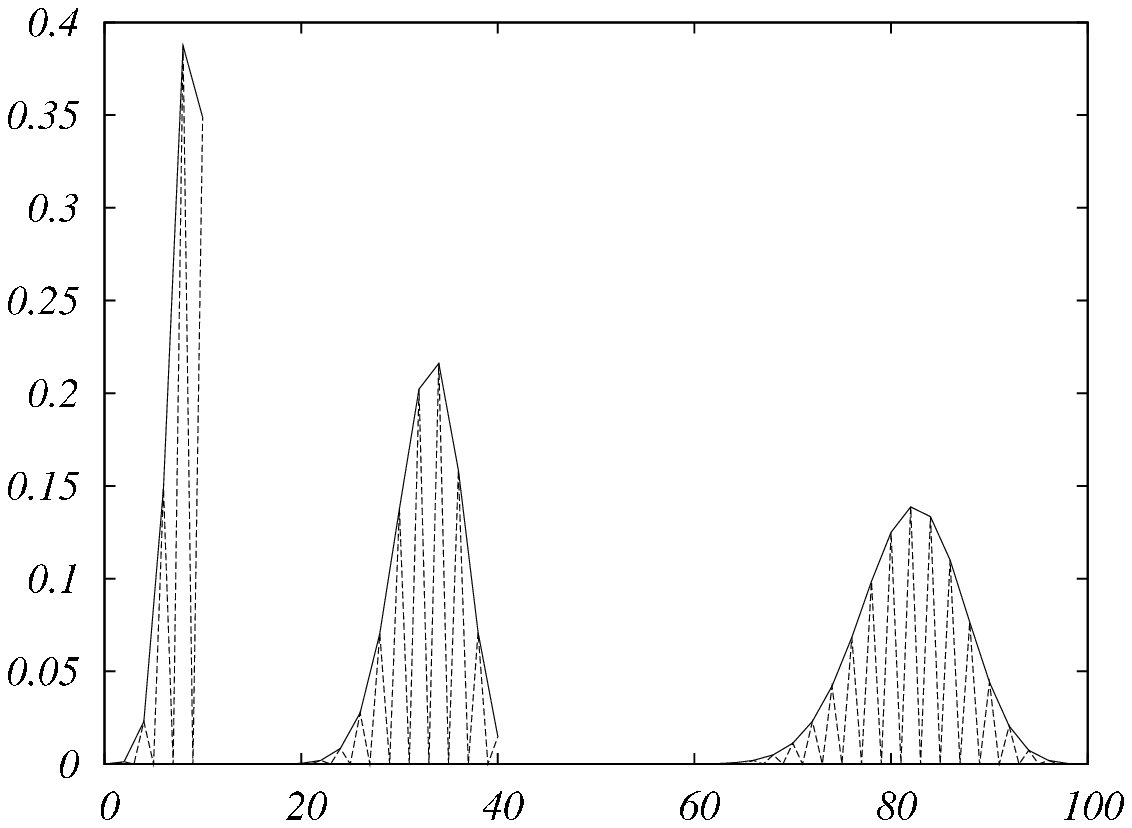}
\includegraphics[width=0.45\linewidth]{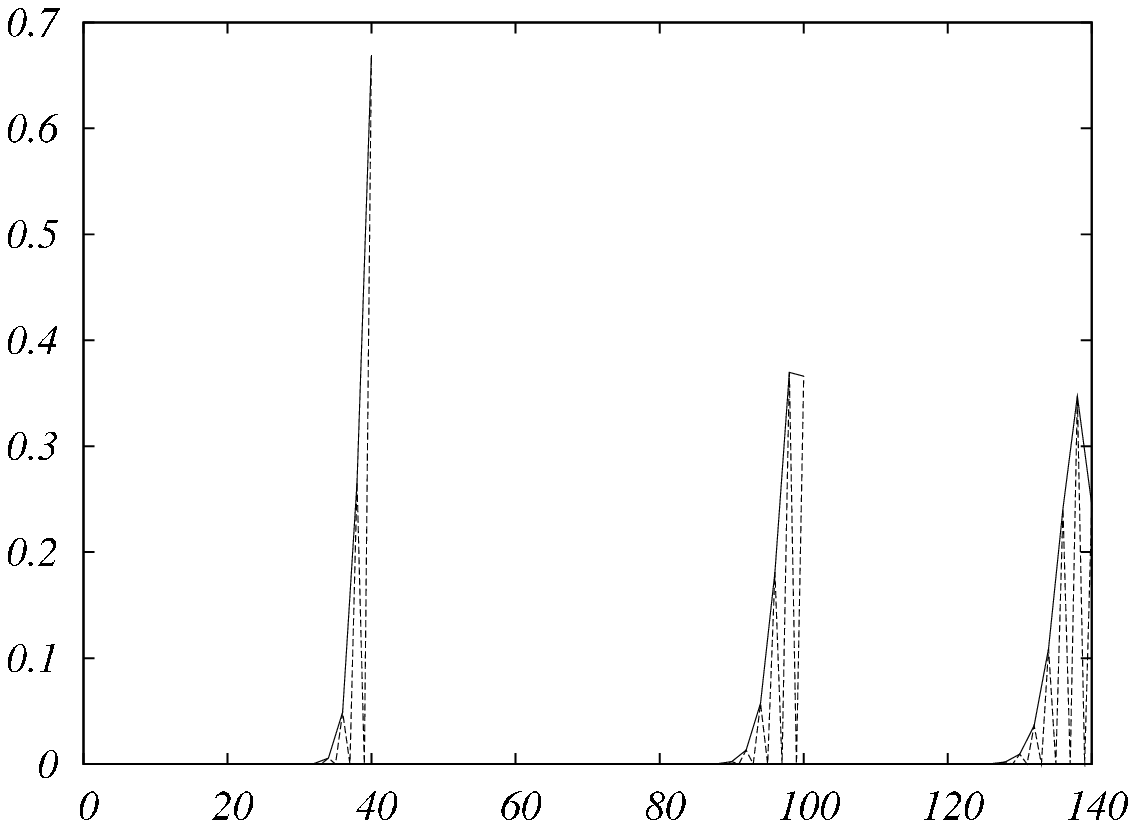}
\put(-100,-10){$i$}
\put(-315,-10){$i$}
\put(-440,100){\rotatebox{90}{$\alpha(i,a)$}}
\caption{On the left: $\alpha_r(i,a)$ for $a=10$, $a=40$ and $a=100$ (from left to right), and $p=0.9$. On the right: $\alpha_r(i,a)$ for $a=40$, $a=100$ and $a=140$ (from left to right) and $p=0.99$. In both figures: the solid lines are the envelopes of $\alpha_r$ and the dotted lines show the alternating behavior of $\alpha_r$ between zero and non-zero values.}
\label{fig_alpha}
\end{figure}

The value of the memory $n$ introduced in section \ref{section_model} is linked to the graph of this function $\alpha_r$. We define $n$ as the smallest integer so as to keep a certain percentage of the total weight of $\alpha_r$ between $t-n$ and $t$. The weight being here the sum of all values of $\alpha_r(i,t)$ for $i \in [1,t]$. In practice, we keep at minimum $99\%$ of the total weight. Furthermore, as seen in Fig. \ref{fig_alpha}, the graph of $\alpha_r$ is such that we assume for all $a$ and all $i < i_n = a - n - 1$ that we have $\alpha_r(i,a) = 0$, indeed, as $a$ grows, the graph of $\alpha_r$ widens and thus $n$ is defined for the worst possible case. It is also shown in Fig. \ref{fig_alpha} how the peak widens as $p$ decreases.\\

%
%
\section{Algorithms} \label{section_algorithms}

In section \ref{section_model} we introduced the general model which we adopt. In this section 
we will introduce four different algorithms to estimate the sequence while achieving the best compromise between precision and complexity.

\subsection{Transfer Matrix algorithm} \label{subsection_algo_tm}

The first algorithm we introduce will be referred to as the \emph{transfer matrix algorithm} (TM). \\

This method relies on the combination of two iterative expressions of the position $a$, one for each direction forward and backward, respectively 
$Z^f_a$ and $Z^b_a$, and which we will refer to as the constrained partition functions. Both are indexed by the sequence $\{ x_{a-n+1}, \ldots , x_a \} \in [1,c]^n$ such that
\begin{eqnarray}
 Z_{a}^f (x_{a-n+1}, \ldots , x_{a} ) & = & \beta (x_a) \sum_{x_{a-n}} Z_{a-1}^f (x_{a-n}, \ldots, x_{a-1} ) \Psi_{a} ( x_{a-n} , \ldots , x_a ) \, , \label{eq_TM_forwards}\\
Z_{a}^b (x_{a-n+1}, \ldots , x_{a} ) &=&  \sum_{x_{a+1}} \beta (x_{a+1}) Z_{a+1}^b (x_{a-n+2}, \ldots, x_{a+1} ) \Psi_{a+1} ( x_{a-n+1} , \ldots , x_{a+1} ) \, . \label{eq_TM_backwards}
\end{eqnarray}
These two functions can then be combined to write the exact marginal of $\{x_{a-n+1}, \ldots ,   x_a \}$ with respect to the distribution (\ref{eq_posterior_prob}) as
\begin{equation}
\nu ( x_{a-n+1}, \ldots , x_{a} ) = \frac{Z_{a}^f (x_{a-n+1}, \ldots , x_{a}  ) Z_{a}^b (x_{a-n+1}, \ldots , x_{a})}{\sum_{\hat{x}_{a-n+1}, \ldots , \hat{x}_{a}} Z_{a}^f (\hat{x}_{a-n+1}, \ldots , \hat{x}_{a}  ) Z_{a}^b (\hat{x}_{a-n+1}, \ldots , \hat{x}_{a}) } \,.
\end{equation}
Therefore, the marginal distribution of variable $x_a$ again with respect to the probability distribution defined in Eq. (\ref{eq_posterior_prob}) is then
\begin{equation}
\nu (x_a)  =  \frac{1}{\mathcal{N}} \sum_{x_{a-n+1}, \ldots , x_{a-1}} Z_{a}^f (x_{a-n+1}, \ldots , x_{a}  ) Z_{a}^b (x_{a-n+1}, \ldots , x_{a}  ) \,,
\end{equation}
where $\mathcal{N}$ is a normalization constant. 

This algorithm corresponds to a reordering of the model defined in section \ref{section_model}. It has complexity of order $\Theta (c^n)$ which is huge in most of the regimes we are interested in. Because of this, we introduce in subsequent sections a set of approximations to reduce this complexity.

\subsection{First order approximation} \label{subsection_algo_mc}

In this section we introduce two algorithms that emerge from the same approximation to the TM algorithm of section \ref{subsection_algo_tm}. They rely on a first order expansion of the constrained partition functions.\\

We assume the constrained partition functions $Z_{a}^* (x_{a-n+1}, \ldots , x_{a} )$ (where the $*$ is either $f$ or $b$) defined in Eqs. (\ref{eq_TM_forwards}) and (\ref{eq_TM_backwards}) factorize approximately
\begin{equation}
Z_a^* (x_{a-n+1}, \ldots , x_{a} ) \simeq \prod_{i=a-n+1}^{a} z_{a,i}^* ( x_i) \,, \label{eq_first_order_approx} 
\end{equation}
where the $ z_{a,i}^* $ are functions of a single variable. The new iterative procedures are one variable nationalizations of the functions in Eqs. (\ref{eq_TM_forwards}) and (\ref{eq_TM_backwards}) and where the functions $Z^*_a$ on the right hand side are replaced by the approximation in Eq. (\ref{eq_first_order_approx}). These can be written, for $i \in [a-n+1,a]$, as
\begin{eqnarray}
 z^f_{a,i} (x_{i}) &= & \frac{z^f_{a-1,i}({x}_i)}{\mathcal{N}^f} \sum_{ R^f_{a,i} } \beta (\hat{x}_{a})  \prod_{ \substack { j=a-n \\ j \neq i} }^{a-1} z^f_{a-1,j}(\hat{x}_j) ~ \Psi_{a} (\hat{x}_{a-n}, \ldots , x_{i} , \ldots , \hat{x}_{a}) \,, \label{eq_first_order_forwards}\\
 z^b_{a,i} (x_{i}) &= & \frac{z^b_{a+1,i}({x}_i)}{\mathcal{N}^b} \sum_{R^b_{a,i} }  \beta (\hat{x}_{a+1})  \prod_{ \substack{  j=a-n+2 \\ j \neq i}}^{a+1} z^b_{a+1,j}(\hat{x}_j) ~ \Psi_{a+1} (\hat{x}_{a-n+1} , \ldots , x_i , \ldots , \hat{x}_{a+1}) \,, \label{eq_first_order_backwards} 
\end{eqnarray}
where the $\mathcal{N}^*$ are normalization constants and where $R^f_{a,i} \equiv [1,c]^n$ (resp. $R^b_{a,i} \equiv [1,c]^n$) is the set of all possible values of $\{ \hat{x}_{a-n}, \ldots , \hat{x}_{i-1}, \hat{x}_{i+1}, \ldots , \hat{x}_{a} \}$ (resp. $\{ \hat{x}_{a-n+1}, \ldots, \hat{x}_{i-1}, \hat{x}_{i+1} , \ldots , \hat{x}_{a+1} \}$). Furthermore, since they were not estimated in any previous step, $z^f_{a-1,a}({x}_{a})$ and $z^b_{a+1,a-n+1}({x}_{a-n+1})$ are both set to $1$ prior to the computation of $z^f_{a,a}({x}_{a})$ and $z^b_{a,a-n+1}({x}_{a-n+1})$ which are the first values to be computed at step $a$ in their respective directions. These values are then reinjected into the subsequent calculations at step $a$ by setting $z^f_{a-1,a}({x}_{a})=z^f_{a,a}({x}_{a})$ and $z^b_{a+1,a-n+1}({x}_{a-n+1})=z^b_{a,a-n+1}({x}_{a-n+1})$. To initiate the backwards iteration, we define $z^b_{a,a} (x_{a}) = \frac{1}{c}$ for $a \in \{ t-n, \ldots, t \}$ to account for the free boundary conditions.

Finally, we have an approximation of the marginal of $x_a$ as
\begin{equation}
 \nu (x_a) = \frac{1}{\mathcal{N}} ~ z^{f}_{a,a} (x_a) ~ z_{a,a}^b (x_{a}), \label{eq_nu_1st}
\end{equation}
where $\mathcal{N}$ is a normalization constant. \\

The sums in Eqs. (\ref{eq_first_order_forwards}) and (\ref{eq_first_order_backwards}) are over $c^n$ terms and thus, this algorithm, as such, has no benefit over the TM algorithm described in section \ref{subsection_algo_tm}. We therefore introduce another couple of approximations to this algorithm. When possible, though, we will wish to compare these approximations to this algorithm which we will refer to as TM.1A.

\subsubsection{First order approximation with Monte Carlo}  \label{subsubsection_mc}

The first of these two approaches will be called \emph{first order approximation with Monte Carlo} (TM.1A.MC) since it relies on random sampling from iteratively computed distributions.\\

At step $a$ of the procedure described in (\ref{eq_first_order_forwards}), we start by estimating $z^f_{a,a}({x}_{a})$ knowing that we have computed the values of $z^f_{b,i} (x_{i})$ for all $b<a$ and in particular the values of $z^f_{a-1,i} (x_{i})$ which are probability distributions over the variables $x_i$. We can therefore use the importance sampling technique. By using these distributions, we generate $N_f$ independent random samples of $n$ independent variables $\{ \hat{x}_{a-n}, \ldots , \hat{x}_{a-1} \}$ which we use to compute the estimation
\begin{equation}
z^f_{a,a}({x}_{a}) \simeq \overline{z^f_{a,a}}({x}_{a}) = \frac{\beta (x_a)}{\mathcal{N}} \sum_{\{ \hat{x}_{a-n}, \ldots , \hat{x}_{a-1} \}}  ~ \Psi_{a} (\hat{x}_{a-n}, \ldots , \hat{x}_{a-1}) \,, \label{eq_importance_sampling_forward}
\end{equation}
where $\mathcal{N}$ is a normalization constant and where we use the same set of samples for all $x_a \in [1,c]$. Other positions $z^f_{a,i}({x}_{i})$ are each computed in the same manner with a new sampling for each one.\\

The backwards iteration is hereby discarded. Indeed, the backwards iteration of Eq. (\ref{eq_first_order_backwards}) will result in a large number of \emph{false positives}. This happens because the $z^b_{a+1,i}({x}_i)$ become very peeked about a mean that is not the correct $x_i$ due to the repercussion of early errors in subsequent iterations. It is therefore necessary to perform the initial sums over a very large number of samples which defeats the purpose of this algorithm. Furthermore, empirical tests show that under the first order approximation, very little information is actually gained by using the backwards algorithm.\\

When all iterations have been computed, we simply equate the $z^f_{t,i}({x}_{i})$ to the marginals, i.e. $\forall a \in [1,t]$ and $x_a \in [1,c]$
\begin{equation}
\nu (x_a) = z^f_{t,a}({x}_{a}) \,.
\end{equation}

The complexity is $\Theta (N_f t n)$ which takes of the order of a second to decode a full chain for typical parameter values.\\

\subsubsection{First order approximation with Gauss} \label{subsection_algo_gauss}

The second first order approximation will be referred to as \emph{first order approximation with Gauss} (TM.1A.G). 

We make the assumption that the variable $X_i^{(a)}=\sum_{  \substack{ j=a-n \\ j \ne i }}^{a} \alpha (j,a) x_j $ present in $\Psi_a$ at step $a$ (Eq. \ref{eq_psi}) can be approximated with a Gaussian random variable. This is possible since the variables $\alpha (j,a) x_j$ are independently drawn from their respective distributions under the approximation of Eq. (\ref{eq_first_order_approx}) and we assume $n$ is \emph{large}.

The complete derivation can be found in appendix \ref{appendix_gauss}, but if we write the mean and variance of $X_i^{(a)}$ as respectively
$\mu_{X_i}$ and $\sigma_{X_i}^2$, the iterated marginal distribution of $x_i$ for $i \in \{a-n, \ldots , a-1 \}$ at step $a$ can be written as
\begin{equation}
\nu_i^{(a)} (x_i) = \frac{\nu_i^{(a-1)} (x_i)}{\mathcal{N}} \exp \left[ - \frac{1}{2 (\sigma^2 + \sigma_{X_i}^2)} \left( Y_a - \mu_{X_i} - \alpha(i,a) x_i  \right)^2    \right] \,, \label{eq_marginal_gauss}
\end{equation}
where $\mathcal{N}$ is a normalization constant and a similar expression for $\nu_a^{(a)} (x_a)$ since $\nu_a^{(a-1)}$ does not exist and is replaced by the prior of $x_a$: $\beta(x_a)$. The final iteration for $a=t$ returns the complete set of marginals $\nu_i (x_i)$ for all $i$. We make the same assumption on the backwards iteration as in section \ref{subsubsection_mc}.\\

\subsection{Two point algorithm} \label{subsection_algo_two_point}

Our final algorithm, which will be called \emph{second order approximation with Gauss} (TM.2A.G) is similar to the algorithm described in section 
\ref{subsection_algo_gauss} but introduces a different factorized approximation.\\

We introduce a second order approximation, which is similar to the factorized expression in Eq. (\ref{eq_first_order_approx}), i.e.
\begin{equation}
 Z_a (x_{a-n+1}, \ldots , x_{a} ) =  \prod_{i=a-n+1}^a z_{a,i}(x_i) \prod_{(i,j)} [1 + w_{ij}(x_i,x_j)] \,, \label{eq_second_order_approx}
\end{equation}
where the functions $w_{ij}(x_i,x_j)$ are very small. This expression introduces two point correlations in the expression of the iterative marginal. 

There is one drawback to the expression in Eq. (\ref{eq_second_order_approx}) and which is that it is required that the functions $w_{ij}(x_i,x_j) \to 0$ and numerically the control of such structures is very difficult. We thus make the assumption that the factor graph is in fact one-dimensional. That is, each variable is connected to exactly two nodes, except for the extremities. By introducing this approximation we can take advantage of a decomposition property for the joint probability of an arbitrary number of variables taken on a tree graph \cite{mezard_montanari.info_phys_computation}. Thus, at any step $a$, for any number of successive variables taken between positions $i$ and $j \leq a$ , as a relationship between the joint probability and the marginals, we have
\begin{equation}
\mu^{(a)} (x_i, \ldots, x_j) = \mu^{(a)}_{i,i+1} (x_i, x_{i+1}) \ldots \mu^{(a)}_{j-1,j} (x_{j-1}, x_{j}) \frac{1}{ \mu^{(a)}_{i+1} (x_{i+1}) \ldots \mu^{(a)}_{j-1} (x_{j-1})} ,
\end{equation}
where the $\mu^{(a)}_{k,k+1} (x_k, x_{k+1})$ and $\mu^{(a)}_{k} (x_{k})$ are true marginals of the approximation $\mu^{(a)}$.\\

By setting interactions only between closest neighbors and taking inspiration from section \ref{subsection_algo_gauss}, we consider the variable $X_{i,i+1}=\sum_{  \substack{ j=a-n \\ j \ne i,i+1 }}^{a} \alpha (j,a) x_j$ at step $a$ to be Gaussian and we assume it has mean and variance respectively $\mu_{X_{i,i+1}}$ and $\sigma_{X_{i,i+1}}^2$. It then comes that the main expression for the two point marginal at step $a$ for the couple $\{ x_i,x_{i+1} \}$ can be written
\begin{equation}
\nu^{(a)}_{i,i+1} (x_{i}, x_{i+1}) = \frac{\nu^{(a-1)}_{i,i+1} (x_{i}, x_{i+1})}{\mathcal{N}} ~ \exp {\left[ - \frac{1}{2 (\sigma^2 + \sigma_{X_{i,i+1}}^2)} [ Y_a - \mu_{X_{i,i+1}} - \alpha(i,a) x_{i} - \alpha(i+1,a) x_{i+1} ]^2    \right]} . \label{eq_z_i_ip1}
\end{equation}
The details of this algorithm can be found in appendix \ref{appendix_two_point}. \\

%
%

\section{Bounds} \label{section_discussion}

In this section we describe how we test the behavior of our different approximated algorithms when it is impossible to compare them to the TM algorithm. This is done by introducing approximations on the probability of error and on the average number of errors that occur when performing estimation. We start by studying the probability of error in the limit of a memoryless channel and then we discuss another approximation for a channel with memory.

\subsection{Memoryless channel} \label{subsection_full_perr} 

In this subsection, we assume that the channel is without memory, i.e. that the parameter $n$ introduced in section \ref{section_model} is zero and thus that Eq. (\ref{eq_y_a_def}) reduces to
\begin{equation}
 Y_a = A_a +  \eta_a \,,
\end{equation}
where $\eta_a$ is the random variable with mean $0$ and variance $\sigma^2$ defined in Eq. (\ref{eq_y_a_def}). The probability of error for the single variable is then
\begin{equation}
 P_{err} = \sum_{x_0=1}^c \beta(x_0) \int \frac{{\rm d} \eta}{\sqrt{2 \pi}} ~ e^{- \frac{1}{2} \eta^2} ~ \mathbb{I} \left[   \argmax[x] \left\{ \beta(x) e^{- \frac{1}{2 \sigma^2} (x_0+\eta_a -x)^2}  \right\} \neq x_0 \right] \,, \label{eq_perr_memoryless}
\end{equation}
where $\mathbb{I}$ is the indicator function. Errors thus happen when the expression $\beta(x) e^{- \frac{1}{2 \sigma^2} (x_0+ \eta_a -x)^2}$ in (\ref{eq_perr_memoryless}) is maximized by a value $x \neq x_0$. If the distribution $\beta$ is uniform this occurs if
\begin{eqnarray}
| \eta | &>& \frac{1}{2 } ~~~~~~ \textrm{ if } x_0 \neq 1,c \,,\\
 \eta  &>& \frac{1}{2 } ~~~~~~ \textrm{ if } x_0 = 1 \,,\\
 \eta  &<& - \frac{1}{2 } ~~~~ \textrm{ if } x_0 = c \,.
\end{eqnarray}
In general, for other expressions of $\beta$, these events provide a lower bound on the probability we seek.
These lead to the very general expression for the probability of error as
\begin{equation}
P_{err} \geq \prob [ x_0 = 1 ] \prob \left[ \eta > \frac{1}{2 } \right] + \prob [ x_0 = c ] \prob \left[ \eta < - \frac{1}{2 } \right] + \sum_{x=2}^{c-1} \prob [ x_0 = x ] \prob \left[ |\eta | > \frac{1}{2 } \right] \,,
\end{equation}
where, $\eta$ being normal, we have 
\begin{equation}
\prob \left[ \eta > \frac{1}{2 } \right] = \prob \left[ \eta < - \frac{1}{2 } \right] = \frac{1}{2} \prob \left[ |\eta | > \frac{1}{2 } \right] = Q \left( \frac{1}{2 \sigma} \right) \,,
\end{equation}
where 
\begin{equation}
Q \left( x \right) = \prob \left[ X \geq x\right] = \frac{1}{\sqrt{2 \pi}}\int_{x}^{+ \infty} {\rm d} \nu ~ e^{-\frac{1}{2} \nu^2} \,.
\label{eq_Q_func}
\end{equation}
Therefore, the probability of error can finally be expressed as
\begin{equation}
P_{err} \geq  \left( \beta ( 1 ) + \beta ( c ) +2 \sum_{x=2}^{c-1} \beta ( x ) \right) Q \left( \frac{1}{2 \sigma} \right) \,,
\end{equation}
where $\beta$ is one of the distributions introduced in section \ref{subsection_beta_distributions}, i.e. depending on which distribution is being studied we will use one of the following expressions
\begin{eqnarray}
P_{err, g} &\geq&  \frac{3}{2} \left( 1 - \frac{1}{2^{c-1}} \right)  ~ Q \left( \frac{1}{2 \sigma} \right) \,, \\
P_{err, u} &=& 2 ~\frac{c-1}{c}  ~ Q \left( \frac{1}{2 \sigma} \right) \,, \\
P_{err, t} &\geq& 3 ~\frac{2^{c-1} - 1}{2^{c}-1} ~ Q \left( \frac{1}{2 \sigma} \right) \,,
\end{eqnarray}
when in $\beta_g$ and $\beta_t$ we have $q = \frac{1}{2}$.

\subsection{Channel with memory}  \label{subsection_mean_num_errors}

Let us now derive a lower bound on the probability of errors in the case of channel with non-zero memory.

To do this, consider the probability density defined in (\ref{eq_posterior_prob}) and write it as follows
\begin{equation}
\prob \left[  \uX | \uY \right]  = \frac{1}{Z} \left(  \prod_{a=1}^t \beta(x_a) \right) ~ e^{ - \frac{1}{2 \sigma^2}   H_{\underline{A}}  (\underline{X})   }\,,
\end{equation}
where $Z$ is a normalization constant and 
\begin{equation}
H_{\underline{A}}  (\underline{X}) = \sum_{a=1}^t \left[    \sum_{i=1}^a \alpha(i,a) ( A_i - x_i )   + \eta_a  \right]^2 \,,
\end{equation}
where all the parameters are the same as in (\ref{eq_y_a_def}).

Using these notations, we can write the block MAP probability of error as
\begin{equation}
P ( \underline{A},\sigma) = \prob \left\{  \exists \underline{X} : H_{\underline{A}}  (\underline{X}) + B(\underline{X})< H_{\underline{A}}  (\underline{A}) + B(\underline{A})    \right\} \,, \label{eq_prob_err_part_func}
\end{equation}
where $B(\underline{X})$ is the prior and takes the form
\begin{equation}
B_g (\underline{X}) = 2 \sigma^2 \ln(2) \sum_{a=1}^t x_a
\end{equation}
 if we consider the geometrical distribution $\beta_g$ with $q=\frac{1}{2}$,
\begin{equation}
B_t (\underline{X}) = 2 \sigma^2 ( \ln(2) \sum_{a=1}^t x_a - t\ln(\gamma_t))
\end{equation} 
if we consider the truncated distribution $\beta_t$ and 
\begin{equation}
B_u (\underline{X}) = B_u = 2 \sigma^2 t \ln(c)
\end{equation}
if we consider the uniform distribution.

For any $\uX = \{ x_1, \ldots , x_t \} \in \mathbb{N}^t$ a vector of $t$ strictly positive integers the probability $\prob \left\{  H_{\underline{A}}  (\underline{X}) + B(\underline{X})< H_{\underline{A}}  (\underline{A}) + B(\underline{A})    \right\}$ is a lower bound of the right hand side of Eq. (\ref{eq_prob_err_part_func}). In order to estimate this probability we write
\begin{eqnarray}
H_{\underline{A}}  (\underline{X}) - H_{\underline{A}}  (\underline{A}) &=& \sum_{a=1}^t \left[  \sum_{i=1}^a \alpha(i,a) (A_i - x_i)  \right]^2 + 2 \sum_{a=1}^t \eta_a \sum_{i=1}^a \alpha(i,a) (A_i - x_i)  \,,\\
B_g (\underline{X}) - B_g (\underline{A}) &=& 2 \sigma^2 \ln(2) \sum_{a=1}^t (x_a - A_a)\,, \\
B_u (\underline{X}) - B_u (\underline{A}) &=& 0 \,,
\end{eqnarray}
and if we define
\begin{eqnarray}
\Sigma_{\uA, \uX} &=& \sum_{a=1}^t \left[  \sum_{i=1}^a \alpha(i,a) (A_i - x_i)  \right]^2 \,, \\
\eta_{\uA, \uX} &=& 2 \sum_{a=1}^t \eta_a \sum_{i=1}^a \alpha(i,a) (A_i - x_i) \,, \\
B_{\uA, \uX} &=& B (\underline{X}) - B (\underline{A}) \,,
\end{eqnarray}
we obtain that
\begin{equation}
P ( \underline{A},\sigma) \geq \prob \left\{ \Sigma_{\uA, \uX} + \eta_{\uA, \uX} + B_{\uA, \uX} < 0 \right\} \,.
\end{equation}
Furthermore, $\eta_{\uA, \uX}$ is a Gaussian random variable of mean $0$ and variance $4 \sigma^2 \Sigma_{\uA, \uX}$, thus, using the same function $Q$ as in (\ref{eq_Q_func}), we have
\begin{equation}
P ( \underline{A},\sigma) \geq Q \left( \frac{\Sigma_{\uA, \uX} + B_{\uA, \uX}}{2 \sigma \sqrt{ \Sigma_{\uA, \uX}}} \right)  \,. \label{P_for_Sigma}
\end{equation}

This expression is a function of both $\uX$ and $\uA$ and thus is still impractical both analytically and numerically. We thus introduce the notation $\underline{X} = \underline{A}^{i,x}$ which differs from $\underline{A}$ at position $i$ only where it takes value $x \neq A_i$, i.e. $\underline{A}^{i,x} = \{ A_1, \ldots, A_{i-1}, x , A_{i+1}, \ldots , A_t \}$. 
That is, we have 
\begin{eqnarray}
P ( \underline{A},\sigma) &\geq & \prob \left\{  \exists i,x : H_{\underline{A}}  (\underline{A}^{i,x}) + B(\underline{A}^{i,x}) < H_{\underline{A}}  (\underline{A}) + B(\underline{A})    \right\} \\ 
&\geq &\max_{i,x} \, \prob \left\{  H_{\underline{A}}  (\underline{A}^{i,x}) + B(\underline{A}^{i,x}) < H_{\underline{A}}  (\underline{A}) + B(\underline{A})    \right\} \\
& \equiv & \max_{i,x} R_{i,x} (\uA ) \,,
\end{eqnarray}
where the expression of $R_{i,x}$ will be given in the following. We use the previous expression and Eq. (\ref{P_for_Sigma}) to write a lower bound on the probability of error
\begin{equation}
P_{err} \geq \max_i \, \E_{\underline{A}} \, \max_x \, R_{i,x} (\uA )
\end{equation}
where $\E_{\underline{A}}$ is the expectation over the distribution of the vectors $\underline{A}$.
We can also write a lower bound of the expectation of the number of errors as
\begin{equation}
\E \{   \# \textrm{errors}   \} \geq   \E_{\underline{A}} \sum_{i=1}^t \max_x R_{i,x} (\uA ) \label{eq_max_num_errors} \,.
\end{equation}

The expression of $R_{i,x} (\uA )$ varies according to the prior distribution $\beta$, we separate the results accordingly. For the geometrical distribution, we have
\begin{equation}
R_{i,x} (\uA ) = Q \left(  \frac{(A_i - x)^2 \sum_{a=i}^t \alpha(i,a)^2  +  2 \sigma^2 (x - A_i) \ln(2)}{2 \sigma \sqrt{ (A_i - x_i)^2 \sum_{a=i}^t \alpha(i,a)^2}}    \right) \,,
\end{equation}
which is maximized by $x=A_i - 1$ if $A_i \neq 1$ and $x=2$ if $A_i = 1$ thus we define the function of a single integer
\begin{eqnarray}
R_{i}(A_i) &=& Q \left(  \frac{\sum_{a=i}^t \alpha(i,a)^2 + 2 \sigma^2  \ln(2)}{ 2 \sigma \sqrt{  \sum_{a=i}^t \alpha(i,a)^2}  } \right) ~~~~~ \textrm{if } A_i=1 \,, \\
 &=& Q \left(  \frac{\sum_{a=i}^t \alpha(i,a)^2 - 2 \sigma^2  \ln(2)}{  2 \sigma \sqrt{  \sum_{a=i}^t \alpha(i,a)^2}  } \right) ~~~~~ \textrm{if } A_i \neq 1\,.
\end{eqnarray}\\
We can therefore estimate the lower bounds when considering the geometrical distribution as
\begin{eqnarray}
P_{err,g} &\geq& \max_i \left( \frac{1}{2} R_{i} (A_i = 1)  + \frac{1}{2} ~ R_{i} (A_i \neq 1) \right) \,,\\
\E_g \{   \# \textrm{errors}   \}  &\geq &  \sum_{i=1}^t \left( \frac{1}{2} R_{i} (A_i = 1)  + \frac{1}{2} R_{i} (A_i \neq 1)  \right) \,. \label{eq_num_errors}
\end{eqnarray}\\

In the case of the uniform distribution we have, again using $R_{i}(A_i)$ as a function of a single integer,
\begin{equation}
\E_{\underline{A}} \max_x R_{i,x} (\uA ) = \frac{1}{c} \sum_{A_i = 1}^{c}  R_{i}(A_i) \,, \label{eq_uniform_num_errors}
\end{equation}
where each $R_{i}(A_i)$ in Eq. (\ref{eq_uniform_num_errors}) is in fact solely a function of $i$ and all have the same expression, thus we define 
\begin{equation}
R (i) = Q \left(  \frac{\sqrt{ \sum_{a=i}^t \alpha(i,a)^2 } }{ 2 \sigma   } \right) \,,
\end{equation}
and thus
\begin{eqnarray}
P_{err,u} &\geq &  \max_i \left[ Q \left(  \frac{ \sqrt{ \sum_{a=i}^t \alpha(i,a)^2 } }{2 \sigma}  \right)  \right]  \,, \\
\E_u \{   \# \textrm{errors}   \} &\geq &   \sum_{i=1}^t R (i) \,.
\end{eqnarray}\\

Finally, the expressions of $R_{i,x} (\uA)$ and $R_i (A_i)$ for the truncated distribution are the same as for the geometrical distribution and 
\begin{eqnarray}
P_{err,t} &\geq & \frac{1}{1-2^{-c}}  \max_i \left[ \sum_{A_i = 1}^{c} \frac{1}{2^{A_i}}   R_{i}(A_i)  \right] \,,
\end{eqnarray}
and 
\begin{eqnarray}
\E_t \{   \# \textrm{errors}   \} &\geq & \frac{1}{1-2^{-c}}  \sum_{i=1}^t \sum_{A_i = 1}^{c} \frac{1}{2^{A_i}}   R_{i}(A_i) \,.
\end{eqnarray}\\

%
%
\section{Results} \label{section_results}

In this section we study the behavior of our various algorithms according to the parameters used. We will separate this section into three separate subsections according to the parameters we use to control our tests.

In general, the behavior of the algorithms is the same if we are to consider either $\beta_g$ or $\beta_t$, thus we will usually only present the results for one of the two except in section \ref{subsection_tvar} where we emphasize the similarities. 

There are in total six algorithms we compare. These were all defined in section \ref{section_algorithms} except for the one referred to as TM.1A.f which is the first order approximation of section \ref{subsection_algo_mc} with the backwards iteration discarded.

\subsection{Noise as parameter} \label{subsection_svar}

\begin{figure}[h!]
\includegraphics[width=0.7\linewidth]{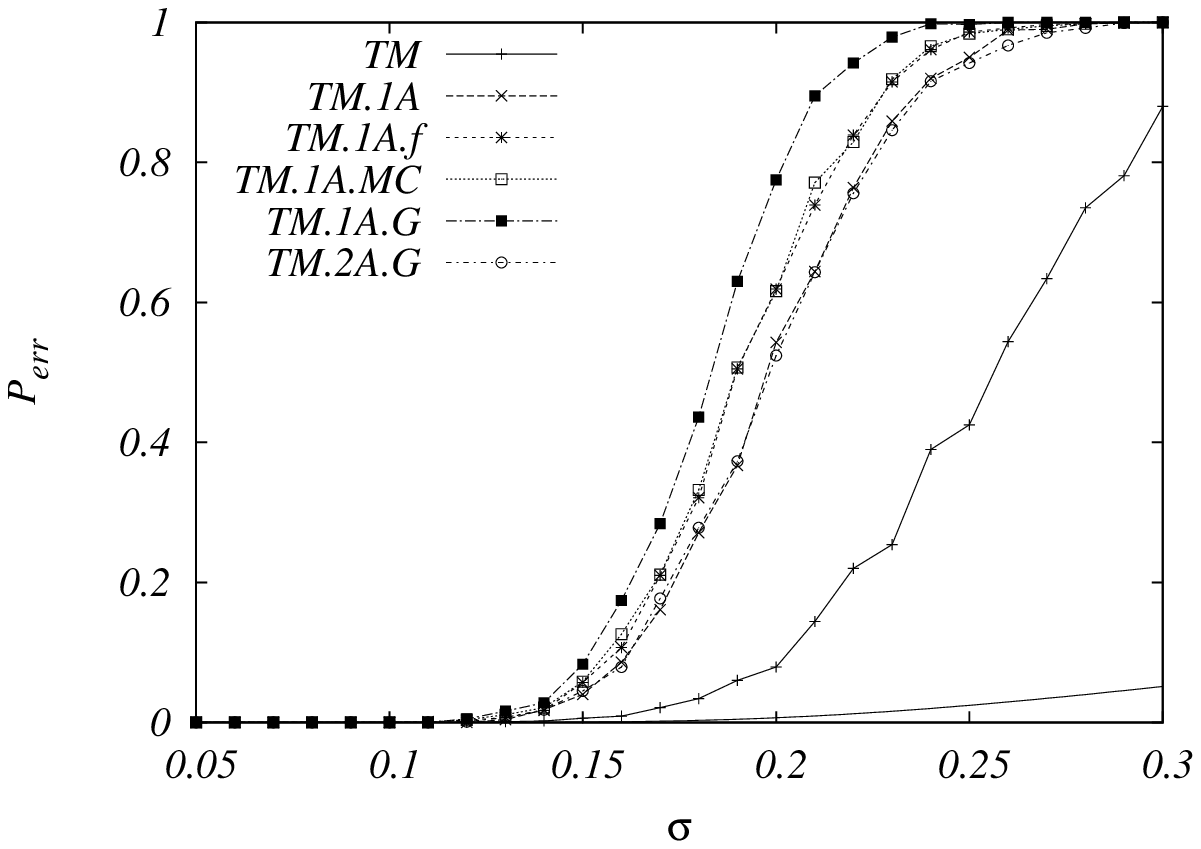}
\includegraphics[width=0.7\linewidth]{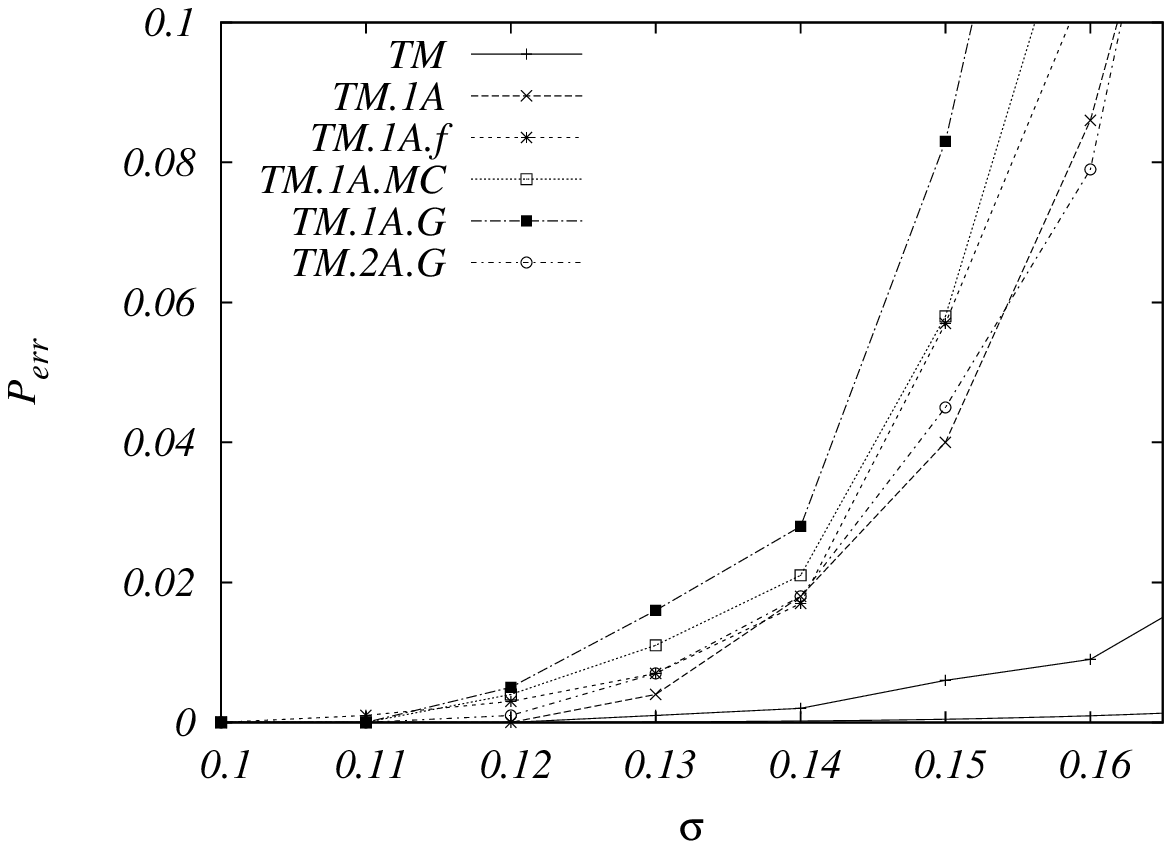}
\caption{Probability of error ($P_{err}$) for various algorithms vs $\sigma$. The full line is the analytical lower bound. The figure on the bottom is a blowup of the one on the top for $\sigma$ small. With $t=100$, $n=3$, $c=15$, using $\beta_t$ and $\alpha_f$, average over $1000$ samples and $N_f=500$ for TM.1A.MC.}
\label{figs_t100_f3_v3_l4_c15_n1000_svar_Pcd}
\end{figure}

\begin{figure}[h!]
\includegraphics[width=0.7\linewidth]{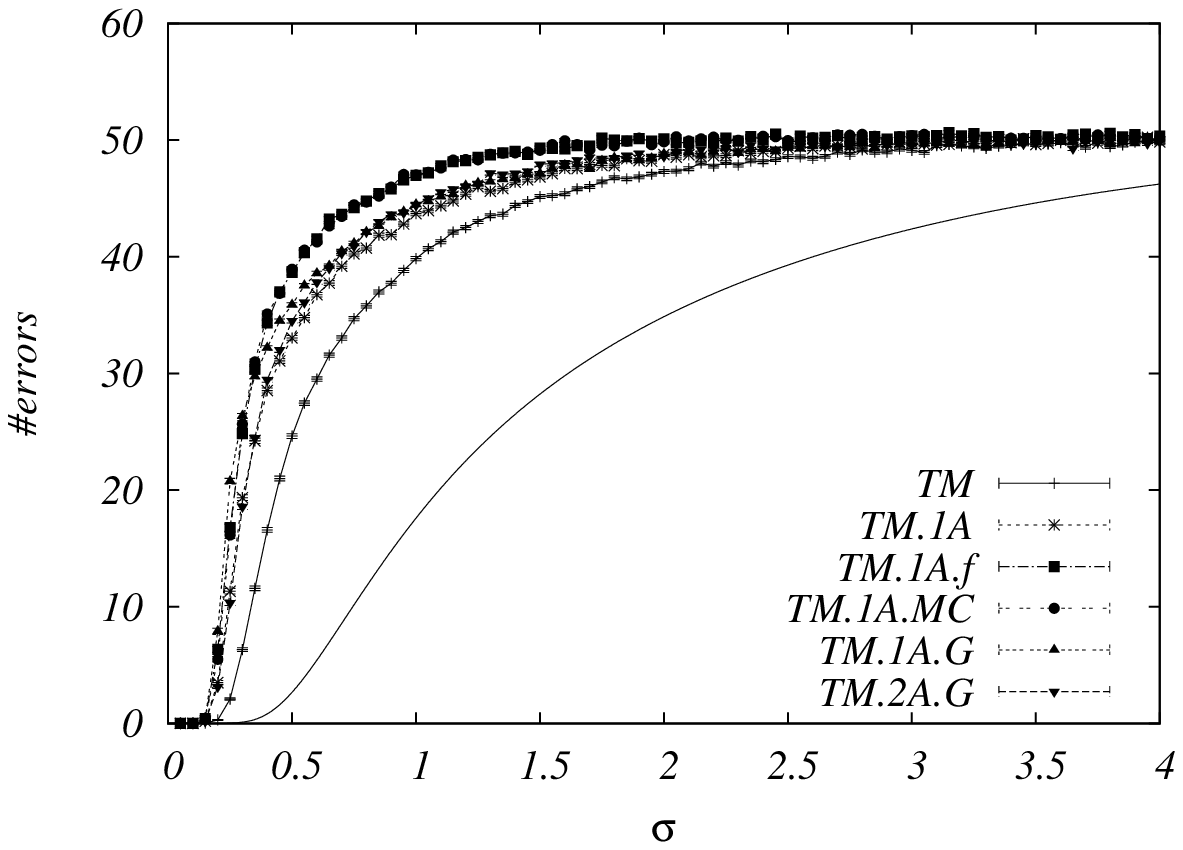}
\includegraphics[width=0.7\linewidth]{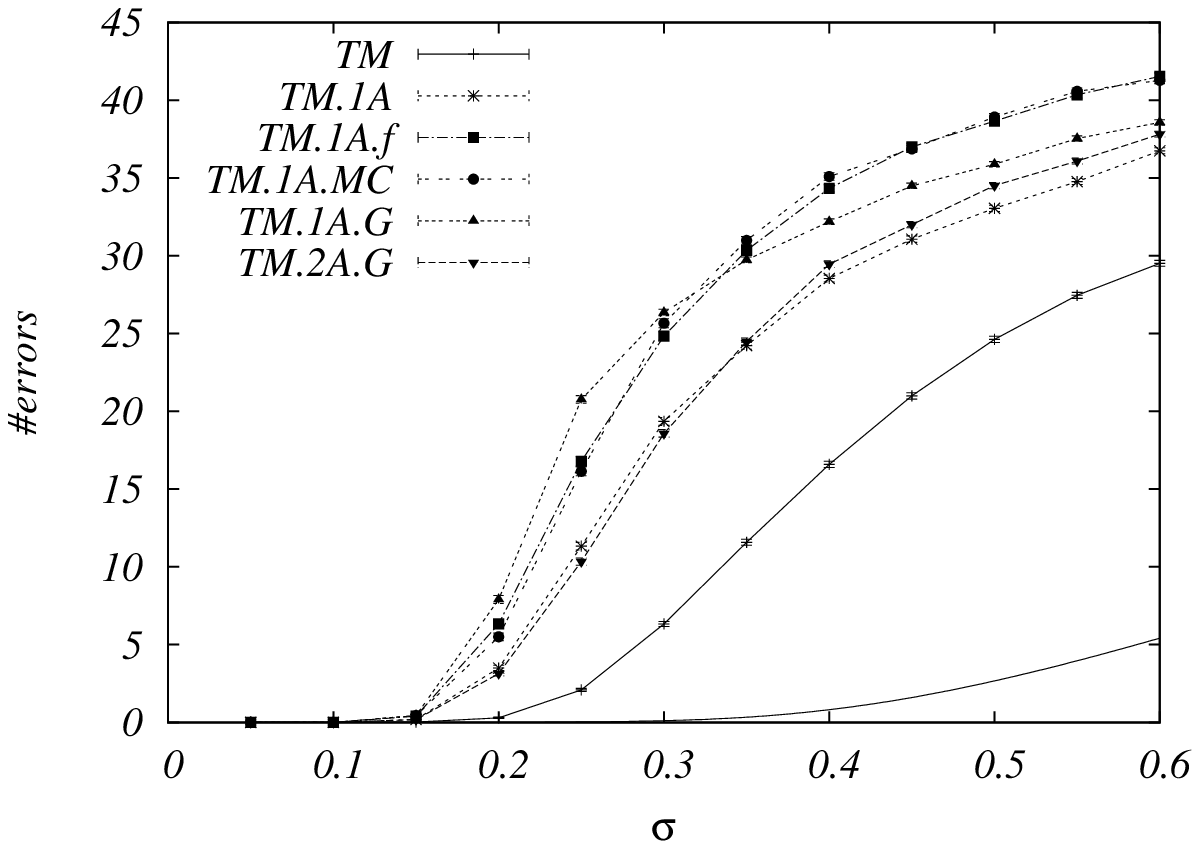}
\caption{Average number of errors ($\#errors$) for various algorithms vs $\sigma$. The full line is the analytical lower bound. The figure on the bottom is a blowup of the one on the top for $\sigma$ small. With $t=100$, $n=3$, $c=15$, using $\beta_t$ and $\alpha_f$, average over $1000$ samples and $N_f=500$ for TM.1A.MC.}
\label{figs_t100_f3_v3_l4_c15_n1000_svar_numerr}
\end{figure}

\begin{figure}[h!]
\includegraphics[width=0.7\linewidth]{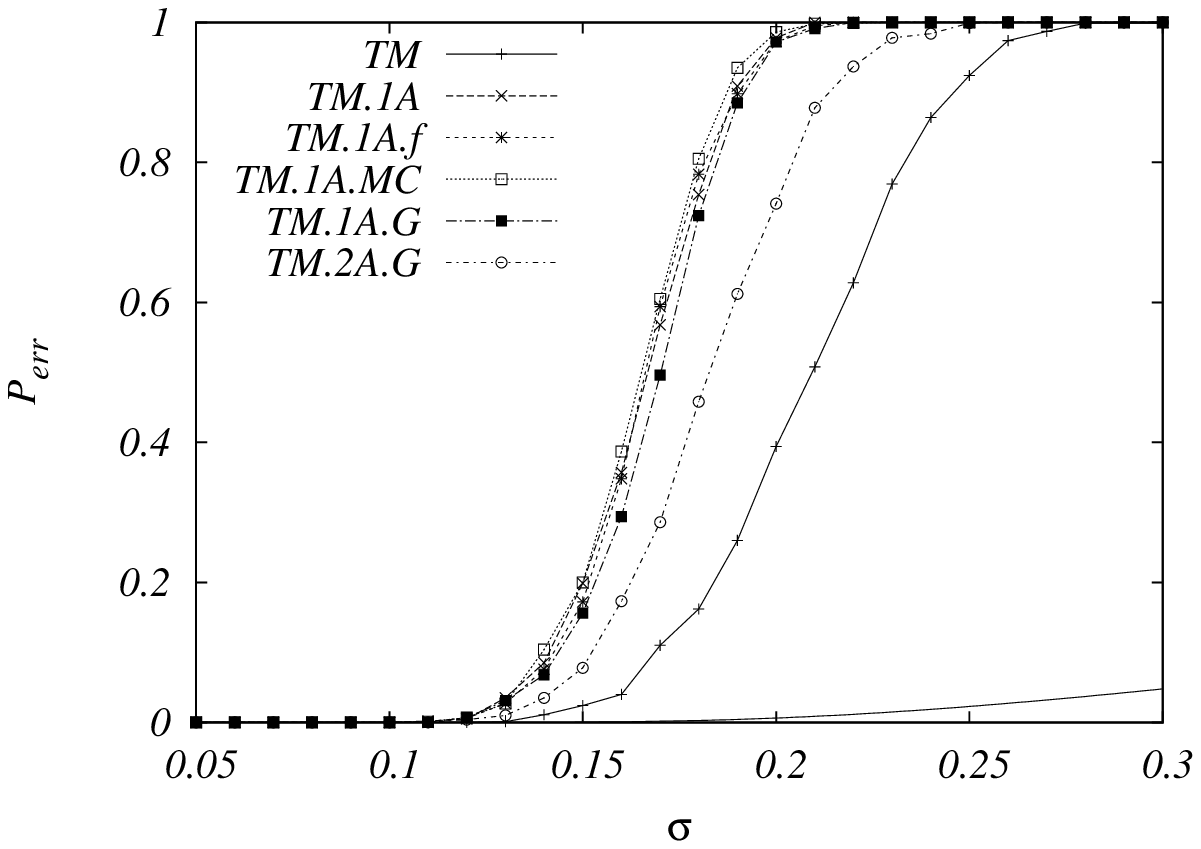}
\includegraphics[width=0.7\linewidth]{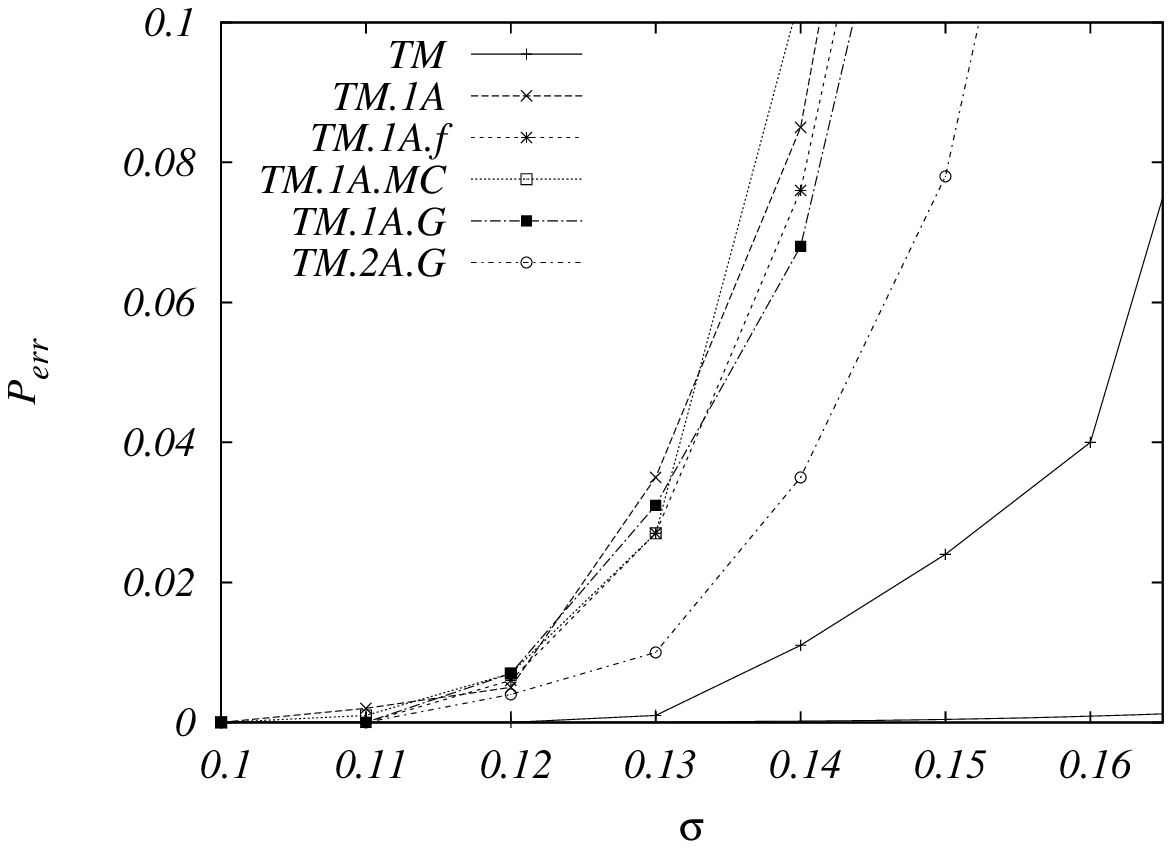}
\caption{Same as \ref{figs_t100_f3_v3_l4_c15_n1000_svar_Pcd} with $t=100$, $n=3$, $c=15$, using $\beta_u$ and $\alpha_f$, average over $1000$ samples and $N_f=500$ for TM.1A.MC.}
\label{figs_t100_f3_v2_l4_c15_n1000_svar_Pcd}
\end{figure}

\begin{figure}[h!]
\includegraphics[width=0.7\linewidth]{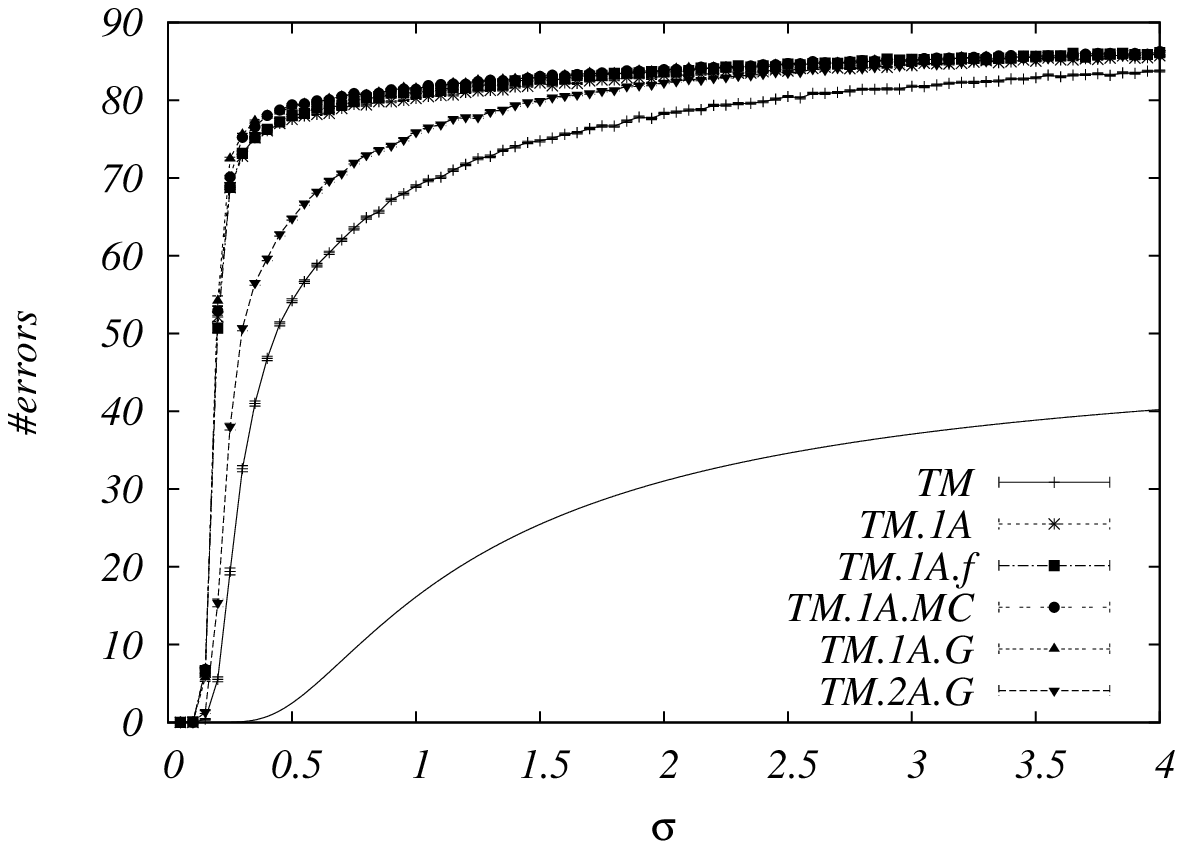}
\includegraphics[width=0.7\linewidth]{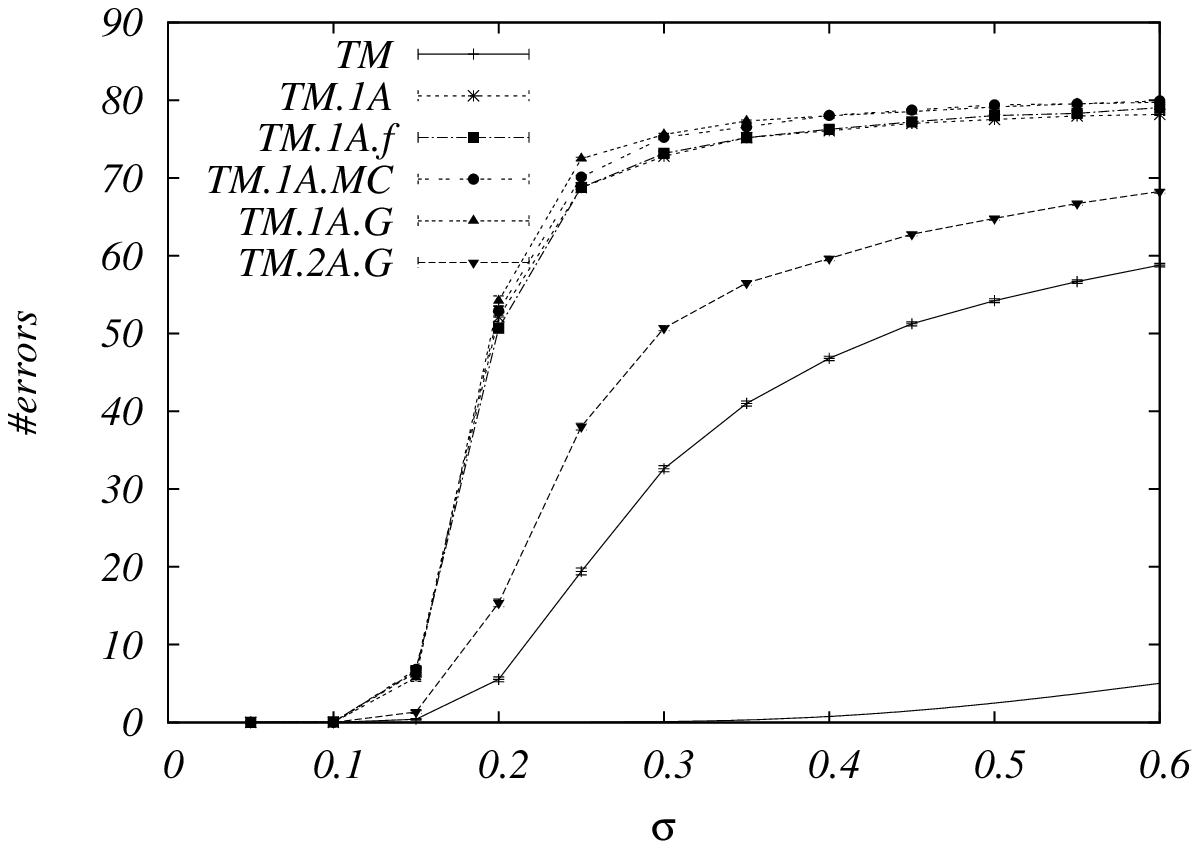}
\caption{Same as \ref{figs_t100_f3_v3_l4_c15_n1000_svar_numerr} with $t=100$, $n=3$, $c=15$, using $\beta_u$ and $\alpha_f$, average over $1000$ samples and $N_f=500$ for TM.1A.MC.}
\label{figs_t100_f3_v2_l4_c15_n1000_svar_numerr}
\end{figure}

\begin{figure}[h!]
\includegraphics[width=0.7\linewidth]{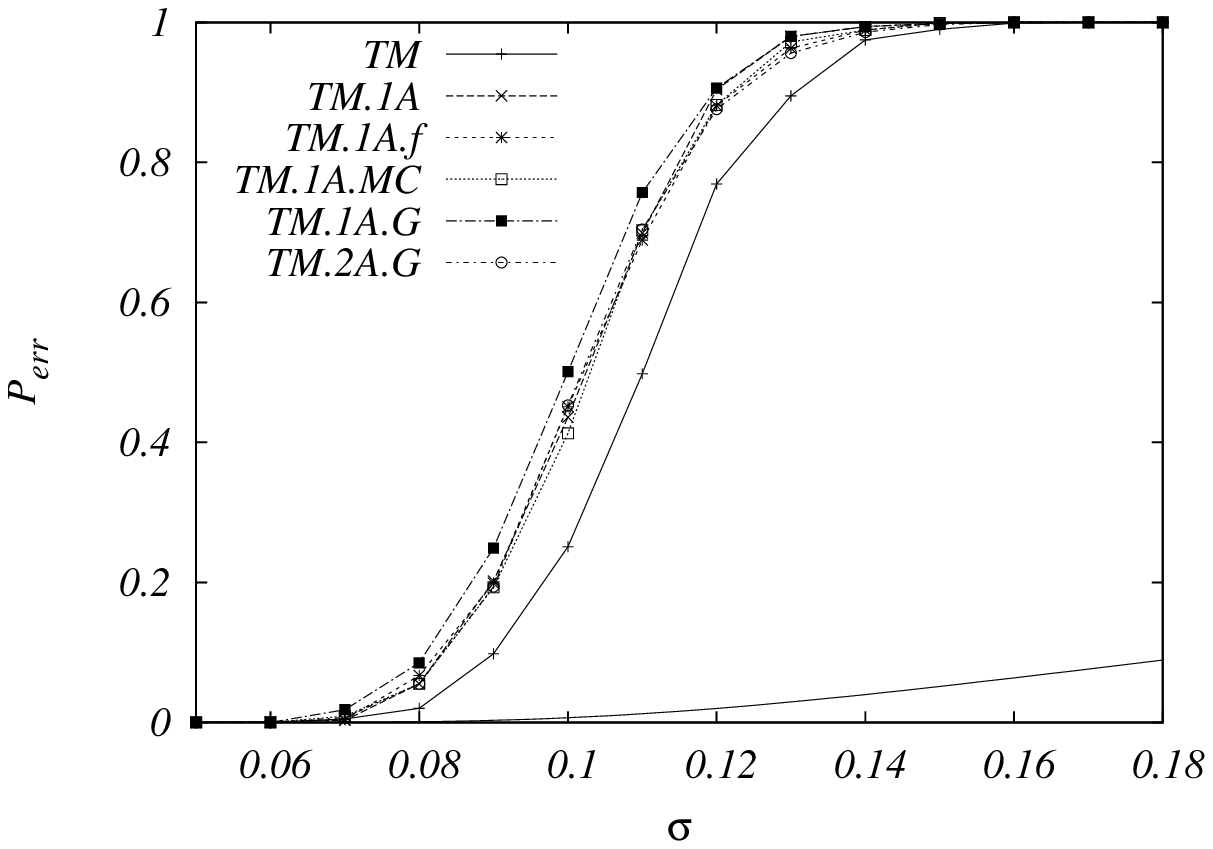}
\includegraphics[width=0.7\linewidth]{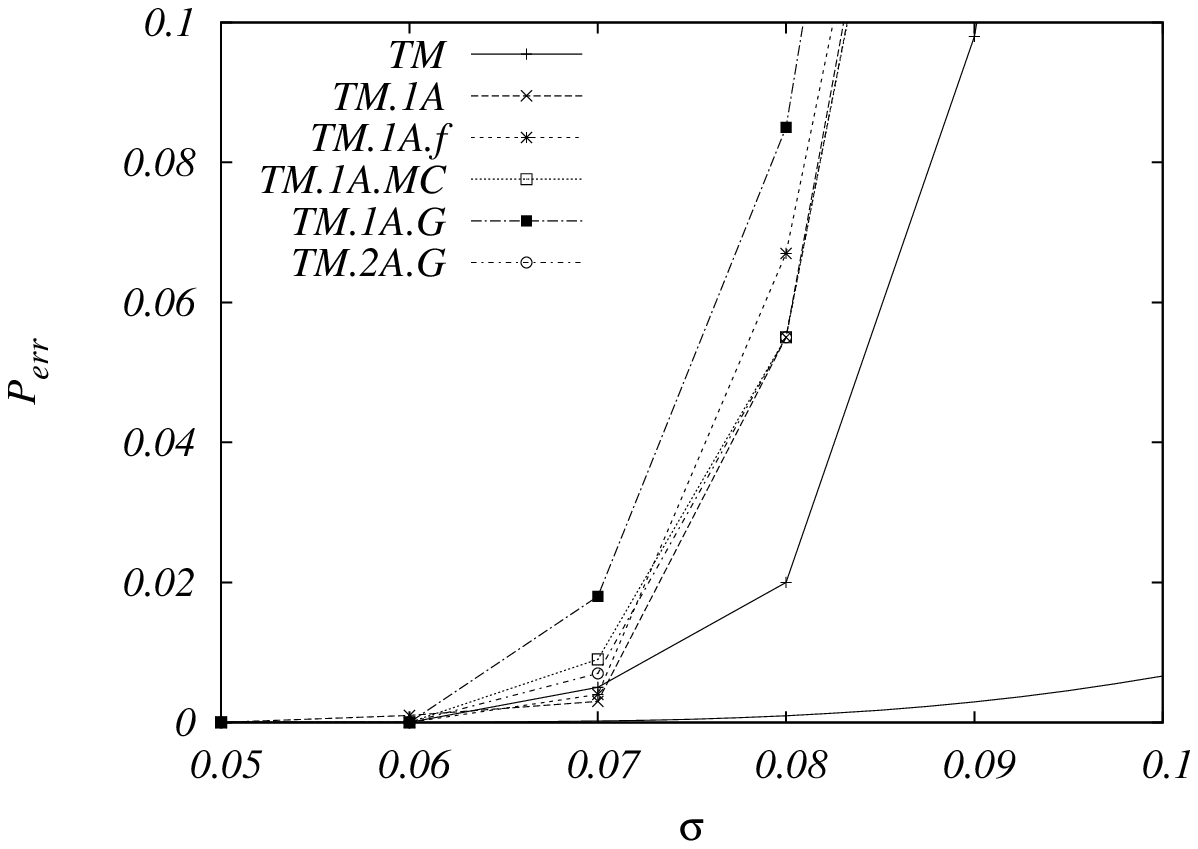}
\caption{Same as \ref{figs_t100_f3_v3_l4_c15_n1000_svar_Pcd} with $t=100$, $n=3$, $c=15$, using $\beta_t$ and $\alpha_h$, average over $1000$ samples and $N_f=500$ for TM.1A.MC.}
\label{figs_t100_f3_v3_l3_c15_n1000_svar_Pcd}
\end{figure}

\begin{figure}[h!]
\includegraphics[width=0.7\linewidth]{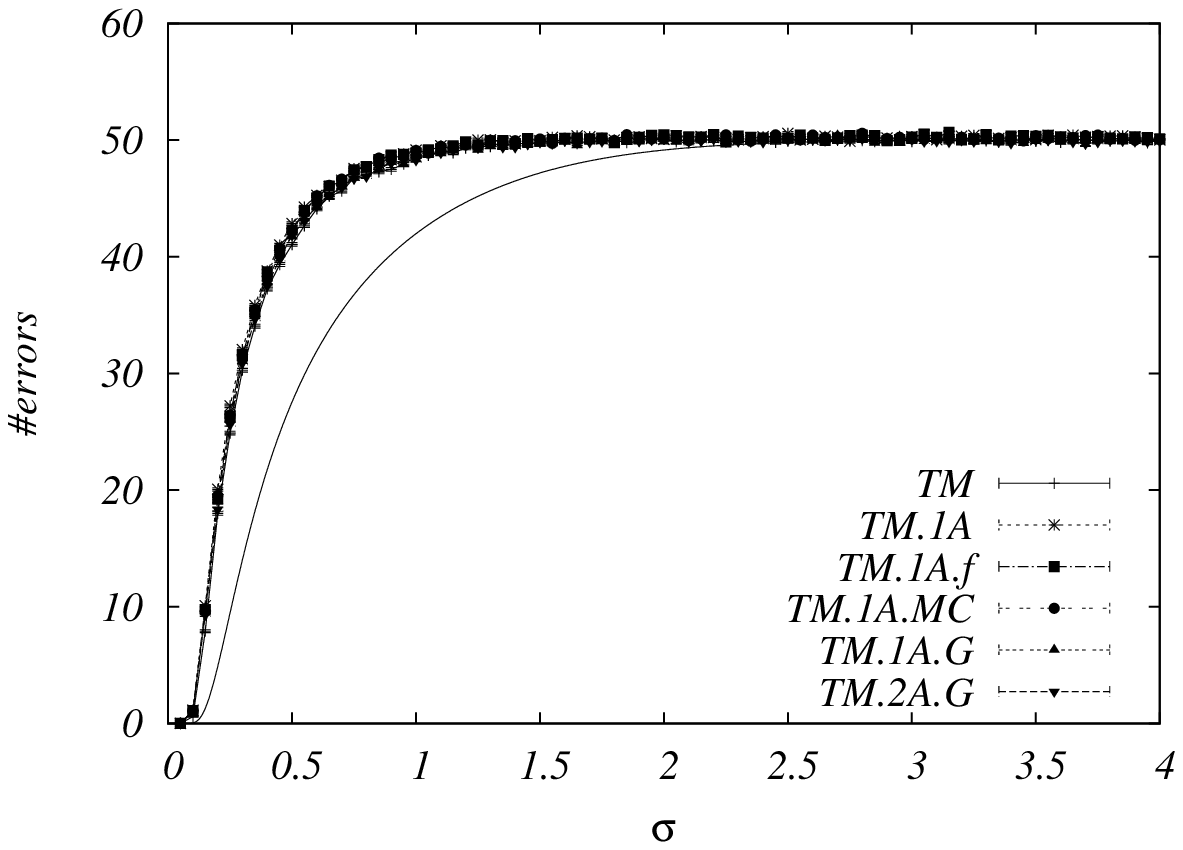}
\includegraphics[width=0.7\linewidth]{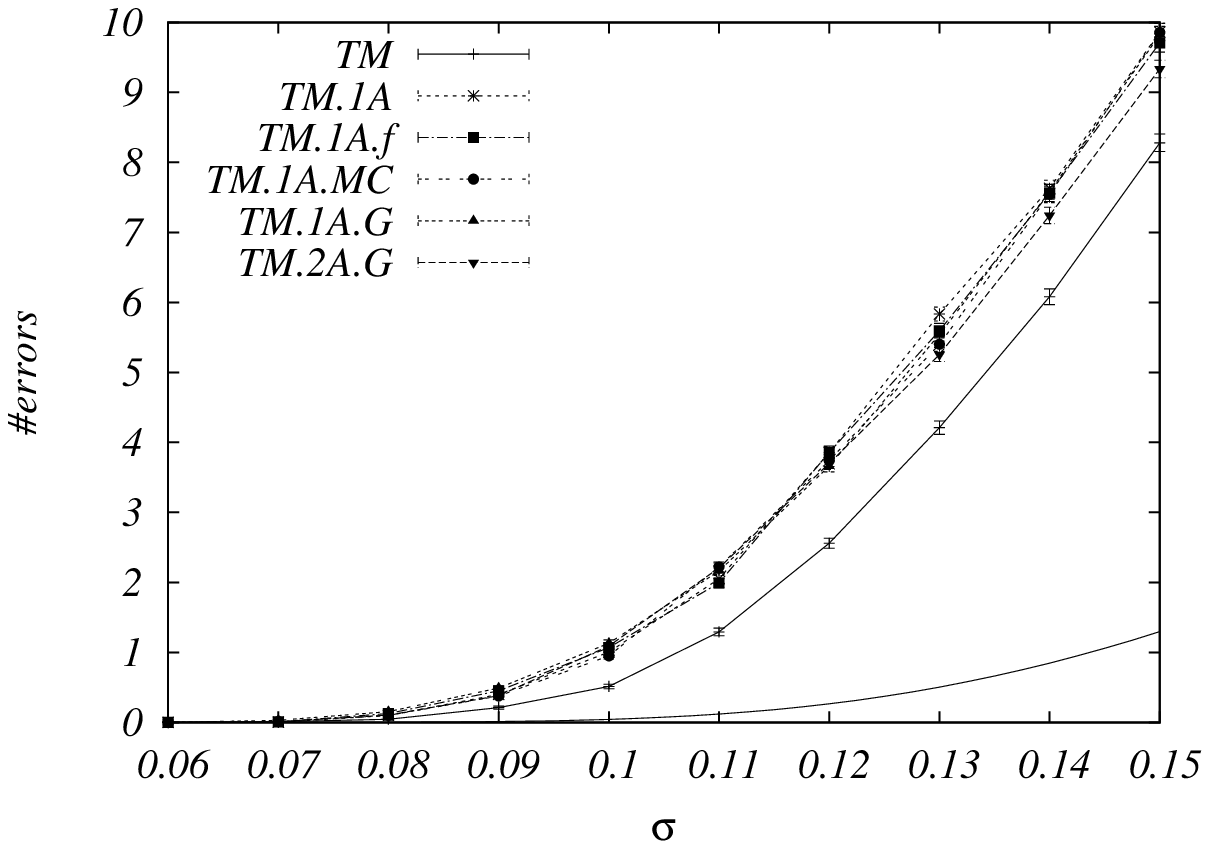}
\caption{Same as \ref{figs_t100_f3_v3_l4_c15_n1000_svar_numerr} with $t=100$, $n=3$, $c=15$, using $\beta_t$ and $\alpha_h$, average over $1000$ samples and $N_f=500$ for TM.1A.MC.}
\label{figs_t100_f3_v3_l3_c15_n1000_svar_numerr}
\end{figure}

\begin{figure}[h!]
\includegraphics[width=0.7\linewidth]{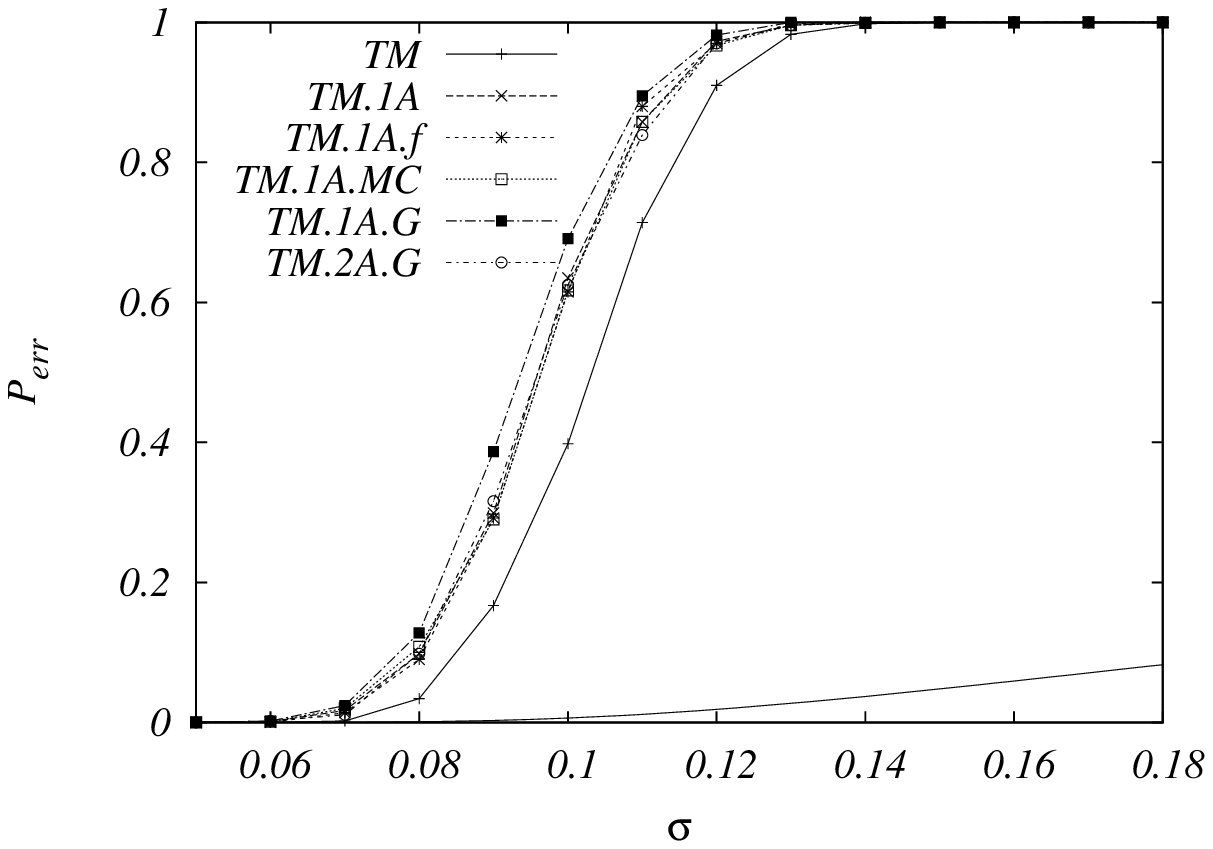}
\includegraphics[width=0.7\linewidth]{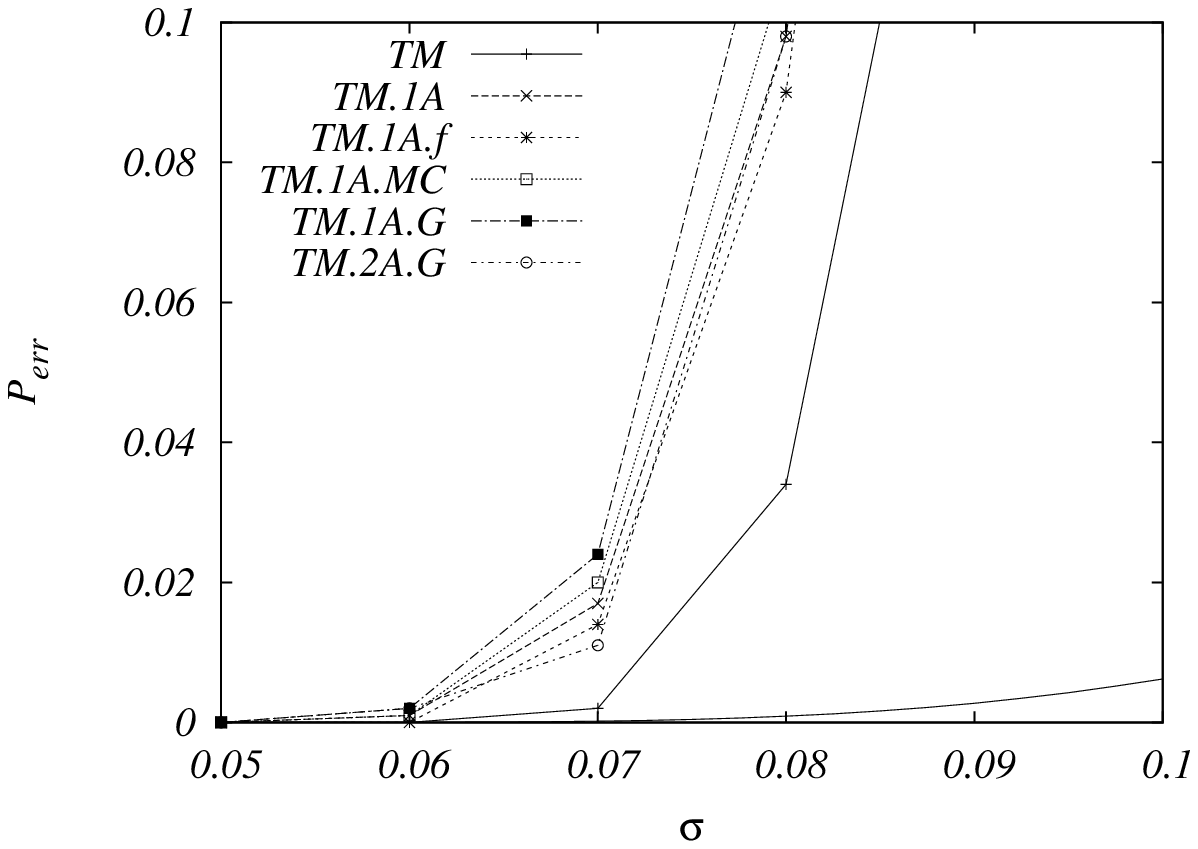}
\caption{Same as \ref{figs_t100_f3_v3_l4_c15_n1000_svar_Pcd} with $t=100$, $n=3$, $c=15$, using $\beta_u$ and $\alpha_h$, average over $1000$ samples and $N_f=500$ for TM.1A.MC.}
\label{figs_t100_f3_v2_l3_c15_n1000_svar_Pcd}
\end{figure}

\begin{figure}[h!]
\includegraphics[width=0.7\linewidth]{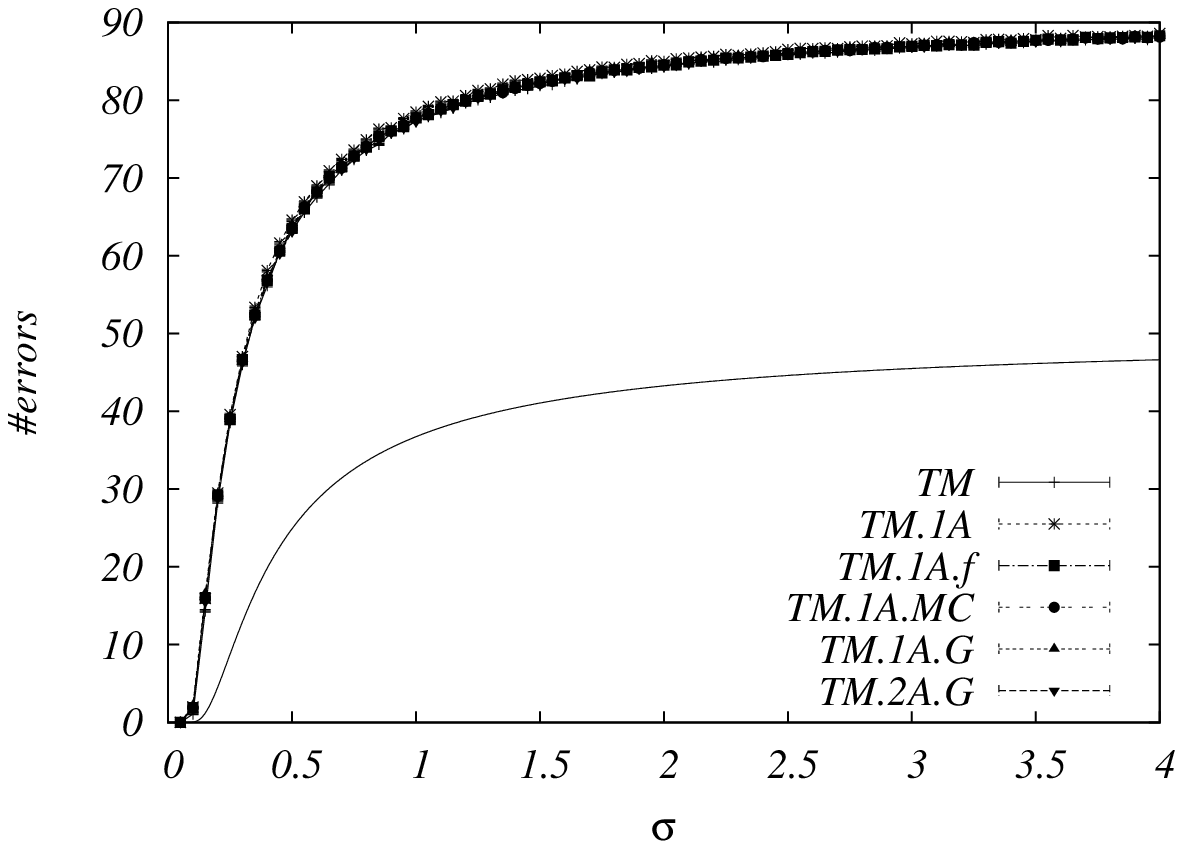}
\includegraphics[width=0.7\linewidth]{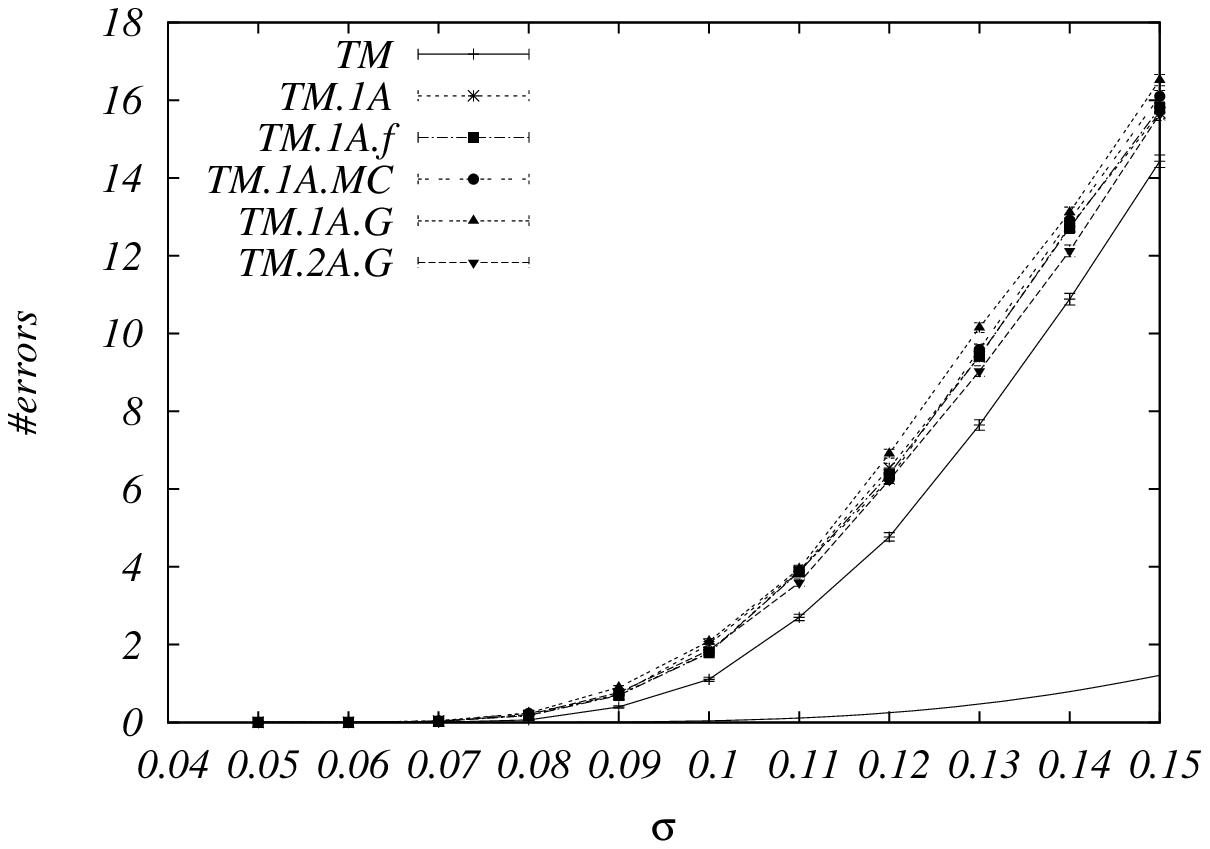}
\caption{Same as \ref{figs_t100_f3_v3_l4_c15_n1000_svar_numerr} with $t=100$, $n=3$, $c=15$, using $\beta_u$ and $\alpha_h$, average over $1000$ samples and $N_f=500$ for TM.1A.MC.}
\label{figs_t100_f3_v2_l3_c15_n1000_svar_numerr}
\end{figure}

In this subsection we vary the noise parameter $\sigma$. We have eight figures (\ref{figs_t100_f3_v3_l4_c15_n1000_svar_Pcd}-\ref{figs_t100_f3_v2_l3_c15_n1000_svar_numerr}) that range over the set of distributions $\beta$ and functions $\alpha$. For each couple $(\beta,\alpha)$ we plot the probability of error $P_{err}$ (Figs. \ref{figs_t100_f3_v3_l4_c15_n1000_svar_Pcd}, \ref{figs_t100_f3_v2_l4_c15_n1000_svar_Pcd}, \ref{figs_t100_f3_v3_l3_c15_n1000_svar_Pcd} and \ref{figs_t100_f3_v2_l3_c15_n1000_svar_Pcd}) and the average number of errors $\#errors$ (Figs. \ref{figs_t100_f3_v3_l4_c15_n1000_svar_numerr}, \ref{figs_t100_f3_v2_l4_c15_n1000_svar_numerr}, \ref{figs_t100_f3_v3_l3_c15_n1000_svar_numerr} and \ref{figs_t100_f3_v2_l3_c15_n1000_svar_numerr}) over a set of independent samples. We recall that the probability of error is the probability that at least one error occurs during the estimation process. 

We emphasize more on the use $\alpha_f$ and $\alpha_h$ since they allow us to control the value of the memory length $n$ and thus allows us to use TM.

The probability of error varies from $0$ to $1$ as $\sigma$ grows from $0$. The rate of increase is dependent on the couple $(\beta,\alpha)$. The average number of errors varies from $0$ to a value that depends on the distribution $\beta$.

There is one expected and obvious trend we can take out of the figures \ref{figs_t100_f3_v3_l4_c15_n1000_svar_Pcd} to \ref{figs_t100_f3_v2_l3_c15_n1000_svar_numerr} and which is that the TM algorithm always, and usually very notably so, outperforms all the other algorithms. Indeed the TM algorithm is an exact implementation of bit MAP decoding. By definition, it minimizes the probability of error over variables.

Also, in general, we can see that the worst performing algorithm is, unsurprisingly, TM.1A.G though it is at par with the other algorithms for $\sigma$ small.

More specifically, in Fig. \ref{figs_t100_f3_v3_l4_c15_n1000_svar_Pcd}, we see for larger $\sigma$ that TM.1A and TM.2A.G perform similarly and better than TM.1A.f and TM.1A.MC which also perform the same. All perform very well for $\sigma<0.11$. In Fig. \ref{figs_t100_f3_v3_l4_c15_n1000_svar_numerr} we see the same threshold of $\sigma=0.11$ below which there is virtually no errors. Above this value we see that TM.1A and TM.2A.G remain very close, though the former slightly outperforms the latter for $\sigma>0.5$. Both TM.1A.f and TM.1A.MC perform very similarly and are outperformed by TM.1A.G for $\sigma>0.3$ though this could be linked to the way algorithms respond to higher values of $\sigma$. For very large values of $\sigma$, the limit value of half the total length (i.e. $50 = \frac{t}{2}$ in the current example) is explained by the fact that when estimation is impossible the algorithms always return the estimate $1$ for all positions which is the maximum of the prior distributions $\beta_t$. In the present case the probability that $A_i = 1$ is very close to $1/2$ thus the probability $1/2$ of ending up with the correct value.

In the case of $(\beta_u,\alpha_f)$, i.e. Figs. \ref{figs_t100_f3_v2_l4_c15_n1000_svar_Pcd} and \ref{figs_t100_f3_v2_l4_c15_n1000_svar_numerr}, we see that TM.2A.G performs very well compared to all the other algorithms which all perform very similarly, except for TM. The limit value in Fig. \ref{figs_t100_f3_v2_l4_c15_n1000_svar_numerr} is the worst case scenario of randomly falling on the correct value. It is the total length times the complementary probability of randomly picking a value and therefore is $\frac{t (c-1)}{c} = 93.33$ with the current parameters. 

By setting the $\alpha$ function to $\alpha_h$ in Figs. \ref{figs_t100_f3_v3_l3_c15_n1000_svar_Pcd} to \ref{figs_t100_f3_v2_l3_c15_n1000_svar_numerr}, we observe a similar behavior independently of the distribution $\beta$. Besides TM and TM.1A.G which behave according to the trends described previously, all algorithms perform very similarly. The only major difference is in the limit value for the average number of errors which is a function of $\beta$.

\begin{figure}[h!]
\begin{center}
\includegraphics[width=0.7\linewidth]{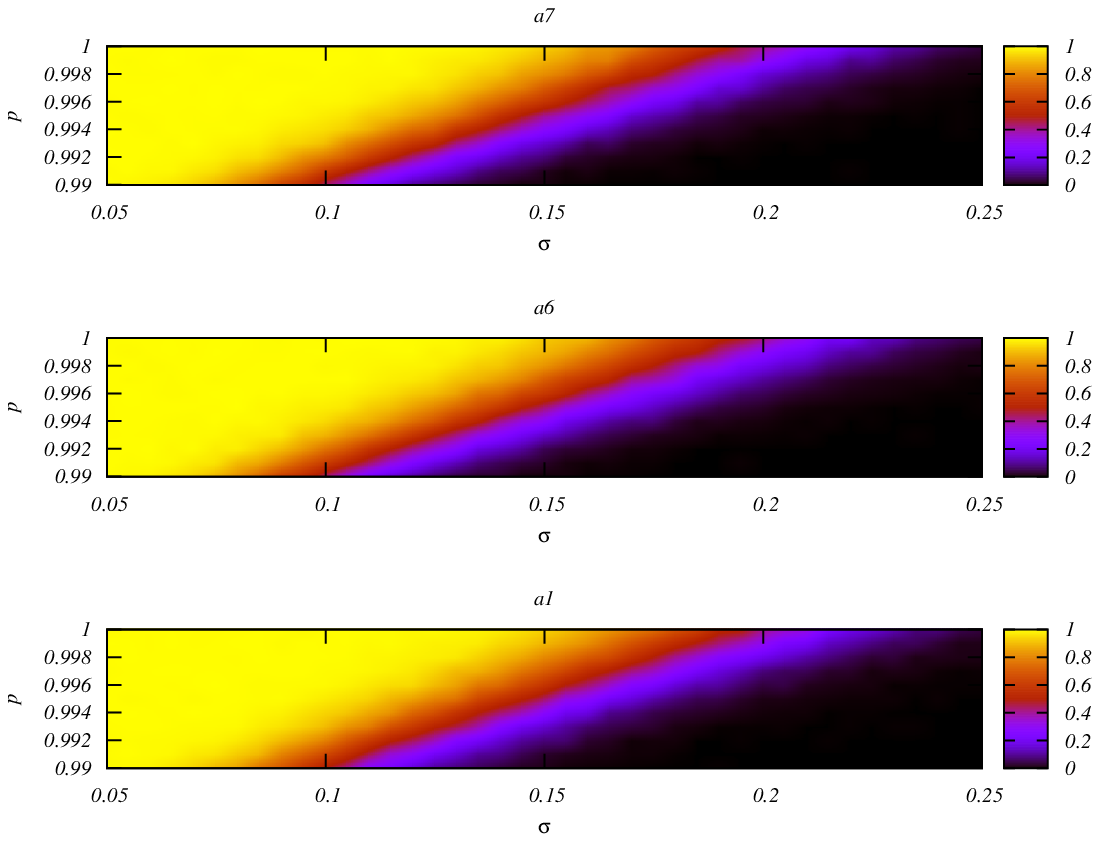}
\caption{Contour lines of the probability of correct decoding ($P_{cd}$) for various algorithms and versus the noise $\sigma$ and the incorporation rate $p$. Parameters are $t= 100$, $c=15$, $n=13$, using $\beta_g$ and $\alpha_r$ average over $1000$ samples and $N_f=500$ for TM.1A.MC.}
\label{figs_multiplot}
\end{center}
\end{figure}

\begin{figure}[h!]
\includegraphics[width=0.7\linewidth]{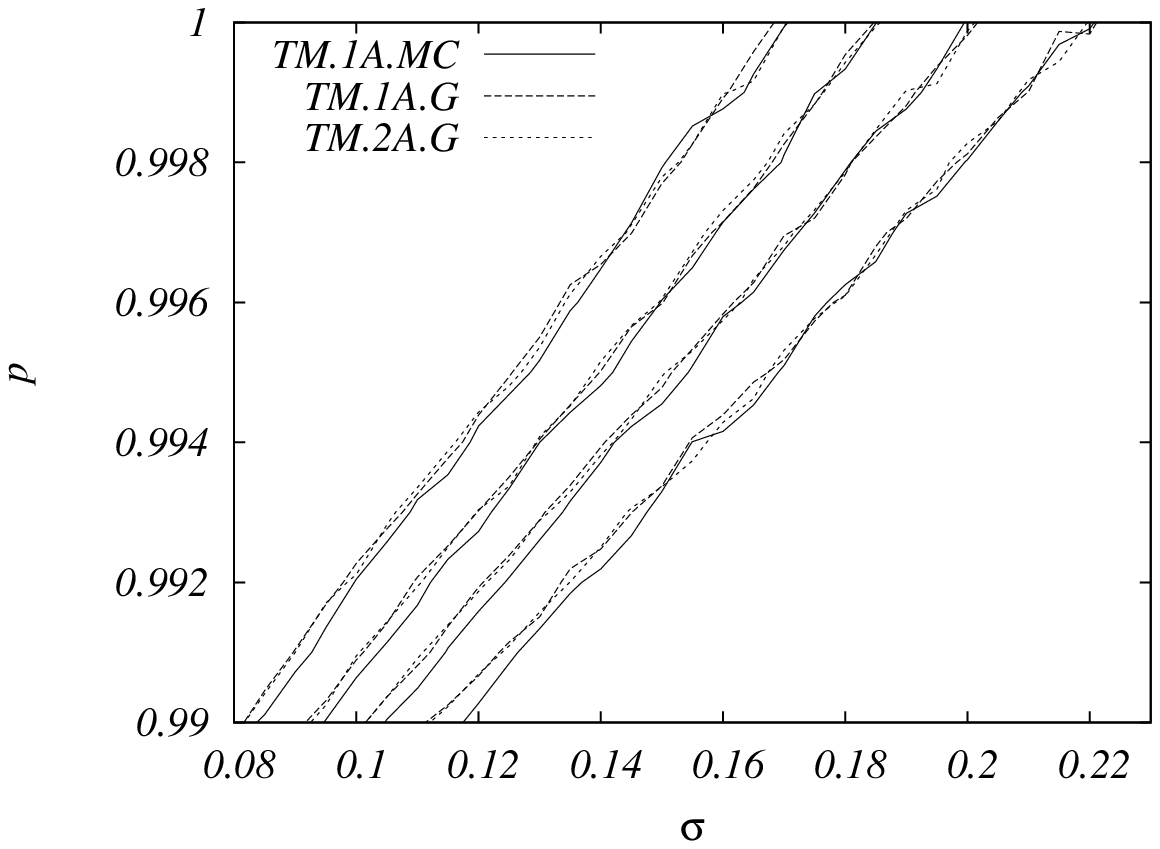}
\caption{Contour lines of $P_{cd}=0.8$, $0.6$, $0.4$ and $0.2$ from left to right for TM.1A.MC, TM.1A.G and TM.2A.G versus the noise $\sigma$ and the incorporation rate $p$. Parameters are $t= 100$, $c=15$, $n=13$, using $\beta_g$ and $\alpha_r$ average over $1000$ samples and $N_f=500$ for TM.1A.MC.}
\label{figs_Pcd_contour_t100_v1_l1_pvar_svar}
\end{figure}

Finally, in the case of $\alpha_r$ described in Eq. (\ref{eq_real_alpha}) and the distribution $\beta_g$ described in (\ref{equation_prob_a_priori}) with $q=1/2$, we have very similar performances for the three algorithms TM.1A.MC, TM.1A.G and TM.2A.G. The algorithms TM, TM.1A and TM.1A.f cannot be used with these parameters. Indeed, we have a relatively large memory $n=13$, which is kept the same for all values of $p$, and therefore TM does not fit in our computer memories and TM.1A and TM.1A.f would take several thousand years to compute a single sample on the computers used. These results are shown in Fig. \ref{figs_multiplot}. In this figure we show the interpolated contour lines of the probability of correct decoding $P_{cd} = 1 - P_{err}$ as a function of both $\sigma$ and the incorporation rate $p$. We show in more detail in Fig. \ref{figs_Pcd_contour_t100_v1_l1_pvar_svar} how performance increases with $p$ and how similar performance is for the three algorithms. It also shows, in this case for smaller values of $p$, that TM.1A.MC has a slight advantage over the two others. This advantage for TM.1A.MC with $\alpha_r$ will be confirmed in \ref{subsection_tvar}.

\begin{figure}[h!]
\includegraphics[width=0.45\linewidth]{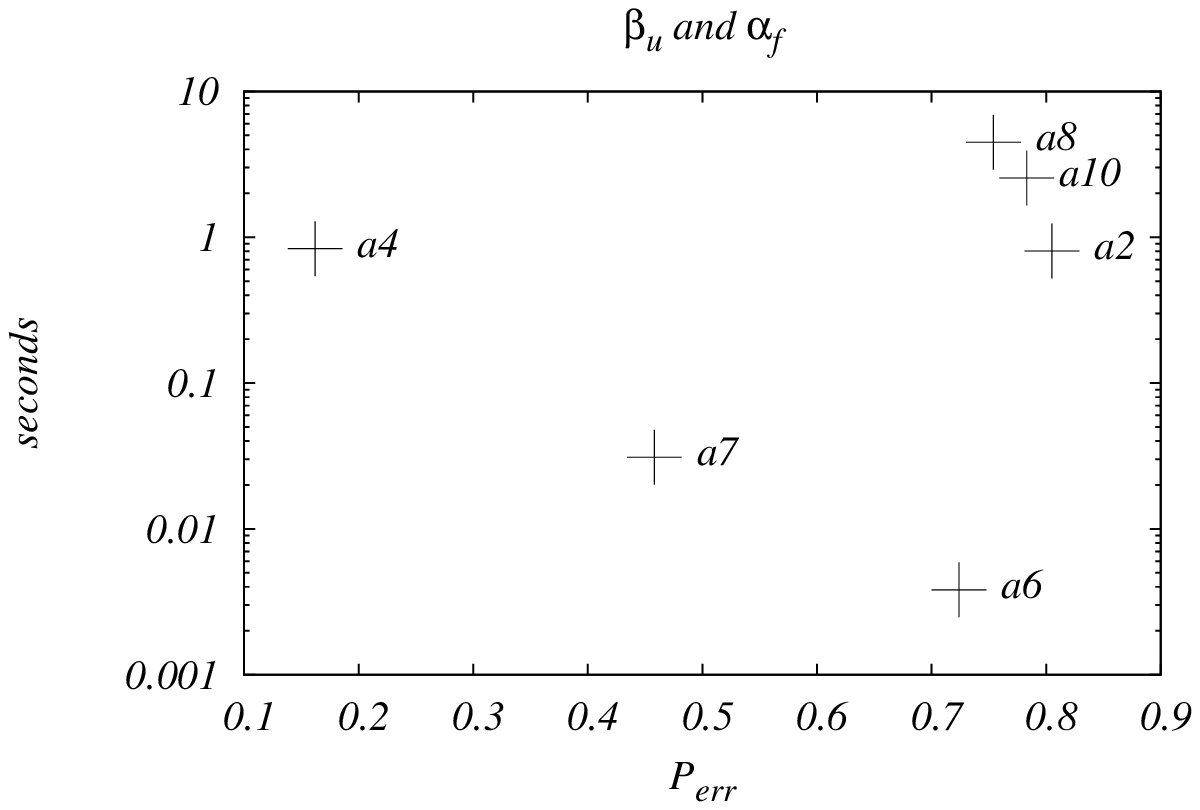}
\includegraphics[width=0.45\linewidth]{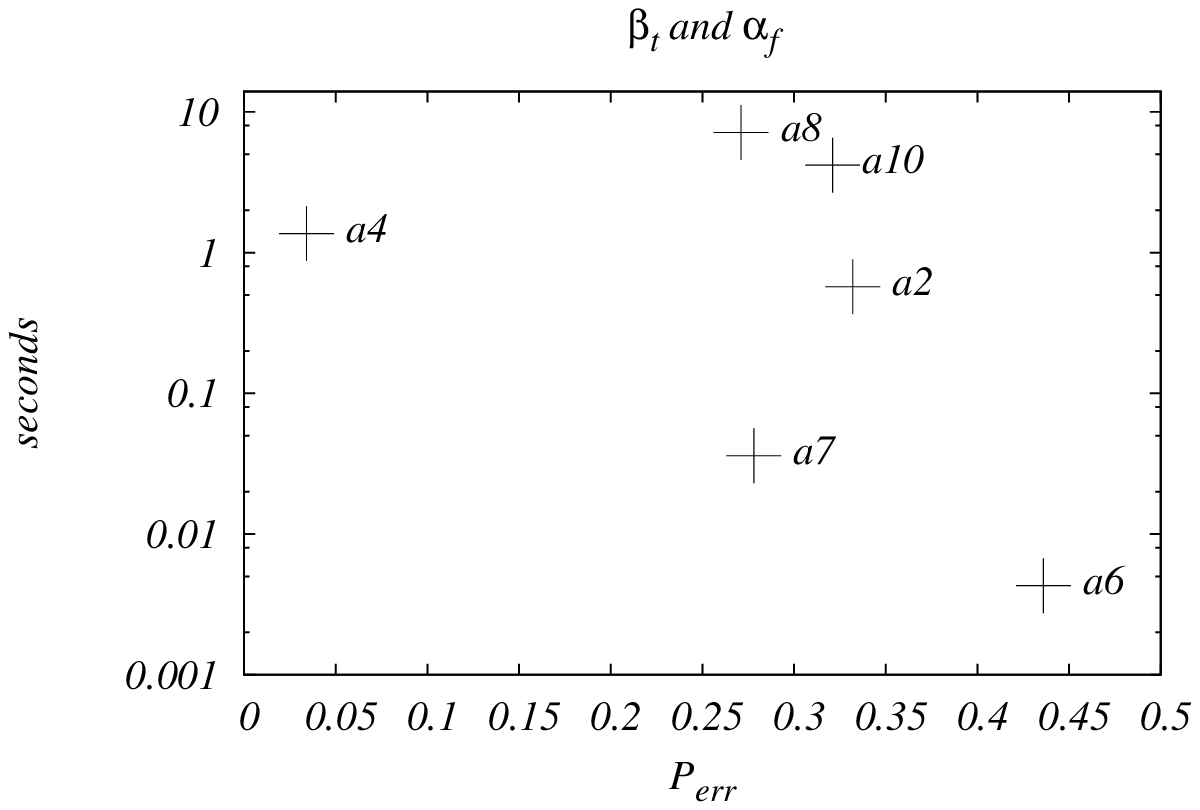}
\includegraphics[width=0.45\linewidth]{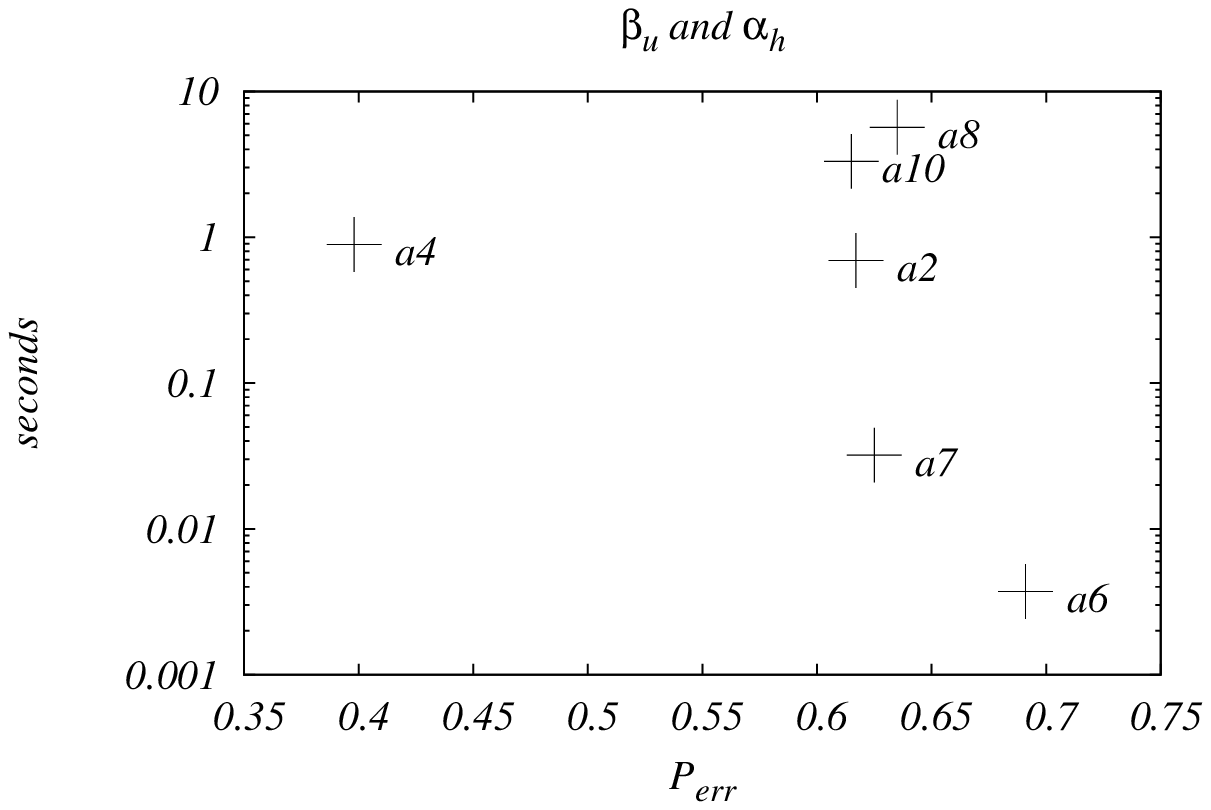}
\includegraphics[width=0.45\linewidth]{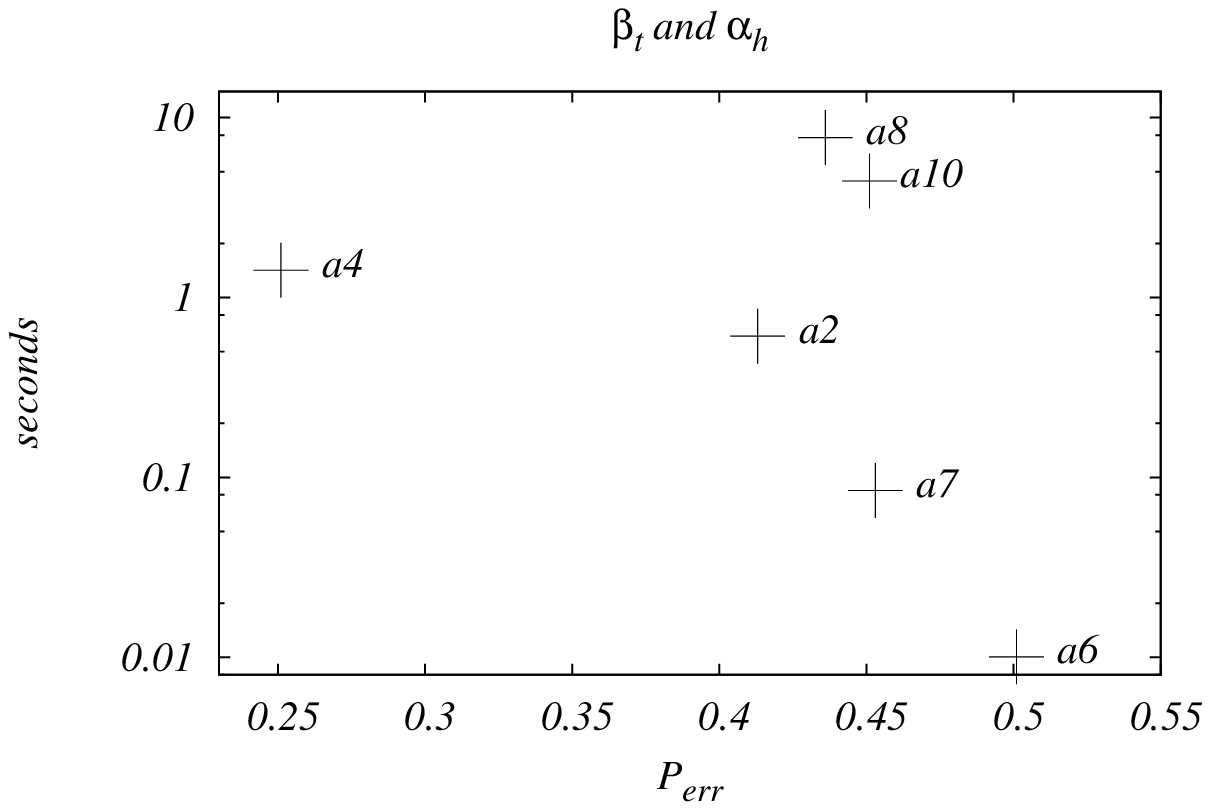}
\caption{Average time per sample on a log scale and in seconds vs $P_{err}$ for the various algorithms. With $\sigma=0.1$ when using $\alpha_h$ and $\sigma=0.18$ when using $\alpha_f$, $t=100$, $n=3$, $c=15$, average over $1000$ samples and $N_f=500$ for TM.1A.MC.}
\label{figs_mean_time_t100_f3_v2_l4_c15_n1000_svar}
\end{figure}

Finally, in Fig. \ref{figs_mean_time_t100_f3_v2_l4_c15_n1000_svar} we show the average computation time\footnote{Computations were performed on one of the following CPUs: Intel Core 2 Duo E6700 at 2.66GHz, Intel Core 2 Quad Q9550 at 2.83GHz and Intel Core 2 Duo E8500 at 3.16GHz. In spite of the variations in CPU speed, the magnitude of the computation times are always the same.} of a single sample according to the algorithm versus the the probability of error at a certain $\sigma$. The differences are huge, thus the log scale, and overall, TM.2A.G seems to give the best performance-time trade-off for $t=100$ and $n=3$. \\

\subsection{Memory length as parameter}  \label{subsection_fvar}

In this subsection, we consider the memory length $n$ as being the control parameter. We will be setting $\sigma$ large so as to never have a completely decoded chain and thus we will be comparing the average number of errors only.

\begin{figure}[h!]
\includegraphics[width=0.45\linewidth]{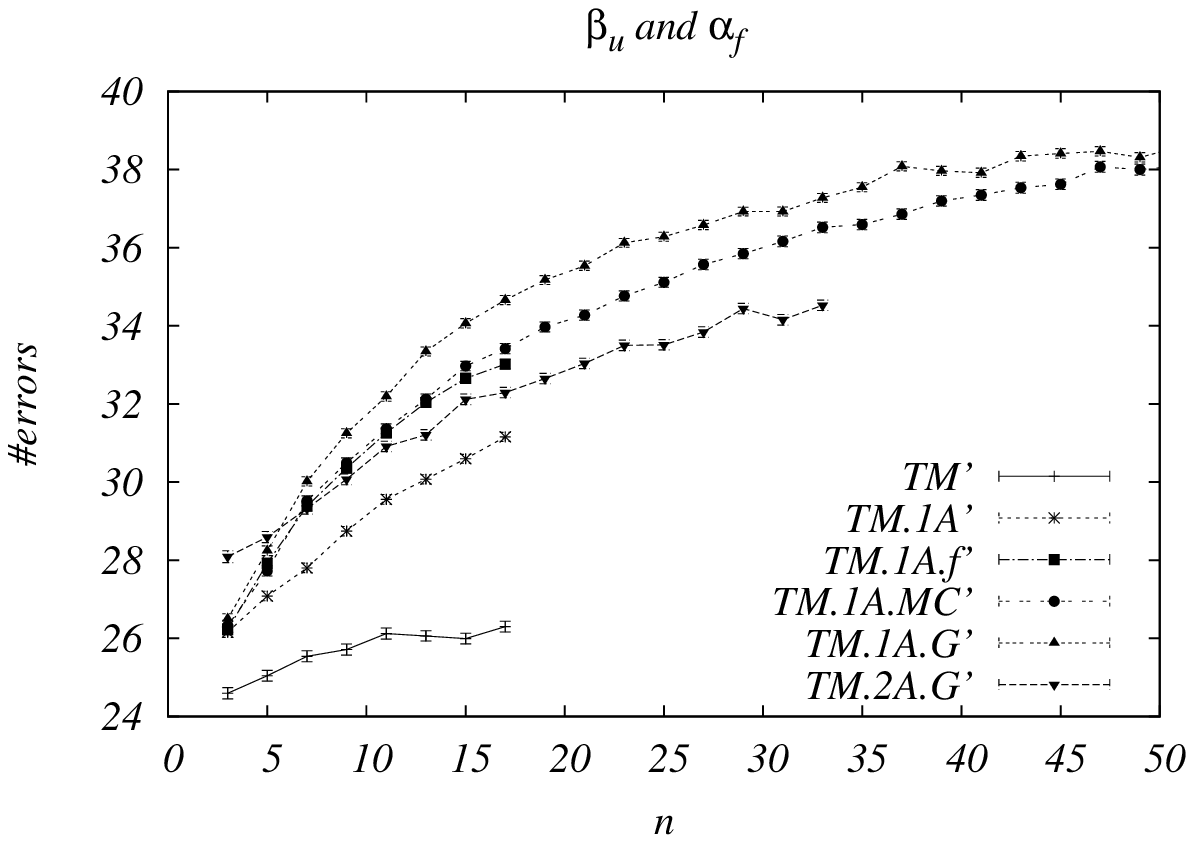}
\includegraphics[width=0.45\linewidth]{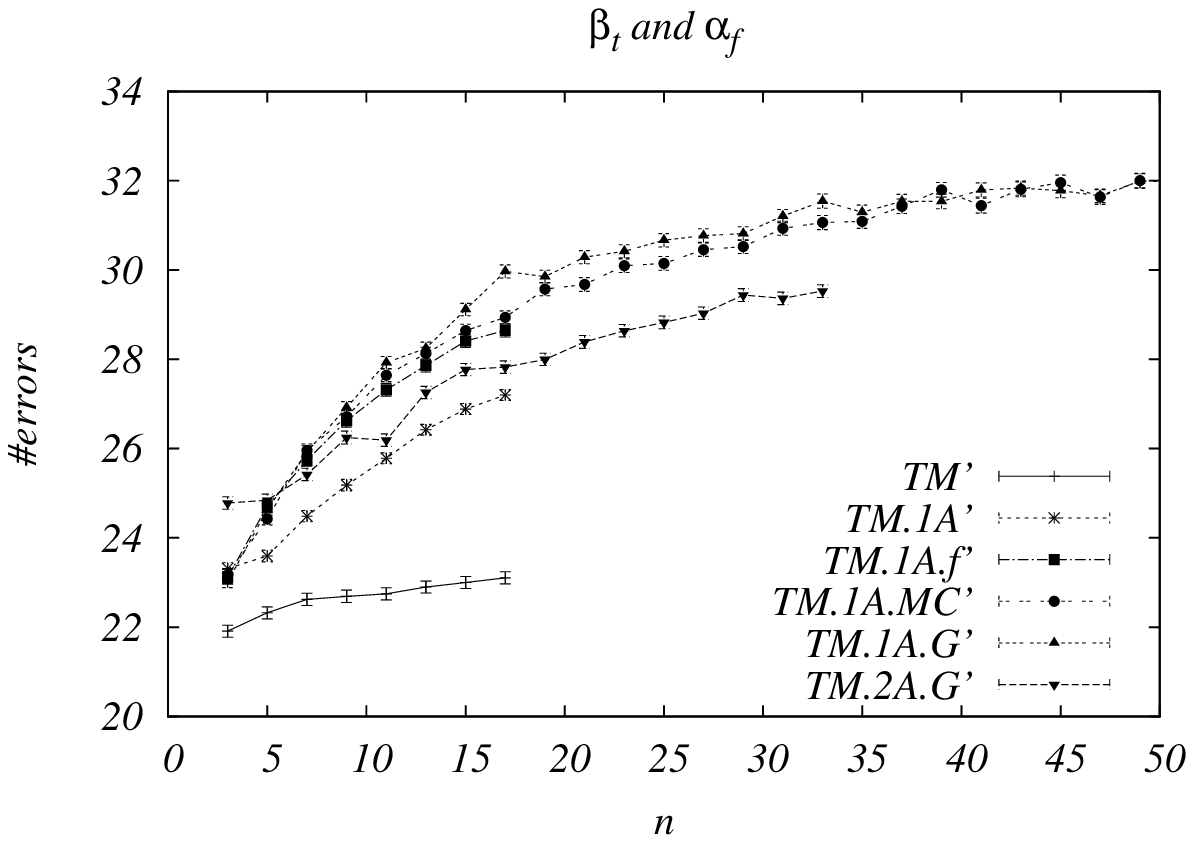}
\caption{Average number of errors ($\#errors$) for various algorithms vs memory length $n$. With $t=100$, $s=1$, $c=2$, average over $1000$ samples and $N_f=500$ for TM.1A.MC. }
\label{figs_t100_s1_c2_fvar}
\end{figure}

We first consider the limit case where the cutoff parameter is $c=2$. This will enable us to compare the results between all algorithms, even the ones exponential in $n$, though only for relatively small values of the parameter. The average number of errors for $\alpha_f$ are shown in Fig. \ref{figs_t100_s1_c2_fvar}. There are very little differences between the algorithms when we consider $\alpha_h$, thus we do not show these figures. 

As in the previous subsection, TM outperforms all other algorithms quite well. Of the other algorithms, TM.1A returns the smallest number of errors though it seems to be caught up by TM.2A.G for larger values of $n$. For both distributions $\beta_u$ and $\beta_t$, TM.1A.f and TM.1A.MC are very similar though there is a slight advantage to TM.1A.f. Finally, TM.1A.G performs very much like TM.1A.MC in the case where the distribution $\beta_t$ is used. On the other hand, it performs quite poorly, compared to TM.1A.MC, when $\beta_u$ is considered.

\begin{figure}[h!]
\includegraphics[width=0.7\linewidth]{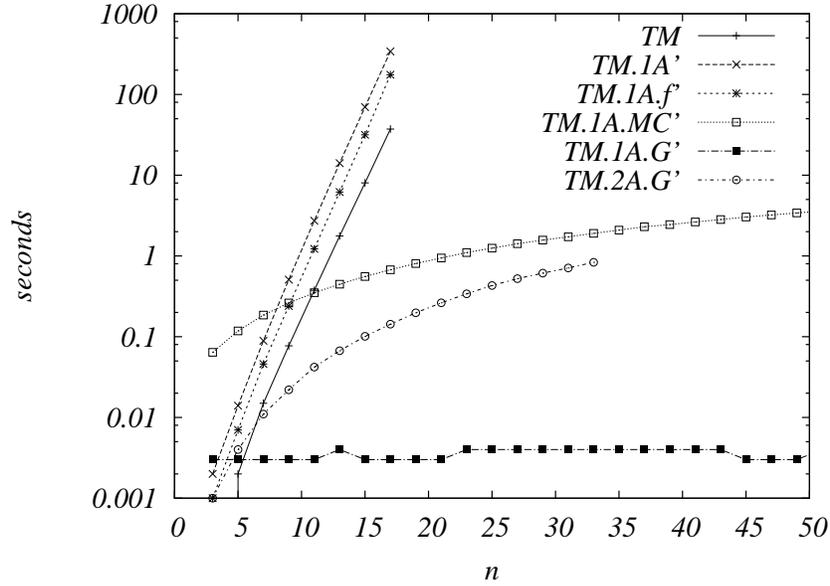}
\caption{Average time per sample on a log scale and in seconds vs memory length $n$. With $t=100$, $s=1$, $c=2$, using $\beta_u$ and $\alpha_f$, average over $1000$ samples and $N_f=500$ for TM.1A.MC. }
\label{figs_time_t100_v2_l4_s1_c2_fvar}
\end{figure}

In Fig. \ref{figs_time_t100_v2_l4_s1_c2_fvar} we show the various times per sample for the different algorithms for $\beta_u$ and $\alpha_f$. The times are exactly the same for the other possible combinations of $\beta$ and $\alpha$. This figure shows why we limit ourselves to $n=17$ as an upper bound for TM, TM.1A and TM.1A.f.

\begin{figure}[h!]
\includegraphics[width=0.45\linewidth]{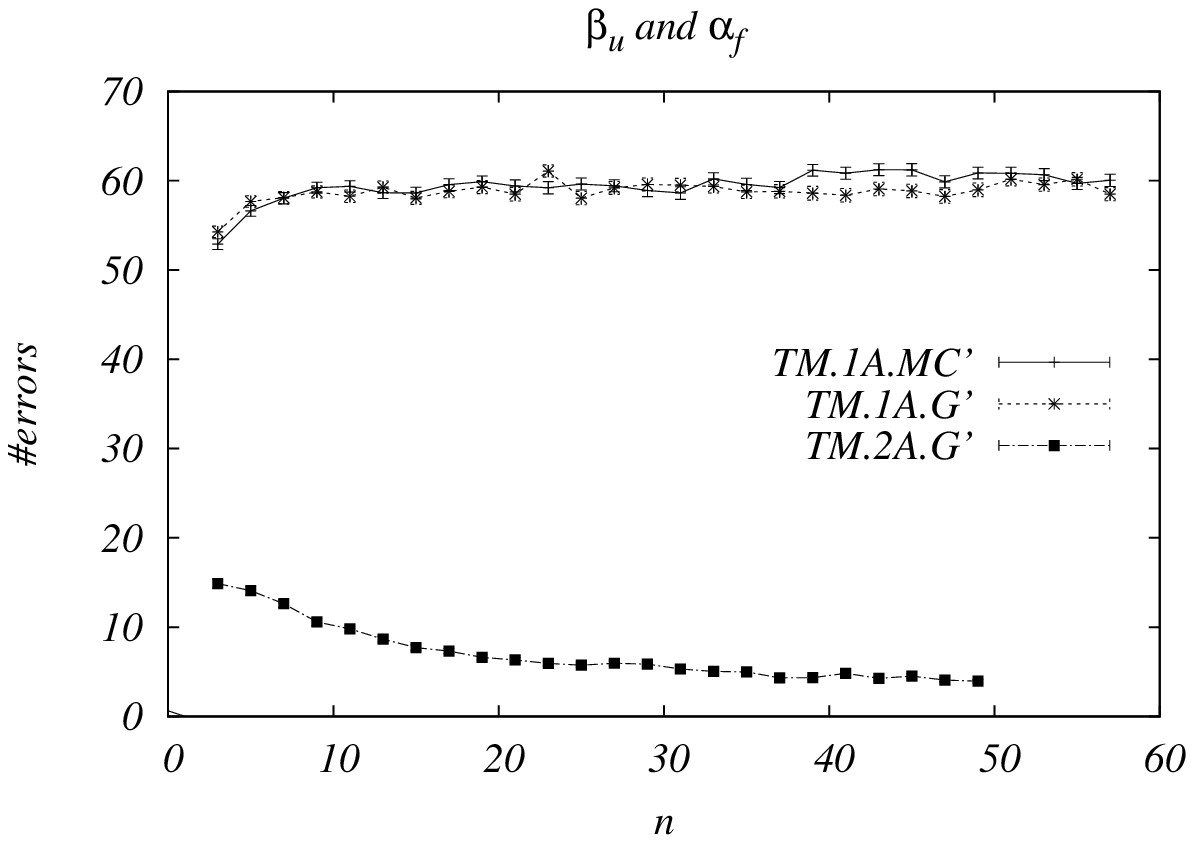}
\includegraphics[width=0.45\linewidth]{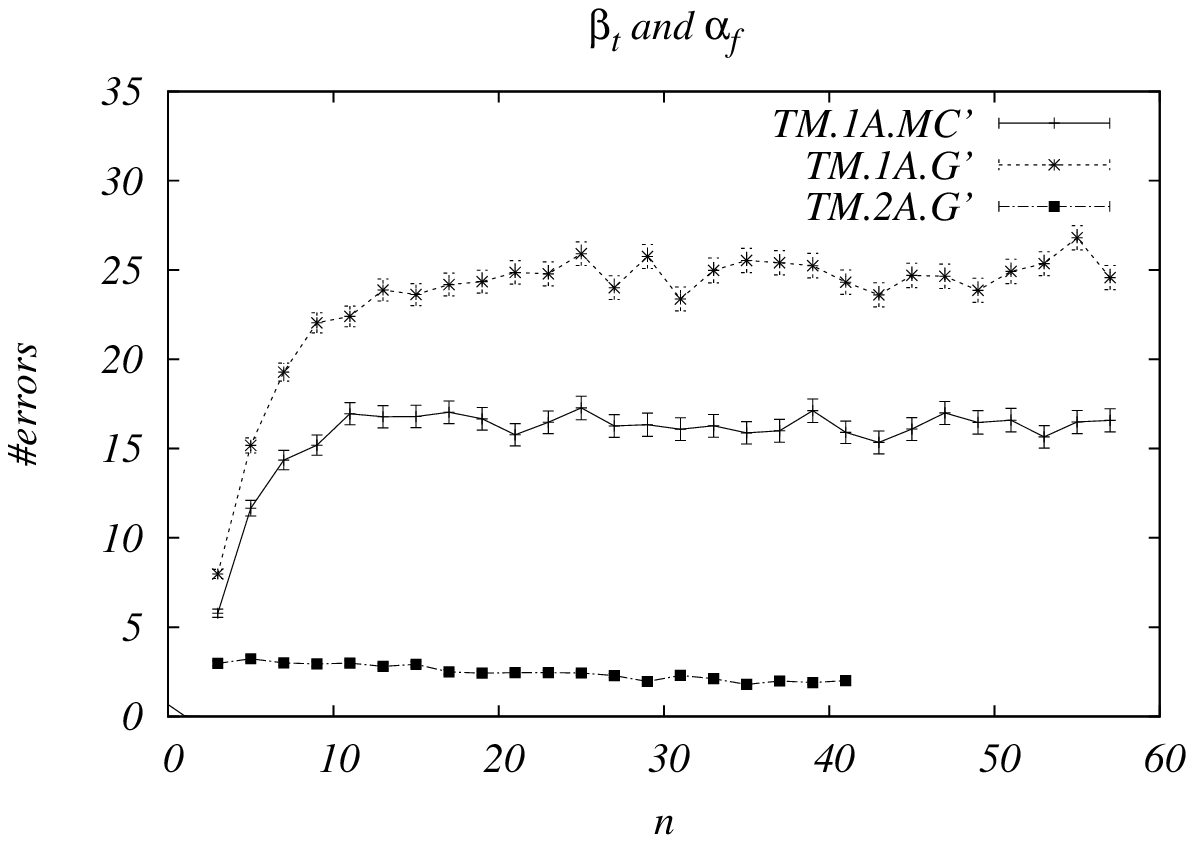}
\includegraphics[width=0.45\linewidth]{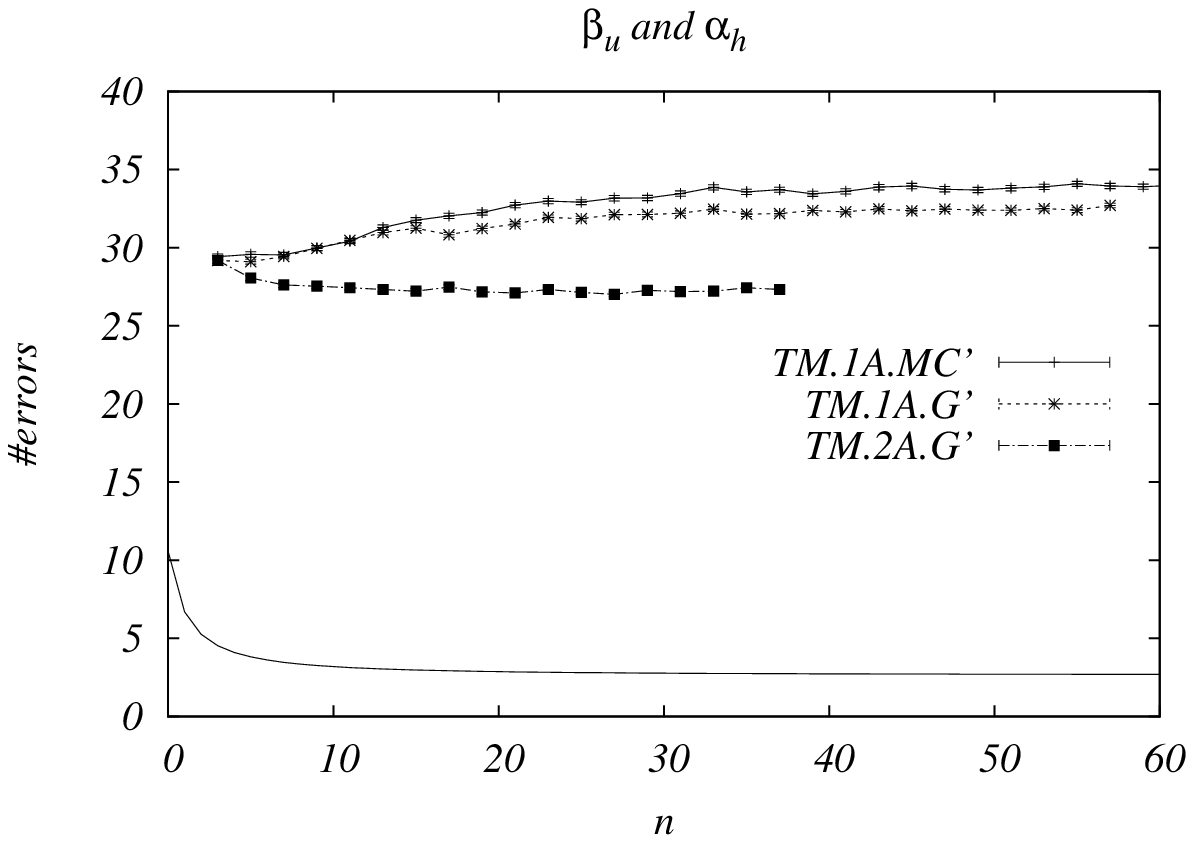}
\includegraphics[width=0.45\linewidth]{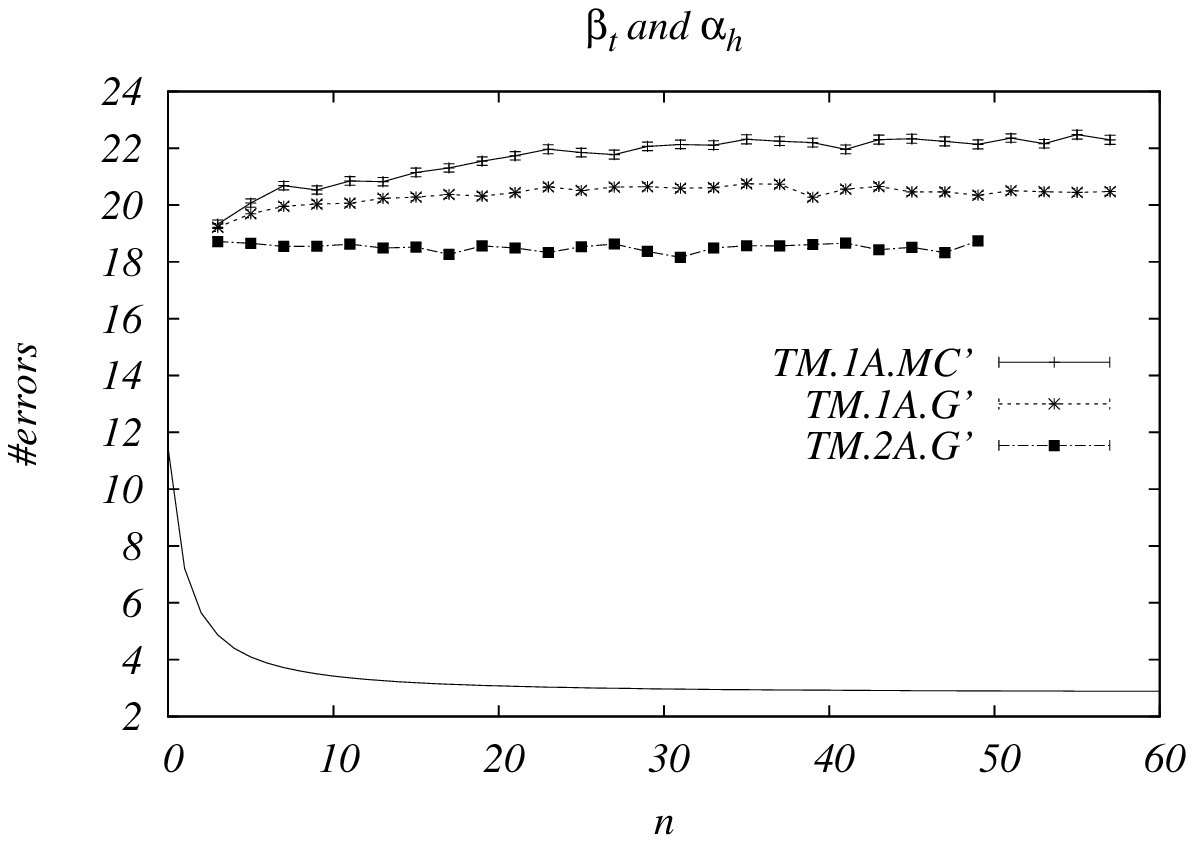}
\caption{Average number of errors ($\#errors$) for various algorithms vs memory length $n$. The full line is the analytical lower bound (almost always equal to zero in the upper two figures). With $t=100$, $s=0.2$, $c=15$, average over $1000$ samples and $N_f=500$ for TM.1A.MC for all figures. }
\label{figs_t100_s0.2_fvar}
\end{figure}

We now consider the case where $c=15$ in Fig. \ref{figs_t100_s0.2_fvar}. We show all four combinations of $\beta_u$ and $\beta_t$ with $\alpha_f$ and $\alpha_h$ for the algorithms TM.1A.MC, TM.1A.G and TM.2A.G. 

In all cases, TM.2A.G outperforms the two others and performs better as $n$ increases. In both cases where we consider the function $\alpha_f$ it performs much better than the two others. With $(\beta_u,\alpha_f)$, TM.1A.MC and TM.1A.G perform about the same, in all other combinations TM.1A.MC performs better. In general, these two algorithms reach a plateau value relatively quickly.

\begin{figure}[h!]
\includegraphics[width=0.7\linewidth]{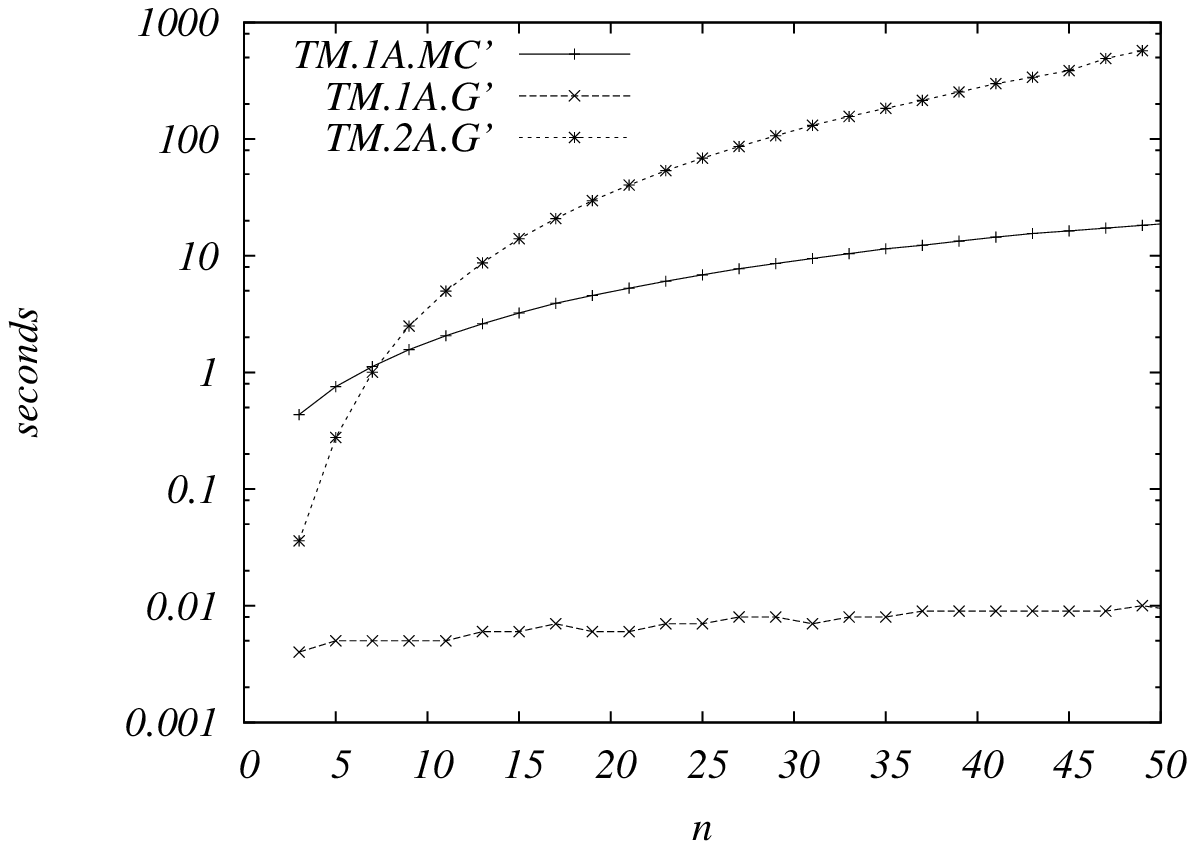}
\caption{Average time per sample on a log scale and in seconds vs memory length $n$. With $t=100$, $s=0.2$, $c=15$, using $\beta_u$ and $\alpha_f$, average over $1000$ samples and $N_f=500$ for TM.1A.MC. }
\label{figs_time_t100_v2_l4_s0.2_c15_fvar}
\end{figure}

In Fig. \ref{figs_time_t100_v2_l4_s0.2_c15_fvar} we have again shown the computation times for the various algorithms. Even though TM.2A.G systematically outperforms the two others in Fig. \ref{figs_t100_s0.2_fvar}, its computation time grows sub-exponentially with $n$, though it does become large.\\

\subsection{Total length as parameter} \label{subsection_tvar}

In this subsection we study the influence of the total length $t$.

\begin{figure}[h!]
\includegraphics[width=0.7\linewidth]{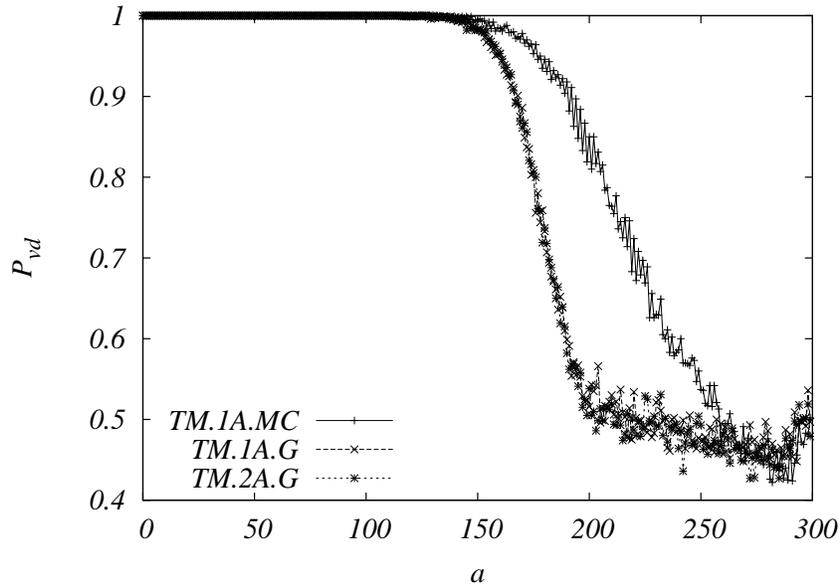}
\caption{Probability of correctly decoding each position $a$ ($P_{vd}$) in a chain of total length $t=300$ for various algorithms. With $p=0.99$, $n=13$, $c=15$, $\sigma=0.04$, using $\beta_t$ and $\alpha_r$, average over $1000$ samples and $N_f=500$ for TM.1A.MC.}
\label{figs_p0.99_v3_l1_f13_Pvd}
\end{figure}

The first figure in this subsection, Fig. \ref{figs_p0.99_v3_l1_f13_Pvd}, shows the probability of correctly decoding each position $a$ ($P_{vd}$) in a chain of total length $t$. We can see that for $a$ bigger than a certain threshold close to $a=150$, $P_{vd}$ rapidly decreases to a value close to $0.5$. This can be explained by looking at the graph of $\alpha_r$ in Fig. \ref{fig_alpha} where we can see that, for $p=0.99$ and $a=140$, the maximum of $\alpha_r(i,a)$ is no longer for $i=a$.

\begin{figure}[h!]
\includegraphics[width=0.7\linewidth]{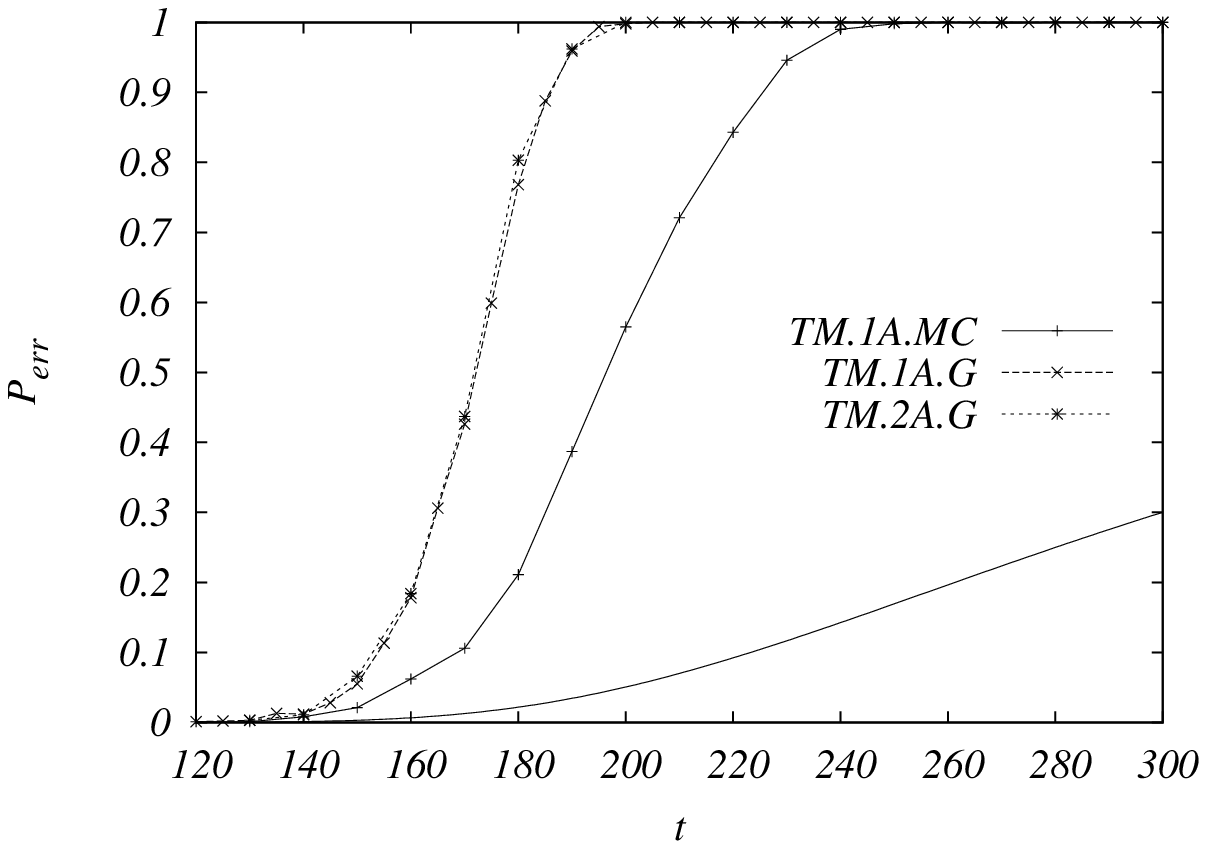}
\caption{Probability of error ($P_{err}$) for various algorithms vs full chain length $t$. The full line is the analytical lower bound. With $p=0.99$, $n=13$, $c=15$, $\sigma=0.04$, using $\beta_t$ and $\alpha_r$, average over $1000$ samples and $N_f=500$ for TM.1A.MC.}
\label{figs_p0.99_v3_l1_f13_tvar}
\end{figure}

\begin{figure}[h!]
\includegraphics[width=0.7\linewidth]{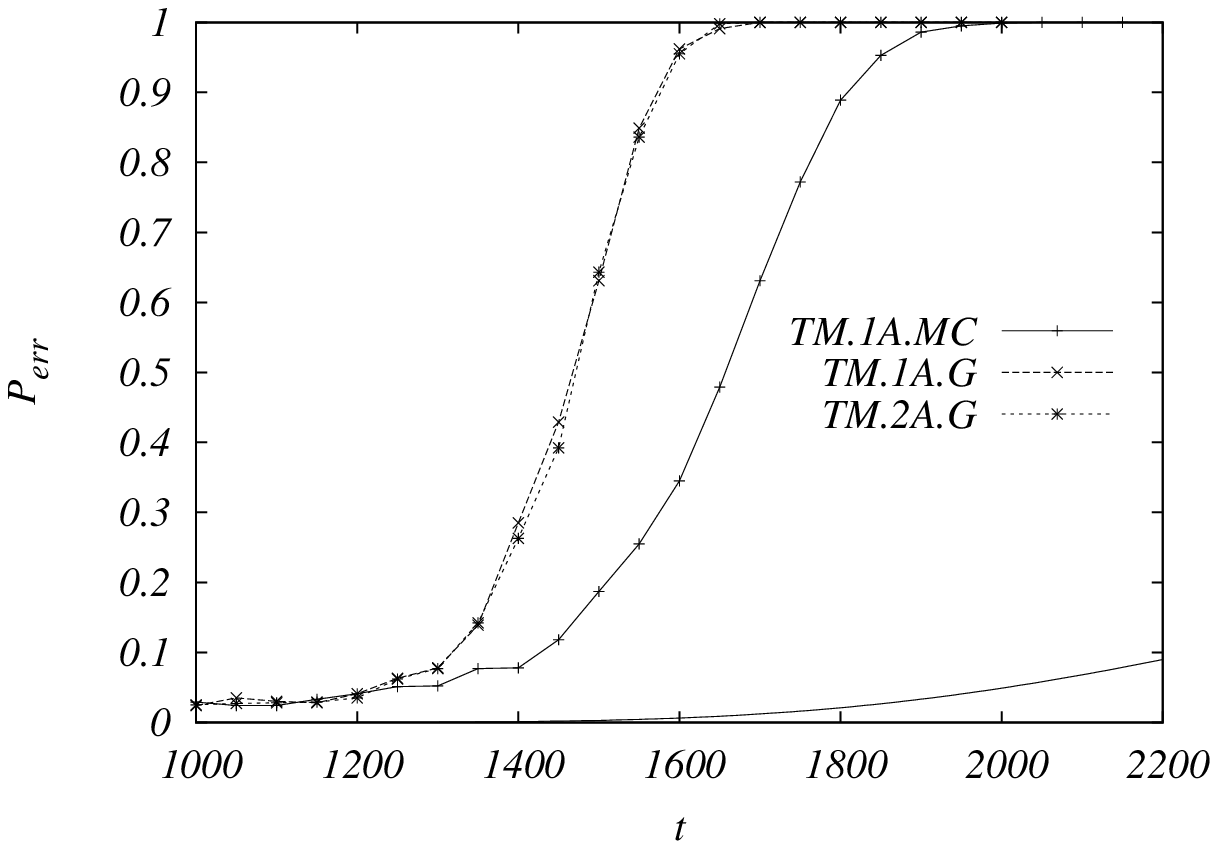}
\caption{Same as \ref{figs_p0.99_v3_l1_f13_tvar} with $p=0.999$, $n=13$, $c=15$, $\sigma=0.04$, using $\beta_g$ and $\alpha_r$, average over $1000$ samples and $N_f=500$ for TM.1A.MC.}
\label{figs_p0.999_v1_l1_f13_Pcd_vs_t}
\end{figure}

In Fig. \ref{figs_p0.99_v3_l1_f13_tvar} we show the probability of error $P_{err}$ for various algorithms with $\beta_t$ and $p=0.99$ as a function of the total length $t$. In Fig. \ref{figs_p0.999_v1_l1_f13_Pcd_vs_t} we show the same thing with $\beta_g$ and $p=0.999$. They confirm what was shown in Fig. \ref{figs_p0.99_v3_l1_f13_Pvd}. In this particular regime, i.e. with $\alpha_r$, TM.1A.MC slightly outperforms TM.1A.G and TM.2A.G. Furthermore, the two latter algorithms seem to behave exactly in the same way to variations of $t$ when $\alpha_r$ is used.

\begin{figure}[h!]
\includegraphics[width=0.7\linewidth]{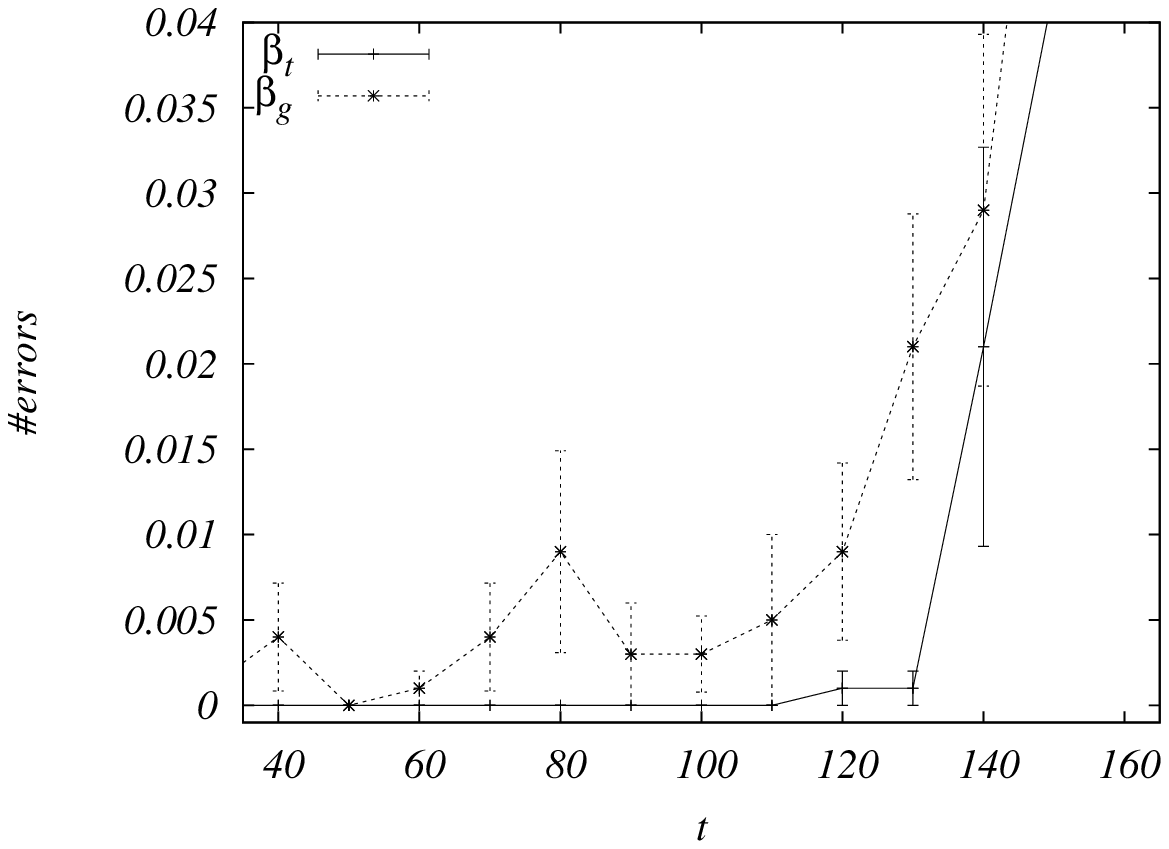}
\caption{Average number of errors ($\#errors$) vs $t$ for TM.1A.MC. With $p=0.99$, $n=13$, $c=15$, $\sigma=0.04$, using $\alpha_r$, average over $1000$ samples and $N_f=500$.}
\label{figs_p0.99_l1_f13_numerr}
\end{figure}

Finally, in Fig. \ref{figs_p0.999_v1_l1_f13_Pcd_vs_t} we see that $P_{err}$ never reaches zero. This behavior is detailed in Fig. \ref{figs_p0.99_l1_f13_numerr} where we see that for very small $t$ the number of errors is non vanishing when $\beta_g$ is used while it is when $\beta_t$ is used. The reason for this is that when the chain is generated using $\beta_g$ then $A_a > c$ with positive probability: the decoder fails in these cases. This is the only notable difference when using $\beta_g$ instead of $\beta_t$.

\begin{figure}[h!]
\includegraphics[width=0.7\linewidth]{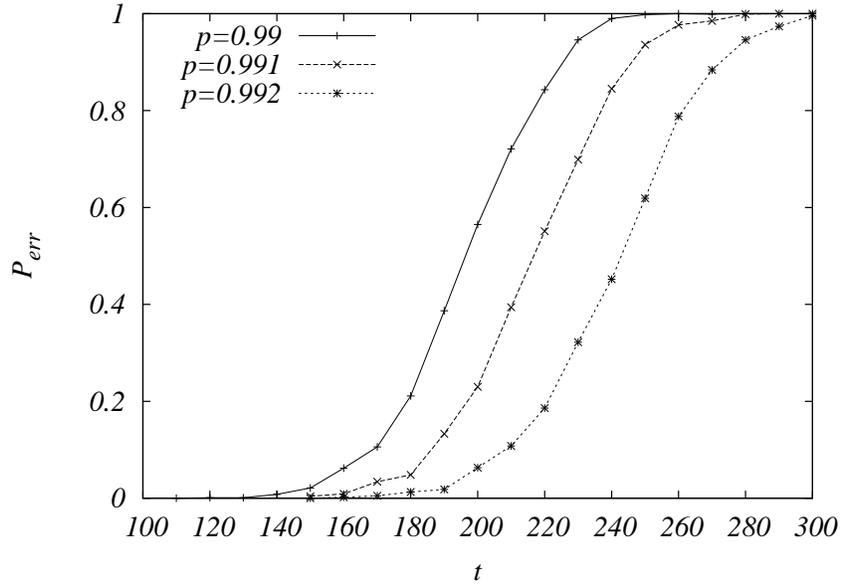}
\caption{Probability of error ($P_{err}$) for TM.1A.MC vs full chain length $t$ for various values of the incorporation rate $p$. With $n=13$, $c=15$, $\sigma=0.04$, using $\beta_t$ and $\alpha_r$, average over $1000$ samples and $N_f=500$.}
\label{figs_a2_v3_l1_f13_tvar_pvar}
\end{figure}

In Fig. \ref{figs_a2_v3_l1_f13_tvar_pvar} we plot the probability of error as a function of $t$ for various values of the incorporation rate $p$ for $\beta_t$ and $\alpha_r$. As $p$ increases, the performance does increase as well.\\

\subsection{Comparison to real data} \label{subsection_real}

In this final section, we use parameter values inspired by real data. These values (used in \cite{eltoukhy_gamal}) are obtained from \cite{eltoukhy.phd} which presents pyrosequencing data generated from two sets of tests and where a maximum likelihood sequence detection (MLSD) algorithm is developed. These sets are obtained from two separate sequencing systems, the first set of tests is obtained on a Pyrosequencing PSQ96MA system for a set of $5$ templates ranging from $55$ to $224$ base sub-sequences. The second test run consisted of $10,000$ data sets extracted using a $454$ Genome Sequencer 20 (GS20) system. These two machines are examples of two separate generations of sequencing systems having different precisions. The mathematical model which was developed to apply the MLSD algorithm was one of the inspirations for this work to which we applied an additional set of approximations and which we generalized.

They estimate that the incorporation rate is $p=0.9955$ for the PSQ96MA system and $p=0.9987$ for the GS20, while the read error introduces a Gaussian noise with a standard deviation of $\sigma = 0.08$ in both cases. In our approach, we have considered a modified base sequence composed of a two letter alphabet in place of the $4$ letters that compose DNA strands. The output being read as the number of repetitions of a single base in both cases (HP sub-sequences), the main difference in both approaches is seen in the modification of the memory function $\alpha$ (\ref{eq_real_alpha}) which has a simplified expression from that of the similar relation defined in \cite{eltoukhy.phd}. Though we estimate that this consideration does not modify significantly the read length and therefore that the behavior of our $2$-base model will be similar to that of the real $4$-base sequences. One thing that might lack is the non-specific incorporation rate that involves false positives. This difference can result in errorless read lengths that are higher than in experimental data, though its estimated value smaller than $0.02$ will result in only a slight increase. These issues will be further addressed in a future communication, here we will limit ourselves to presenting a few results on the behavior of our model using the parameters introduced at the beginning of this paragraph.

\begin{figure}[h!]
\includegraphics[width=0.7\linewidth]{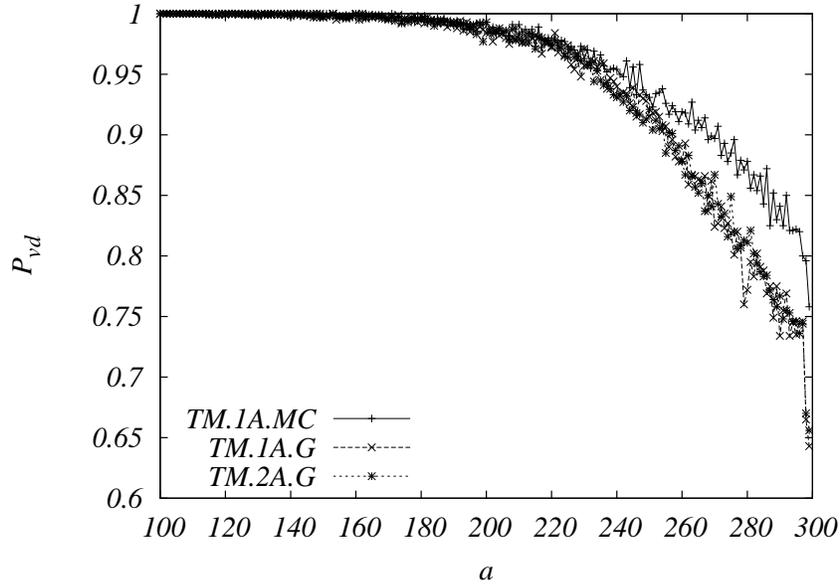}
\caption{Probability of correctly decoding each position $a$ ($P_{vd}$) in a chain of total length $t=300$ for various algorithms. With $p=0.9955$, ($n=11$), $c=15$, $\sigma=0.08$, using $\beta_g$ and $\alpha_r$, average over $1000$ samples and $N_f=500$ for TM.1A.MC.}
\label{figs_p0.9955_s0.08_t300_i500_n1000}
\end{figure}

\begin{figure}[h!]
\includegraphics[width=0.7\linewidth]{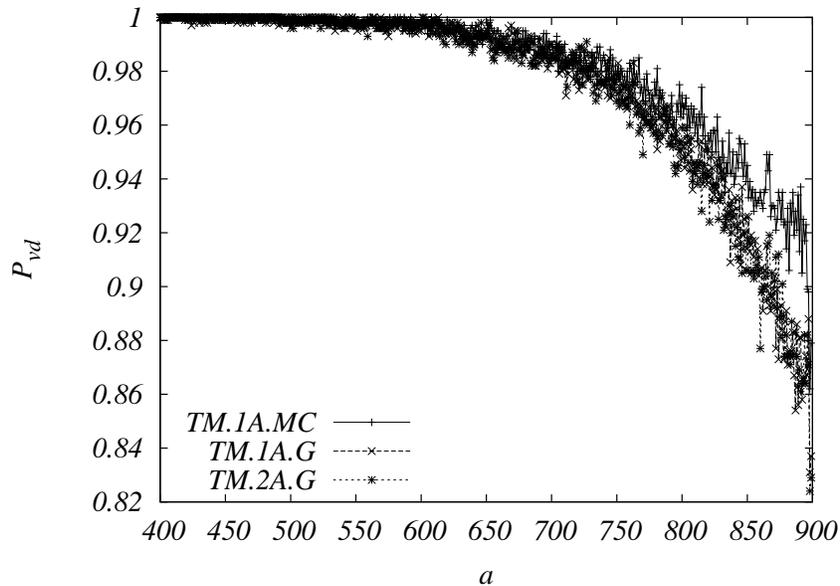}
\caption{Same as \ref{figs_p0.9955_s0.08_t300_i500_n1000} with $p=0.9987$, ($n=9$).}
\label{figs_p0.9987_s0.08_t900_i500_n1000}
\end{figure}

In figure \ref{figs_p0.9955_s0.08_t300_i500_n1000} we have plotted the probability of correctly decoding each position of a chain of length $t=300$ for $p=0.9955$ for three algorithms. This would emulate extracting sequences using the PSQ96MA system. As already mentioned in previous sections, TM.1A.MC performs better than the Gauss algorithms with these parameters for higher values of the position $a$ within the chain. In figure \ref{figs_p0.9987_s0.08_t900_i500_n1000} we plot the same probability for a chain of total length $t=900$ with $p=0.9987$ and which, in this case, would emulate extraction on a GS20 system. In \cite{eltoukhy.phd}, the MLSD algorithm was capable of correctly reading $170$ base sub-sequences out of $208$ and $205$ out of $228$ of the longest two templates ran on the PSQ96MA and all bases for templates of length $168$ on the GS20. By comparing these results to our own, we can say that they are at least consistent. Indeed, in figure \ref{figs_p0.9955_s0.08_t300_i500_n1000} we have $P_{vd} > 0.9$ for values of $a<230$. We can say this since the presented algorithms do not depend directly on the total length $t$ but only on the position $a$. Furthermore, for a total length of $t=300$ the average number of errors encountered during the decoding process of the TM.1A.MC algorithm is $ 7.815 \pm 0.153$ over $1000$ samples. Similar results are also true concerning data presented in figure \ref{figs_p0.9987_s0.08_t900_i500_n1000} although all we can say in comparison to \cite{eltoukhy.phd} is that we do indeed decode correctly chains of total length $168$. Also, they predict that correct decoding is possible for sequences of length greater than $500$ which we easily obtain (we have an average number of errors of $9.044 \pm 0.164$ in decoding chains of length $t=900$ with TM.1A.MC).

We conclude that our data is in good concordance with results obtained experimentally in \cite{eltoukhy.phd}, though we emphasize again that direct testing of our algorithms still needs to be performed.\\

%
%

\section{Conclusion} \label{section_conclusion}

We defined three low complexity algorithms based on the original high complexity transfer matrix algorithm (TM) described in \ref{subsection_algo_tm}. Of these three, two are first order approximations: the Monte Carlo algorithm (TM.1A.MC) and the Gauss algorithm (TM.1A.G) described respectively in \ref{subsubsection_mc} and \ref{subsection_algo_gauss}. The final algorithm is a second order approximation that is based on the Gauss algorithm (TM.2A.G) and is described in \ref{subsection_algo_two_point}. The performances of these algorithms were studied in the previous section \ref{section_results} where we show that the second order approximation with Gauss (TM.2A.G) is the best performing algorithm when the memory $n$ is small and the first order approximation with Monte Carlo (TM.1A.MC) performs best when the memory $n$ is large.\\

Besides its direct application to hidden Markov Models of higher order, the description introduced in this paper has yet to be tested on real pyrosequencing data. The context being different and the variations that need to be brought to our approach being quite substantial, this issue will be addressed directly in a future communication though preliminary comparison to similar work is promising.\\

%
%
\appendix

\section{Gauss algorithm} \label{appendix_gauss}

As stated in section \ref{subsection_algo_gauss}, we assume that the variable $X_i^{(a)}=\sum_{  \substack{ j=a-n \\ j \ne i }}^{a} \alpha (j,a) x_j $ present in $\Psi_a$ at step $a$ (Eq. \ref{eq_psi}) can be approximated with a Gaussian random variable. 

There are two steps for each iteration to derive the marginal distribution of $\underline{X} = \{ x_1, \ldots , x_t \}$ using this method.

We suppose the iterative process of the transfer matrix has brought us to position $a$. 
Then, the first step is to calculate the mean and variance of the variable $X_a=\sum_{i=a-n}^{a-1} \alpha(i,a) x_{i}$ under the Gaussian approximation.
 These can be written, respectively, as
\begin{eqnarray}
 \mu_X = \E ( X ) &=& \sum_{i=a-n}^{a-1} \alpha(i,a) ~ \E^{(a-1)} ( x_{i} ) \,, \\
\sigma_X^2 = \var (X) &=& \sum_{i=a-n}^{a-1} \alpha(i,a)^2 ~ \var^{(a-1)} ( x_i ) \,,
\end{eqnarray}
where $\E^{(a-1)} ( x_{i} )$ and $\var^{(a-1)} ( x_i )$ are the expectation and variance calculated at position $i$ using the distributions $\nu_i^{(a-1)} (x_i)$ which where obtained at the previous iteration.

We can then define the Gaussian variable $x$ such that its probability density function is
\begin{equation}
 P (x) =  \frac{1}{\sqrt{ 2 \pi \sigma_X^2 }} \exp \left[  - \frac{1}{2 \sigma_X^2} (x - \mu_X )^2 \right] ,
\end{equation}
and finally, the probability of having value $x_a$ at step $a$ is
\begin{equation}
 \nu_a^{(a)} (x_a) = \frac{\beta (x_a)}{\mathcal{N}} \int dx P(x) \exp \left[  - \frac{1}{2 \sigma^2} [ y_a - x - \alpha(a,a) x_a ]^2 \right], \label{eq_integral_marg_first_step}
\end{equation}
which yields
\begin{equation}
 \nu_a^{(a)} (x_a) = \frac{\beta (x_a)}{\mathcal{N}}  exp \left[  - \frac{1}{2 ( \sigma^2 + \sigma_X^2 ) } [ Y_a - \alpha(a,a) x_a - \mu_X ]^2  \right],
\label{equation_gauss_Z_distribution}
\end{equation}
where $\mathcal{N}$ is a normalization constant.

We then keep 
\begin{eqnarray}
 \E^{(a)} (x_a) &=& \sum_{x=1}^c x  \nu_a^{(a)} (x) ,  \\
\var^{(a)} (x_a) &=& \sum_{x=1}^c x^2  \nu_a^{(a)} (x)  - \E^{(a)} (x_a)^2 ,
\end{eqnarray}
for the next step and for subsequent iterations. \\

The second step is then to notice that for all $i \in \{ a-n, \ldots , a-1 \}$ we can rewrite Eq. (\ref{eq_first_order_forwards}) as
\begin{equation}
 \nu_i^{(a)} (x_i) = \frac{\nu_i^{(a-1)} (x_i)}{\mathcal{N}} \left( \prod_{\substack{ j=1 \\ j \ne i }  }^{a-1} \sum_{x_j} \nu_j^{(a-1)} (x_j)  \right) \sum_{x_a} \beta (x_a) \exp \left[ - \frac{1}{2 \sigma^2} \left( Y_a - \alpha (i,a) x_i - \sum_{  \substack{ j=1 \\ j \ne i }}^{a} \alpha (j,a) x_j    \right)^2 \right] ,
\end{equation}
where $\mathcal{N}$ is a normalization factor. 

Then we define step and position dependent mean and variance of the variables 
$X_i^{(a)} = \sum_{  \substack{ j=1 \\ j \ne i }}^{a} \alpha (j,a) x_j $  as
\begin{eqnarray}
 \mu_{X_i} = \E ( X_i ) &=& \sum_{  \substack{ j=a-n \\ j \ne i } }^{a} \alpha(j,a) ~ \E_j^{(a-1)} ( x_{j} ) \,, \\
\sigma_{X_i}^2 = \var (X_i) &=& \sum_{  \substack{ j=a-n \\ j \ne i } }^{a} \alpha(j,a)^2 ~ \var_j^{(a-1)} ( x_j ) \,,
\end{eqnarray}
where we artificially set $\E_a^{(a-1)} ( x_{a} ) = \E_a^{(a)} ( x_{a} )$ and the same for the variance. 
From these and a similar expression to Eq. (\ref{eq_integral_marg_first_step}) we recover Eq. (\ref{eq_marginal_gauss}).

For subsequent iterations, we keep $\E_i^{(a)} (x_i)$ and $\var_i^{(a)} (x_i)$. \\

\section{Two point algorithm} \label{appendix_two_point}

As stated in section \ref{subsection_algo_two_point} we consider two point interactions only over closest neighbors. Since we can reconstruct one-point marginals as marginals of the two-point marginals, we need only compute the latter. We use a similar method to the one used in section \ref{subsection_algo_gauss} through approximating our sum variable $X$ with a Gaussian random variable with mean and variance which are expressed in the following two step procedure.

We start by calculating $\mu_{X_{a-1,a}}$ and $\sigma^2_{X_{a-1,a}}$, respectively mean and variance of $X_{a-1,a}=\sum_{i=1}^{a-2} \alpha(i,a) x_{i}$ :
\begin{eqnarray}
 \mu_{X_{a-1,a}} = \E ( X_{a-1,a} ) &=& \sum_{   k=1  }^{a-2} \alpha(k,a) ~ \E^{(a-1)} [ x_{k} | x_{a-1} ] \,, \label{eq_mu_Z_tp}\\
\sigma_{X_{a-1,a}}^2 = \var (X_{a-1,a}) &=& \sum_{  k=1 }^{a-2} \alpha(k,a)^2 ~ \var^{(a-1)} ( x_k | x_{a-1} ) + \nonumber \\
& &+2 \sum_{  1 \leq k < k' < a-1 } \alpha(k,a) \alpha(k',a) ~ \operatorname{Cov}^{(a-1)} (x_k x_{k'} |x_{a-1} )  \,,
\end{eqnarray}
where
\begin{eqnarray}
\var^{(a-1)} ( x_k | x_{a-1} ) &=& \E^{(a-1)} [ x_{k}^2 | x_{a-1} ] - \E^{(a-1)} [ x_{k} | x_{a-1}]^2 , \\
\operatorname{Cov}^{(a-1)} (x_k x_{k'} |x_{a-1} )  &= & E^{(a-1)} [ x_{k} x_{k'} | x_{a-1} ] - \E^{(a-1)} [ x_{k} | x_{a-1}] \E^{(a-1)} [ x_{k'} | x_{a-1}] \,,
\end{eqnarray}
and which yield
\begin{equation}
\nu^{(a)}_{a-1,a} (x_{a-1}, x_{a}) = \frac{\nu_{a-1}^{(a-1)} (x_{a-1}) \beta (x_a)}{\mathcal{N}} ~ e^{\left[ - \frac{1}{2 (\sigma^2 + \sigma_{X_{a-1,a}}^2)} [ Y_a - \mu_{X_{a-1,a}} - \alpha(a-1,a) x_{a-1} - \alpha(a,a) x_a ]^2    \right]} \,. \label{eq_z_am1_a}
\end{equation}

Once these values obtained, we define $\nu^{(a)}_{a} (x_{a}) = \sum_{x_{a-1} =1}^c \nu^{(a)}_{a-1,a} (x_{a-1}, x_{a})$ and 
\begin{eqnarray}
m_a &=& \sum_{x_a=1}^c  x_a ~ \nu^{(a)}_{a} (x_{a}) \,, \\
v_a &=& \sum_{x_a=1}^c  x_a^2 ~ \nu^{(a)}_{a} (x_{a}) - m_a^2 \,,
\end{eqnarray}
which are injected into the second step of the procedure which is to compute $\nu^{(a)}_{i,i+1} (x_{i}, x_{i+1})$ for $i < a-1$. Again, we approximate the sum variable $X_{i,i+1} = \sum_{ \substack{j=1 \\ j \ne i,i+1 }}^{a} \alpha(j,a) x_{j}$ with a Gaussian random variable with mean and variance respectively
\begin{eqnarray}
\mu_{X_{i,i+1}} = m_a + \E ( X_{i,i+1} ) &=&  m_a + \sum_{  \substack{ k=1 \\ k \neq i,i+1} }^{a-1} \alpha(k,a) ~ \E^{(a-1)} [ x_{k} | x_{i}, x_{i+1} ] \,, \\ \label{eq_mean_two_point}
\sigma_{X_{i,i+1}}^2 = v_a + \var (X_{i,i+1}) &=& v_a + \sum_{  \substack{ k=1 \\ k \neq i,i+1} }^{a-1} \alpha(k,a)^2 ~ \var^{(a-1)} ( x_k | x_{i}, x_{i+1} ) + \nonumber \\ 
& &+2 \sum_{  \substack{ 1 \leq k < k' \leq a \\ k,k' \neq i,i+1 } } \alpha(k,a) \alpha(k',a) ~ \operatorname{Cov}^{(a-1)} (x_k x_{k'} |x_{i}, x_{i+1} )  \,, \label{eq_var_two_point}
\end{eqnarray}
which enable us to compute
\begin{equation}
\nu^{(a)}_{i,i+1} (x_{i}, x_{i+1}) = \frac{\nu^{(a-1)}_{i,i+1} (x_{i}, x_{i+1})}{\mathcal{N}} ~ e^{\left[ - \frac{1}{2 (\sigma^2 + \sigma_{X_{i,i+1}}^2)} [ Y_a - \mu_{X_{i,i+1}} - \alpha(i,a) x_{i} - \alpha(i+1,a) x_{i+1} ]^2    \right]} \,, \label{eq_z_i_ip1_appendix}
\end{equation}
which is the expression in (\ref{eq_z_i_ip1}).\\

In writing these equations (\ref{eq_mu_Z_tp} - \ref{eq_z_i_ip1_appendix}) we used several expressions that we need to give more detail to. First of all, in the situation where a distribution can be decomposed on one dimensional factor graph, we can express the conditional probabilities as
\begin{equation}
\mu ( {x_j | x_i } ) = \sum_{x_{i+1}, \ldots , x_{j-1}} \mu (x_{i+1}|x_i) \ldots \mu (x_{j}|x_{j-1}) \,,
\end{equation}
where $\mu$ expresses probabilities in general.

We can then use the previous expression to compute the necessary expectations for Eqs. (\ref{eq_z_am1_a}) and (\ref{eq_z_i_ip1_appendix}), in all following cases $i<j$ and $k<k'$ when necessary,
\begin{eqnarray}
\E^{(a)} [x_k | x_i ] &=& \sum_{x_k = 1}^c x_k ~ \nu^{(a)} ( x_k | x_i) \,, 
\end{eqnarray}
\begin{eqnarray}
\E^{(a)} [x_k x_{k'} | x_i ] &=& \E^{(a)} [x_k | x_i ] \E^{(a)} [x_{k'} | x_i ] ~~~~~~~~~~~~~~~~~~~~~~~~~\textrm{ if } k<i<k'  \,, \\
 &=& \sum_{x_k, x_{k'} = 1}^c x_k x_{k'}  ~ \nu^{(a)} ( x_i | x_k) \nu^{(a)} ( x_k | x_{k'}) ~~~~~~ \textrm{ if } i<k\,, \\
 &=& \sum_{x_k, x_{k'} = 1}^c x_k x_{k'}  ~ \nu^{(a)} ( x_i | x_{k'}) \nu^{(a)} ( x_{k'} | x_{k}) ~~~~~ \textrm{ otherwise}\,, 
\end{eqnarray}
\begin{eqnarray}
 \E^{(a)} [x_k | x_i,x_j  ] &=& \E^{(a)} [x_k | x_i ] ~~~~~~~~~~~~~~~~~~~~~~~~~~~~~~~ \textrm{ if } k<i\,, \\
 &=& \E^{(a)} [x_k | x_j ] ~~~~~~~~~~~~~~~~~~~~~~~~~~~~~~~ \textrm{ if } k>j, \\
 &=& \sum_{x_k = 1}^c x_k ~ \frac{\nu^{(a)} ( x_i | x_k) \nu^{(a)} ( x_k | x_{j})}{\nu^{(a)} ( x_i | x_j) } ~~~~~~ \textrm{ otherwise}\,, 
 \end{eqnarray}
 \begin{eqnarray}
 \E^{(a)} [x_k x_{k'} | x_i,x_j  ] &=&  \E^{(a)} [x_k x_{k'} | x_i ] ~~~~~~~~~~~~~~~~~~~~~~~~~~~~~~~~~~~~~~~ \textrm{ if } k'<i\,, \\
 &=&  \E^{(a)} [x_k x_{k'} | x_j ] ~~~~~~~~~~~~~~~~~~~~~~~~~~~~~~~~~~~~~~~ \textrm{ if } k>j\,, \\
 &=& \E^{(a)} [x_k | x_i ] \E^{(a)} [x_{k'} | x_i,x_j  ] ~~~~~~~~~~~~~~~~~~~~~~~~~ \textrm{ if } k<i<k'<j\,, \\
 &=&  \E^{(a)} [x_k | x_i,x_j  ] \E^{(a)} [x_{k'} | x_j ] ~~~~~~~~~~~~~~~~~~~~~~~~~ \textrm{ if } i<k<j<k'\,, \\
 &=&  \E^{(a)} [x_k | x_i,x_j  ] \E^{(a)} [x_{k'} | x_i,x_j  ] ~~~~~~~~~~~~~~~~~~~~~ \textrm{ if } k<i<j<k'\,, \\
 &=& \sum_{x_k, x_{k'} = 1}^c x_k x_{k'}  ~  \frac{\nu^{(a)} ( x_i | x_{k}) \nu^{(a)} ( x_k | x_{k'})  \nu^{(a)} ( x_{k'} | x_{j}) }{ \nu^{(a)} ( x_i | x_{j}) } \\
 & & ~~~~~~~~~~~~~~~~~~~~~~~~~~~~~~~~~~~~~~~~~~~~~~~~~~~~~~~~~~ \textrm{ if } i<k<k'<j \,. \nonumber
\end{eqnarray}\\

Finally, in most cases, the values of $\nu^{(a)}_{i,i+1} (x_{i}, x_{i+1})$ obtained numerically, though they do contain the information we are looking for, are not a distribution. That is, more often than not do we get $\sum_{x_{i+1}} \nu^{(a)}_{i,i+1} (x_{i}, x_{i+1}) \neq \sum_{x_{i-1}} \nu^{(a)}_{i-1,i} (x_{i-1}, x_{i})$. To overcome this we introduce a new set of \emph{real} distributions $\mu^{(a)}_{i,i+1} (x_{i}, x_{i+1})$ such that the Kullback-Leibler divergences, regarded as distances, between the $\mu^{(a)}_{i,i+1}$ and the $\nu^{(a)}_{i,i+1}$ are minimized. That is, we wish to minimize the quantity
\begin{equation}
\sum_i D \left( \mu^{(a)}_{i,i+1} || \nu^{(a)}_{i,i+1} \right) = \sum_i \sum_{x_i,x_{i+1}} \mu^{(a)}_{i,i+1} (x_{i}, x_{i+1}) \log \left( \frac{\mu^{(a)}_{i,i+1} (x_{i}, x_{i+1})}{\nu^{(a)}_{i,i+1} (x_{i}, x_{i+1})} \right) \,,
\end{equation}
given the constraints
\begin{equation}
\sum_{x_{i+1}} \mu^{(a)}_{i,i+1} (x_{i}, x_{i+1}) = \sum_{x_{i-1}} \mu^{(a)}_{i-1,i} (x_{i-1}, x_{i}) ~~, \forall i , x_i \,. \label{eq_two_point_lagrange_constraint}
\end{equation} 
That is, by using Lagrange multipliers, we wish to minimize
\begin{equation}
\Lambda = \sum_{i} \left[ D \left( \mu^{(a)}_{i,i+1} || \nu^{(a)}_{i,i+1} \right) - \sum_{x_i}  \lambda_{i} (x_i) \left( \sum_{x_{i+1}}  \mu^{(a)}_{i,i+1} (x_i, x_{i+1}) - \sum_{x_{i-1}} \mu^{(a)}_{i-1,i} (x_{i-1}, x_{i}) \right) \right] \,,
\end{equation}
which, after differentiation, results in
\begin{equation}
\log \left( \frac{\mu^{(a)}_{i,i+1} (x_{i}, x_{i+1})}{\nu^{(a)}_{i,i+1} (x_{i}, x_{i+1})} \right) =  \lambda_{i} (x_i)  - \lambda_{i+1} (x_{i+1}) \,,
\end{equation}
for all $i$ and all couples $\{ x_i, x_{i+1}\}$. We define $\gamma_i (x_i) = e^{ \lambda_{i} (x_i)}$, hence we have
\begin{equation}
\mu^{(a)}_{i,i+1} (x_{i}, x_{i+1}) = \frac{\gamma_i (x_i)}{ \gamma_{i+1} (x_{i+1})} \nu^{(a)}_{i,i+1} (x_{i}, x_{i+1}) \,.
\end{equation}
Furthermore
\begin{eqnarray}
\sum_{x_{i-1}} \mu^{(a)}_{i-1,i} (x_{i-1}, x_{i}) &=& \frac{1}{\gamma_i (x_i)} \sum_{x_{i-1}}  \nu^{(a)}_{i-1,i} (x_{i-1}, x_{i}) \gamma_{i-1} (x_{i-1}) \label{eq_two_point_gamma_im1_i} \,,\\
\sum_{x_{i+1}} \mu^{(a)}_{i,i+1} (x_{i}, x_{i+1}) &=& \gamma_i (x_i) \sum_{x_{i+1}} \frac{\nu^{(a)}_{i,i+1} (x_{i}, x_{i+1})}{\gamma_{i+1} (x_{i+1}) } \label{eq_two_point_gamma_ip1_i} \,,
\end{eqnarray}
where the left hand side of Eqs. (\ref{eq_two_point_gamma_im1_i}) and (\ref{eq_two_point_gamma_ip1_i}) are equal by the constraint (\ref{eq_two_point_lagrange_constraint}) and thus by equating the right hand sides we have
\begin{equation}
\gamma_i (x_i)^2 = \frac{\sum_{x_{i-1}} \nu^{(a)}_{i-1,i} (x_{i-1}, x_{i}) \gamma_{i-1} (x_{i-1})}{ \sum_{x_{i+1}} \nu^{(a)}_{i,i+1} (x_{i} , x_{i+1}) / \gamma_{i+1} (x_{i+1}) } \,.
\end{equation}
We initiate the procedure by setting all values of $\gamma_i (x_i)$ in the right hand side to the value $1$. The values obtained on the left hand side are then reinjected into the right hand side iteratively until the values of the $\gamma_i (x_i)$ of all $i$ and $x_i$ are equal on both sides of the equation. To prevent the appearance of static cycles, we produce the update at each step by randomly choosing the order in which we take the functions $\gamma_i$ for all $i$.

Finally, the $\mu^{(a)}_{i,i+1}$ are injected into the next step $a+1$, taking the place of the $\nu^{(a)}_{i,i+1}$.

%
%

\begin{acknowledgements}
I wish to thank Andrea Montanari and Guilhem Semerjian for their unbounded help and support with this work.
\end{acknowledgements}

\bibliographystyle{ieeetr}

\makePDFTitles{Recovering the state sequence of hidden Markov models using mean-field approximations}{Antoine Sinton}
\end{document}